\documentclass[12pt]{article}
\pdfoutput=1

\usepackage{tabu}
\usepackage{geometry}

\usepackage{putex}
\usepackage{psfrag}
\usepackage{amssymb}
\usepackage{amsmath}
\usepackage{empheq}
\usepackage{epsf}
\usepackage{graphicx}
\usepackage{epstopdf}

\usepackage{enumerate}
\usepackage{cite}
\usepackage{caption}
\usepackage{subcaption}

\usepackage[utf8]{inputenc}
\usepackage{braket}
\usepackage{titlesec}

\usepackage{psfrag}
\usepackage{color}
\definecolor{darkblue}{rgb}{0.1,0.1,.7}
\usepackage[colorlinks, linkcolor=darkblue, citecolor=darkblue, urlcolor=darkblue, linktocpage]{hyperref}

\hypersetup{breaklinks}
\usepackage{comment}

\def\<{\langle}
\def\>{\rangle}
\newcommand \s {\sigma}
\newcommand   \f  {\phi}
\newcommand   \p  {\psi}
\newcommand{\bea}{\begin{eqnarray}}
\newcommand{\eea}{\end{eqnarray}}

\def\sec{\section}

\def\O{{\cal O}}

\def\ssec{\subsection}

\def\sec{\section}

\newcommand{\grp}[1]{\mathrm{#1}}

\newcommand{\grSU}{\grp{SU}}
\newcommand{\grSO}{\grp{SO}}

\newcommand{\grSL}{\grp{SL}}

\newcommand {\be} {\begin {equation}}
\newcommand {\ee} {\end {equation}}

\newcommand {\bes} {\begin {equation*}}
\newcommand {\ees} {\end {equation*}}

\newcommand{\beq}{\begin{equation}}
\newcommand{\eeq}{\end{equation}}

\def\be{ \begin{equation} }
\def\ee{ \end{equation} }

 \newmuskip\pFqmuskip

\newcommand*\pFq[6][8]{%
  \begingroup 
  \pFqmuskip=#1mu\relax
  \mathcode`\,=\string"8000
  \begingroup\lccode`\~=`\,
  \lowercase{\endgroup\let~}\pFqcomma
  {}_{#2}F_{#3}{\left[\genfrac..{0pt}{}{#4}{#5};#6\right]}%
  \endgroup
}
\newcommand{\pFqcomma}{\mskip\pFqmuskip}

\makeatletter
\renewcommand{\@maketitle}{
\newpage
 \begin{center}%
  {\large\bfseries \@title \par}%
 \end{center}%
 \par} \makeatother

\numberwithin{equation}{section}

\begin{document}

\institution{CT}{Walter Burke Institute for Theoretical Physics, California Institute of Technology, \cr Pasadena, CA, 91125}
\institution{Yale}{Department of Physics, Yale University, New Haven, CT 06511}

\title{Higher Spin ANEC and the Space of CFTs}

\authors{David Meltzer\worksat{\CT,\Yale}}
\abstract{We study the positivity properties of the leading Regge trajectory in higher-dimensional, unitary, conformal field theories (CFTs). These conditions correspond to higher spin generalizations of the averaged null energy condition (ANEC). By studying higher spin ANEC, we will derive new bounds on the dimensions of charged, spinning operators and prove that if the Hofman-Maldacena bounds are saturated, then the theory has a higher spin symmetry. We also derive new, general bounds on CFTs, with an emphasis on theories whose spectrum is close to that of a generalized free field theory. As an example, we consider the Ising CFT and show how the OPE structure of the leading Regge trajectory is constrained by causality. Finally, we use the analytic bootstrap to perform additional checks, in a large class of CFTs, that higher spin ANEC is obeyed at large and finite spin. In the process, we calculate corrections to large spin OPE coefficients to one-loop and higher in holographic CFTs.}

\date{ }

\maketitle
\setcounter{tocdepth}{2}
\tableofcontents

\sec{Introduction}
\label{sec:Intro}

In this paper, we will study positivity conditions obeyed by the leading Regge trajectory in $d>2$, Lorentzian conformal field theories (CFTs). The leading Regge trajectory is defined to be the set of operators with the smallest scaling dimension, $\Delta$, for each even-spin $s\geq 2$ \cite{Cornalba:2007fs,Costa:2012cb,Caron-Huot:2017vep}. A universal operator which appears on this trajectory is the stress-energy tensor, $T_{\mu\nu}$, which has spin $s=2$ and saturates the unitarity bound, $\Delta_{T}=d$. In all QFTs the light-ray integral of the stress-tensor also obeys a positivity condition: the averaged null energy condition (ANEC). The ANEC states that the following operator is positive:
\bea
\mathcal{E}=\int\limits_{-\infty}^{\infty} dx^{-} T_{--}(x^{-},0),
\eea
where the integral is over a complete, null line.

This positivity condition was first studied extensively for CFTs in \cite{Hofman:2008ar} to derive universal bounds on three-point functions involving $T_{\mu\nu}$. The ANEC has since been proven via two different methods, through causality and OPE arguments in \cite{Hartman:2016lgu} and through monotonicity of relative entropy in \cite{Faulkner:2016mzt}. Here, we will be interested in exploring the results of \cite{Hartman:2016lgu}, where the proof of the ANEC also revealed an infinite set of new, higher spin positivity conditions\footnote{See \cite{Komargodski:2016gci,Hofman:2016awc} for previous work on higher spin sum rules.}. Higher spin ANEC, or HS ANEC, says the positivity of $\int \mathrm{d}x^{-}T_{--}$ generalizes straightforwardly to the entire $T_{\mu\nu}$ Regge trajectory. More precisely, the following operator is positive:
\bea
\mathcal{E}^{(s)}=\int\limits_{-\infty}^{\infty}dx^{-} J^{(s)}_{--...-}(x^{-},0), \qquad s\geq2 \ \& \ s\text{ even,}
\eea
where $J^{(s)}_{\mu_{1}...\mu_{s}}$ is the lightest spin-$s$ operator in a reflection positive OPE. When $s=2$ this reduces to the ANEC operator.

Why might we be interested in studying the positivity properties of higher spin operators? The first, most basic, motivation is we want to use the fundamental principles of causality, unitarity and locality to map out the space of consistent quantum field theories. HS ANEC follows from the axioms of conformal field theory \cite{Hartman:2016lgu} and gives new bounds on CFT data which have not been fully explored. Specifically, it singles out the operators with low twist, $\tau=\Delta-s$, which govern the lightcone OPE \cite{light1,light2}. Understanding positivity conditions on the set of CFT data underlies the success of the conformal bootstrap program \cite{Poland:2018epd,Simmons-Duffin:2016gjk}, so it is natural to expect that this infinite set of positive operators will give a new analytic window into the space of CFTs.

As an illustrative case, by studying HS ANEC we can derive new bounds on how two scalar operators couple to the leading Regge trajectory. The corresponding ANEC bound is trivial by CFT Ward identities, while the HS ANEC bound is non-trivial for general CFTs. Therefore, while the first four-point function which is related to the ANEC in a non-trivial way involves spinning operators \cite{Li:2015itl,krav,Dymarsky:2017xzb}, HS ANEC can be related to a simpler four-point function consisting solely of scalars. These constraints, which have not been used thus far, can be straightforwardly applied to the study of mixed correlator systems.

In free CFTs, operators on the leading Regge trajectory also play an enhanced role as the generators of a higher spin symmetry. The presence of a single, conserved, higher spin current is enough to prove the existence of an infinite-dimensional, higher spin symmetry, which in turn completely fixes the OPE of the higher spin currents \cite{Maldacena:2011jn,Boulanger:2013zza,Alba:2015upa}. In addition, theories like Chern-Simons vector models \cite{Aharony:2011jz,Giombi:2011kc,Maldacena:2012sf} are also tightly constrained by a slightly broken, higher spin symmetry\footnote{See also \cite{Aharony:2018npf,Turiaci:2018nua} for a bootstrap approach to these theories.}. For weakly coupled CFTs, the HS ANEC operators are then natural objects to study as they are both manifestly positive and sensitive to the emergence of an infinite-dimensional symmetry. 

Finally, the Lorentzian inversion formula \cite{Caron-Huot:2017vep,ssw} also guarantees CFT data organizes nicely into analytic families parameterized by the scaling dimension and spin. Analyticity in spin follows from the fact that individual conformal blocks with spin $\ell>1$ diverge in the Regge limit, while four-point functions in a unitary CFT are bounded in this regime\cite{Maldacena:2015waa,Hartman:2015lfa}. Therefore, we cannot independently vary the OPE data for a single, high-spin operator without spoiling boundedness in the Regge limit. Moreover, in \cite{Kravchuk:2018htv} it was shown that the analytic continuation in spin can also be done at the level of the light-ray transformed operators themselves. These results imply that three-point functions of operators on the leading Regge trajectory cannot be completely independent. As an example of this phenomena, we will use HS ANEC to put bounds on $\<TJ^{(4)}T\>$, $\<J^{(4)}TJ^{(4)}\>$, and $\<J^{(4)}J^{(4)}J^{(4)}\>$ in terms of $\<TTT\>$. Using the AdS/CFT dictionary\cite{Maldacena:1997re,Witten:1998qj,Gubser:1998bc}, this corresponds to bounds on cubic interactions for AdS theories with many light, higher spin particles \cite{Vasiliev:1990en,Vasiliev:1999ba,Sezgin:2002rt,Vasiliev:2003ev,Klebanov:2002ja}\footnote{The leading Regge trajectory we discuss here is the exact trajectory of the CFT and does not always correspond to the leading single-trace trajectory in large $N$ CFTs \cite{Cornalba:2007fs,Costa:2012cb}.}.
\ssec{Summary}

This work is organized as follows. In section \ref{sec:ANEC_Review}, we review the lightcone OPE for CFTs in $d>2$, the proof of HS ANEC, the symmetry properties of light-ray operators, and the behavior of the leading trajectory in CFTs. We will also establish notation and introduce the states used to derive the optimal bounds. 

In section \ref{sec:HScouplings}, we will present new constraints for two- and three- point functions from HS ANEC. To start, in section \ref{sec:ChargedBounds} we prove the twist of a charged, spin-$s$ operator which appears in a reflection positive OPE is bounded below by the twist of the lightest, uncharged, spin-$s$ operator that appears in the same OPE if $s\geq2$ and even. In other words, for generic CFTs, the leading Regge trajectory is necessarily composed of uncharged operators. In section \ref{sec:GeneralBounds}, we consider simple examples of HS ANEC, with an emphasis on matrix elements involving a scalar operator. Here HS ANEC strongly constrains three-point functions in theories whose spectrum is close to a generalized free field spectrum.

In section \ref{sec:Spin24_Systems}, we study HS ANEC in states created by the stress-tensor and the lightest spin-4 operator. In section \ref{sec:SatANEC}, we prove that if the ANEC bounds for $\<TTT\>$ are saturated, then the CFT has a higher spin symmetry. We also show how saturation of ANEC implies saturation of HS ANEC. For practical applications, in section \ref{sec:ising} we focus on $3d$ CFTs with an Ising-like spectrum and derive bounds on three-point functions involving the spin-$4$ operator.

In section \ref{sec:LCBootstrap}, we discuss the relation between HS ANEC and the analytic bootstrap. In CFTs, the higher spin positivity conditions bound OPE coefficients which are also computable using large spin expansions. We find that the exchange of isolated operators and towers of double-twist operators in one channel always yield results consistent with HS ANEC in the dual channel, at finite and asymptotically large spin respectively. In the context of large $N$ CFTs, this implies AdS theories with only cubic interactions are consistent with HS ANEC at tree and one-loop level, with the corresponding restrictions on spin. We also consider examples where HS ANEC can na\"{\i}vely be violated at finite spin if we do not include non-perturbative effects in the large spin expansion. Of independent interest, we also present new results for large spin OPE coefficients to all orders in $1/N$ for holographic CFTs. In the dual AdS theory, these OPE coefficients can be found through conformal block decompositions of ladder diagrams.

In section \ref{sec:Conclusion}, we give a brief conclusion and discuss future directions. The appendices contain various technical details used throughout the paper. Appendix \ref{app:integrals} describes the basis of conformally invariant three-point structures and the integrals used to calculate HS ANEC matrix elements. Appendix \ref{app:ConsWard} includes solutions to the conservation conditions and integrated Ward identities. In Appendix \ref{app:spin2_4}, we give examples of relevant (HS) ANEC matrix elements. In Appendix \ref{app:RegDT} we give an example, relevant to section \ref{sec:LCBootstrap}, where sums over double-twist operators can yield negative corrections to large spin OPE coefficients. 

\sec{Review of the Lightcone OPE and HS ANEC}
\label{sec:ANEC_Review}

In this section, we will give a brief overview of the lightcone OPE and how it leads to HS ANEC \cite{Hartman:2016lgu}\footnote{See \cite{Kravchuk:2018htv} for a generalization to continuous spin.}. We will also review the behavior of the leading Regge trajectory in general CFTs and the properties of light-ray operators.

For a CFT in flat space, we can always write down the OPE of identical scalars, $\p$, as:
\begin{equation}
\p(x)\p(-x)=\sum_{\substack{\O}}c_{\p\p\O}x^{-2\Delta_{\p}+\Delta_{\O}-s_{\O}}x_{\mu_{1}}...x_{\mu_{s}}\O_{\Delta,s}^{\mu_{1}...\mu_{s}}.\quad
\end{equation}
where $\O_{\Delta,s}$ is a symmetric, traceless tensor of spin-$s$ and we have not used conformal symmetry to relate primaries and descendents.

As realized in \cite{light1,light2}, if we work in Lorentzian signature and take the limit $x^{+}\rightarrow 0$, the dominant operator is the one with the minimal twist, $\tau=\Delta-s$. For a spin-$s$ operator, $\O_{\Delta,s}$, the leading contribution comes specifically from $\O_{\Delta,s,-...-}$ plus all descendents generated by acting with $\partial_{-}$. At the level of the four-point function, $\<\p\p\p\p\>$, the conformal blocks $g_{\Delta,\ell}(z,\bar{z})$ reduce to a sum of $\grSL(2,\mathbb{R})$ blocks. Here $\grSL(2,\mathbb{R})$ is the group which leaves the light-ray connecting the two, null-separated operators invariant. 

At the level of the OPE, we can write the contribution of the primary and all its minimal twist descendents as an integral over a null line. It was shown in \cite{Hartman:2016lgu} that when $|x^{+}|\ll \frac{1}{|x^{-}|}\ll 1$, such that $x^{+}x^{-}<0$, the OPE becomes:
\begin{equation}
\p(x)\p(-x)\big|_{J^{(s)}}=(-x^{+})^{\frac{\tau_{s}}{2}}(x^{-})^{\frac{\tau_{s}}{2}+s-1}\<\p(x)\p(-x)\>\frac{2^{\tau_s+s}c_{\p\p s}\Gamma\left(\frac{\tau_{s}+2s+1}{2}\right)}{\sqrt{\pi}C^{(s)}_{J}\Gamma\left(\frac{\tau_s+2s}{2}\right)}\int d y^{-} J^{(s)}_{--...-}(y,0), \label{eq:LCOPE}
\end{equation}
where we have isolated the contribution of a given operator, $J^{(s)}$, to the OPE and kept its normalization, $C^{(s)}_{J}$, arbitrary. In this form, we also see how the operators with the minimal twist dominate the lightcone OPE. This is not the entire contribution of the $J^{(s)}$ multiplet to the OPE, but if we insert $\p(x)\p(-x)$ between two states, this integral captures the leading behavior in the above lightcone limit. For the remainder of this work, $J^{(s)}$ will always denote the operator with the smallest twist for a given spin-$s$. For OPE coefficients, we also use the label $``s"$ as a shorthand for $J^{(s)}$.

In order to prove ANEC and its higher spin generalization, \cite{Hartman:2016lgu} used the positivity properties of Rindler symmetric correlation functions in Minkowski space. The Rindler reflection for scalars is defined by:
\begin{equation}
\bar{x}=(-t^{*},-y^{*},\vec{x}), \qquad \overline{\O}(x)=\O^{\dagger}(\bar{x}),
\end{equation}
and maps operators in one wedge to the other. Rindler positivity\footnote{See also \cite{Maldacena:2015waa,Casini:2010bf} for a derivation of Rindler positivity.} states the following correlation function is positive:
\bea
\<\overline{\O}_1\overline{\O}_{2}\O_1\O_2\>\geq0,
\eea 
where the unbarred operators are inserted in the right Rindler wedge and Rindler reflection does not reverse the order of operators. To make a connection with ANEC, we consider the states:
\bea
A=\O\p(-x), \qquad B=\p(-x)\O, \label{eq:statesAB}
\eea
where $A$ and $B$ are defined on the right Rindler wedge and $\p$ is real. Using Rindler positivity to define an inner product, the Cauchy-Schwarz inequality implies: 
\bea
|\<\overline{A}B\>|=|\<\overline{\O}\p(x)\p(-x)\O\>|\leq \left(\<\overline{A}A\>\<\overline{B}B\>\right)^{\frac{1}{2}} \label{eq:CauchyAB}
\eea
Using this inequality and analyticity properties of the four-point function, \cite{Hartman:2016lgu} derived a sum rule for the normalized correlator:
\begin{equation}
G=\frac{\<\overline{\O}\p(x)\p(-x)\O\>}{\<\overline{\O}\O\>\<\p\p\>},
\end{equation}
which is most clearly stated if we introduce the variables $\eta= -x^{+}x^{-}$ and $\sigma=1/x^{-}$. The sum rule is then:
\bea
Re \oint d\sigma\s^{s-2}\left(1-G(\eta,\s)\right)=0.
\eea
The integral runs over a semi-circle of radius $R$ in the lower $\s$ plane, just below the origin. They also take $\eta\ll R\ll1$, such that the correlator on the arc is well approximated by the lightcone OPE. When we perform the OPE on the arc, the fact $\eta\ll1$ means we are projecting onto the minimal-twist operators, while the factor of $\s^{s-2}$ projects onto the spin-$s$ operators. The final result is:
\begin{align}
(-1)^{\frac{s}{2}}c_{\p\p s}\<\overline{\O}\mathcal{E}^{(s)}\O\>\propto \underset{R\rightarrow 0}{\lim} \ \underset{\eta\rightarrow 0}{\lim} \ \eta^{-\tau_{s}/2}Re\int_{-R}^{R}d\s \s^{s-2}(1-G(\eta,\s))\geq 0,&\label{eq:arcOPE}
\\
\mathcal{E}^{(s)}=\int dx^{-}J^{(s)}_{--...-}(x^{-},0),&
\end{align}
where the right hand side of (\ref{eq:arcOPE}) is positive because the positive ordered correlation functions on the right hand side of (\ref{eq:CauchyAB}) factorize at small $\s$\cite{Hartman:2015lfa,Hartman:2016lgu}. In sum, they derived the positivity condition:
\begin{equation}
(-1)^{\frac{s}{2}}c_{\p\p s}\mathcal{E}^{(s)}\geq 0. \label{eq:sumrule}
\end{equation}

When $s=2$, this gives the averaged null energy condition (ANEC) since the stress-tensor is always the lightest spin-two operator in any positive OPE\footnote{We will always assume our CFT does not have multiple decoupled sectors and the stress-tensor is the unique, conserved, spin-two operator.}. In the language of \cite{Kravchuk:2018htv}, $\mathcal{E}^{(s)}$ is the light-ray transform of a local operator on the leading Regge trajectory. In sections \ref{sec:HScouplings} and \ref{sec:Spin24_Systems} we will assume the probe operator, $\p$, is chosen such that $(-1)^{\frac{s}{2}}c_{\p\p s}>0$, while in section \ref{sec:LCBootstrap} it is more natural to keep the product of OPE coefficients explicit.

Next, we consider the behavior of the leading Regge trajectory in generic CFTs. From the work of \cite{light1,light2}, we know that for CFTs in $d>2$, the large spin sector has the structure of a generalized free field theory. They showed that for any two operators, $\O_{1}$ and $\O_{2}$, with twists $\tau_i$, there always exists an infinite tower of double-twist operators, $[\O_1\O_2]_{n,\ell}$, such that as $\ell\rightarrow\infty$ the twists $\tau_{n,\ell}\rightarrow \tau_1+\tau_2+2n$. In a generalized free field theory, these operators are given by \cite{Heemskerk:2009pn}:
\begin{equation}
[\O_1\O_2]_{n,\ell}=\O_1\partial_{\mu_{1}}...\partial_{\mu_{\ell}}\partial^{2n}\O_{2}-(traces)
\end{equation}

Furthermore, it was shown in \cite{light2,Komargodski:2016gci,Costa:2017twz} that the leading Regge trajectory in a reflection positive OPE, $\O^{\dagger}\O$, is a monotonically increasing, convex function of the spin. Since every local CFT contains the stress-energy tensor, which has twist $\tau_{T}=d-2$, we find the following bound for the twists, $\tau_{s}$, of the leading trajectory: 
\begin{eqnarray}
d-2 \ \leq \ \tau_{s} \ \leq \ 2(d-2). \label{eq:convexity}
\end{eqnarray}

If there exists a light scalar, $\phi$, with $\Delta_{\f}<d-2$, then we can replace the upper bound with $2\Delta_{\f}$. We will generally assume the lower bound is not saturated for $\ell>2$ so the theory is not free \cite{Maldacena:2011jn,Boulanger:2013zza,Alba:2015upa}. For generic CFTs, we expect that the leading Regge trajectory is composed of double-twist operators, as is seen for example in the Ising CFT \cite{Simmons-Duffin:2016wlq}. There are counterexamples, e.g. weakly coupled CFTs with a gauge theory description have low spin operators with twist $\tau_{s}\approx d-2+\gamma_{s}$ and $\gamma_{s}\ll1$ without having a low dimension scalar in the spectrum. To apply HS ANEC we do not need to make any assumptions on which scenario is realized, while when applying the analytic bootstrap we will assume the trajectory consists of double-twist operators.

In deriving HS ANEC, it is important that the operator $J^{(s)}$ is the lightest, spin-$s$ operator in a given positive OPE. It is not a priori clear that the same Regge trajectory gives the leading contribution in every positive OPE. In section \ref{sec:ChargedBounds}, we will rule out a wide class of possible counterexamples by showing the leading trajectory must be in the singlet representation of any internal global symmetry.

Finally, we will review the structure of three-point functions involving light-ray operators. To derive the optimal bounds from the positivity of $\mathcal{E}^{(s)}$, our states will always be momentum eigenstates:
\bea
|\O(q,\lambda)\>=\mathcal{N}\int d^{d}x e^{-i q\cdot x}\lambda\cdot \O(x)|0\>.
\eea
We will set $\mathcal{N}=1$ and use the mostly plus convention for the metric. It is also convenient to define a covariant version of the (HS) ANEC operators as \cite{zhib}:
\bea
\mathcal{E}^{(s)}(n)=\int_{-\infty}^{\infty}d(x\cdot n)\underset{x\cdot\bar{n}\rightarrow\infty}{\lim}(x\cdot\bar{n})^{\tau_s}J^{(s)}_{\mu_{1}...\mu_{s}}(x\cdot n,x\cdot\bar{n},0)\bar{n}^{\mu_1}...\bar{n}^{\mu_s}, \nonumber
\\ n=(1,\vec{n}), \quad \bar{n}=(-1,\vec{n}).
\eea
We will always choose $n$ and $q$ such that $n\cdot q = q^{2}=-1$.

We then want to calculate $\<\O(q,\lambda)|\mathcal{E}^{(s)}(n)|\O(q,\lambda)\>$ and impose positivity for all $\lambda$. To organize the bounds, we classify how $\lambda$ transforms under the residual $\grSO(d-2)$ symmetry which leaves $n$ and $q$ fixed. If $\O_{\Delta,\ell}$ is a symmetric, traceless operator of spin $\ell$, then $\lambda$ can transform in the spin $0,1,...,\ell$ representations of $\grSO(d-2)$.

In practice, it is convenient to construct the polarization tensors from the set of vectors $\{q,n,e_{i}\}$, where $e_{i}\cdot e_{j}=\delta_{ij}$ and $e_{i}\cdot n=e_{i}\cdot q=0$. For the spin-$j$ bound we fix a set $\{e_{1},...,e_{j}\}$ and consider all polarization tensors of the form: 
\bea
\lambda^{(\ell,j),\mu_{1}...\mu_{\ell}}_{k}=e_{1}^{(\mu_{1}}...e_{j}^{\mu_{j}}q^{\mu_{j+1}}...q^{\mu_{j+k}}n^{\mu_{j+k+1}}...n^{\mu_{\ell})}-(traces). \label{eq:polbasis}
\eea

Then the general matrix $\<\O_{\Delta,\ell}(q,\lambda)|\mathcal{E}^{(s)}(n)|\O_{\Delta,\ell}(q,\lambda)\>$ becomes a block diagonal matrix, where for each $j$ we obtain a $(\ell-j+1) \times (\ell-j+1)$ positive matrix:
\begin{eqnarray}
\hspace{-.5in}\<\O_{\Delta,\ell}|\mathcal{E}^{(s)}(n)|\O_{\Delta,\ell}\>=
\left( \begin{array}{cccc}
\mathcal{E}_{\O\O}^{(s,0)} & 0 & 0 & 0 \\
0 & \mathcal{E}_{\O\O}^{(s,1)}  & 0 &0 \\
0 & 0 & ... & 0\\
0 & 0 & 0 & \mathcal{E}_{\O\O}^{(s,\ell)} 
 \end{array} \right) \succeq0,\hspace{.15in} 
\end{eqnarray}
\begin{center}
\vspace{-.25in}
\begin{equation}
\mathcal{E}_{\O\O,ab}^{(s,j)}=\<\O_{\Delta,\ell}\big(q,\lambda^{(\ell,j)}_{a}\big)|\mathcal{E}^{(s)}(n)|\O_{\Delta,\ell}\big(q,\lambda^{(\ell,j)}_{b}\big)\>.
\end{equation}
\end{center}

For low spacetime dimensions or high external spin, some of these polarization choices are not possible, e.g. if $\ell>d-2$ we can not find $d$ vectors, $e_{i}$, orthogonal to $\{q,n\}$ such that $e_i\cdot e_j=\delta_{ij}$. Finally, if the external operators are conserved we can always eliminate $q$, and we instead have $\ell+1$ linear bounds from $\mathcal{E}^{(s,j)}_{\O\O,00}\geq0$. When it is clear from the context, we will drop the indices for the matrix $\mathcal{E}^{(s,j)}_{\O\O}$. 

If we set $\O=T$, then there are three linear bounds for $d\geq4$ and two linear bounds in $d=3$. We can always write $\<TTT\>$ in terms of the tensor structures which appear in a free field theory:
\begin{equation}
\<TTT\>=n_{B}\<TTT\>_{B}+n_{F}\<TTT\>_{F}+n_{T}\<TTT\>_{T}. \label{eq:TTTFF}
\end{equation}
Here $B$, $F$, and $T$ refer to a free theory of scalars, fermions, and tensors\footnote{In even dimensions there exist free field theories of $\frac{d-2}{2}$ forms. For $d>3$ and odd, such free field theories do not exist, but the corresponding tensor structure still does.}, respectively. Then the ANEC yields:
\begin{align}
&\mathcal{E}^{(2,0)}_{TT}\geq 0 \ \Longrightarrow \ n_{B}\geq 0,
\\ &\mathcal{E}^{(2,1)}_{TT}\geq 0 \ \Longrightarrow \ n_{F}\geq 0,
\\ &\mathcal{E}^{(2,2)}_{TT}\geq 0 \ \Longrightarrow \ n_{T}\geq 0. \label{eq:TTTnvbound}
\end{align}
These are also known as the Hofman-Maldacena bounds \cite{Hofman:2008ar}. In $d=3$ we only have the first two bounds.

Generically, conservation of the stress-tensor implies the ANEC bounds will be stronger than the corresponding HS ANEC bounds. For the three-point function $\<\O_{1}T\O_{2}\>$, conservation of the stress-tensor at non-coincident points implies relations between the OPE coefficients, while the integrated CFT Ward identities relate this three-point function to the two-point function $\<\O_{i}\O_{j}\>\propto  \delta_{ij}$\cite{Osborn:1993cr}. For interacting field theories, similar identities do not hold if we replace $T$ by a higher spin operator.

One related benefit of studying the ANEC operator is it makes solving the Ward identities simpler. From its definition we have the following property \cite{Hofman:2008ar}:
\begin{equation}
\int_{S^{d-2}} d\vec{n} \<\O(q,\lambda)|\mathcal{E}(n)|\O(q,\lambda)\> = 2^{d}q^{0}\<\O(q,\lambda)|\O(q,\lambda)\>, \label{eq:WardGen}
\end{equation} 
where we integrate $\vec{n}$ over the sphere at infinity\footnote{The extra factor of $2^{d}$ is due to our unconventional normalization of $\mathcal{E}^{(2)}$.}. We will use this equation to solve the Ward identity constraints on $\<J^{(4)}TJ^{(4)}\>$.

\sec{Bounds on Higher Spin Couplings}
\label{sec:HScouplings}

\ssec{Bounds for Charged Operators}
\label{sec:ChargedBounds}

By studying CFT four-point functions and HS ANEC, we will prove a lower bound on the dimensions of spinning, charged operators in terms of the uncharged operators. More precisely, we show that if an operator $\O$ satisfies the following properties:
\begin{itemize}
\item $\O$ has spin-$s$ with $s\geq 2$ and even,
\item $\O$ appears in a reflection positive OPE, e.g. $\p^{\dagger}\p$,
\item $\O$ transforms in a non-trivial representation, $\mathcal{R}$, of some internal, global symmetry group,
\end{itemize}
then $\tau_{\O}\geq \tau^{\mathbb{I}}_{s,min}$ where $\tau^{\mathbb{I}}_{s,min}$ is the twist of the lightest, uncharged operator with spin-$s$. This implies that for generic CFTs, the leading Regge trajectory in any positive OPE must be composed of uncharged operators\footnote{For free CFTs with a global symmetry, we can have multiple, degenerate Regge trajectories with different global symmetry properties.}. For $s=2$, this is trivially satisfied since we assume there is a unique, conserved, spin-two operator. However, for all $s\geq 4$ this statement is non-trivial and gives lower bounds on the dimensions of spinning, charged operators. 

A simple way to motivate this bound is to consider a four-point function of scalar operators in the fundamental representation of $\grSU(N)$. To make contact with \cite{Hartman:2016lgu}, we will also insert them symmetrically with respect to the Rindler wedges. The $t$-channel, or $1\times 4\rightarrow 3\times 2$, conformal block decomposition now has the form:
\begin{equation}
\begin{aligned}
G(x_{1},x_{2},\epsilon)&=\epsilon^{*,i_{2}\bar{i}_1}\<\O^{\dagger}_{\bar{i}_{1}}(\overline{x}_1)\psi_{i_2}(\overline{x}_2)\psi^{\dagger}_{\bar{i}_{3}}(x_2)\O_{i_4}(x_1)\>\epsilon^{\bar{i}_3 i_4} \nonumber
\\ &=Tr(\epsilon^{*}\cdot \epsilon) I_t(\overline{x}_1,\overline{x}_2,x_2,x_{1})+\bigg(Tr(\epsilon)Tr(\epsilon^{*})-\frac{1}{N}Tr(\epsilon^{*}\cdot \epsilon)\bigg)Adj_t(\overline{x}_1,\overline{x}_2,x_2,x_{1}), \label{eq:SUNscalars}
\end{aligned}
\end{equation}
where $\{I,Adj\}$ denotes the global symmetry representations of the exchanged operators. The points $x_1$ and $x_2$ will lie in the right Rindler wedge.

When we take the pair of operators $\psi(\bar{x}_{2})\psi^{\dagger}(x_{2})$ to be light-like separated, light-ray operators will contribute to both $I_{t}$ and $Adj_{t}$. Within the spin-$s$ sector, the leading contribution to the lightcone OPE can come from either $I_{t}$ or $Adj_{t}$, depending on the twists of the exchanged operators. In either case, the HS ANEC sum rule gives a sign constraint on the corresponding OPE coefficients.

If we assume the minimal twist, light-ray operators are uncharged and appear in $I_{t}$, the proof of HS ANEC is unchanged as the dependence on the $\grSU(N)$ polarizations is trivial. On the other hand, if we assume for some spin-$s$ the lightest operator is in the adjoint representation, we see an immediate problem: the prefactor in front of $Adj_{t}$ is not sign-definite. The sum rule (\ref{eq:arcOPE}) now becomes:
\begin{equation}
\hspace{-.1in}\bigg(Tr(\epsilon)Tr(\epsilon^{*})-\frac{1}{N}Tr(\epsilon^{*}\cdot \epsilon)\bigg)(-1)^{\frac{s}{2}}c_{\p\p^{\dagger} s}\<\overline{\O}\mathcal{E}_{Adj}^{(s)}\O\>\propto \underset{R\rightarrow 0}{\lim} \ \underset{\eta\rightarrow 0}{\lim} \ \eta^{-\tau_{s}/2}Re\int_{-R}^{R}d\s \s^{s-2}(1-G(\eta,\s,\epsilon))\geq 0,\label{eq:arcOPEch}
\end{equation}
where the right hand side is still positive by the Cauchy-Schwarz inequality and factorization. 

By choosing different $\grSU(N)$ polarizations, the left hand side can take either sign and the HS ANEC sum rule implies $c_{\p\p^{\dagger} s}c_{\O^{\dagger}\O s}=0$. This is a contradiction since we assumed the lightest spin-$s$ operator in the $\p\p^{\dagger}$ and $\O^{\dagger}\O$ OPEs was charged. Therefore, either the leading spin-$s$ operator is in the singlet representation or we have two spin-$s$ operators with degenerate twist, $J^{(s)}_{\mathbb{I}}$ and $J^{(s)}_{Adj}$, and the OPE coefficients for the charged operator are bounded in terms of the uncharged one. We can always derive the HS ANEC bound on the singlet operators alone by choosing the probe operator, $\p$, to be uncharged.

To prove this in general, we can use some simple properties of the Clebsch-Gordon coefficients, or equivalently the 6j-symbol. We will consider a general four-point function of scalars:
\begin{align}
G=\epsilon^{*,\bar{j}\bar{i}}\<\bar{\f}_{1,\bar{i}}\f^{\dagger}_{2,\bar{j}}(\overline{x}_2)\f_{2,k}(x_2)\f_{1,r}\>\epsilon^{kr},
\end{align}
where we take $\f_{i}$ to transform in the representation $\mathcal{R}_{i}$ of the global symmetry. The operator $\f_1$ can be smeared or inserted at a single point.

We will also assume a single operator, $\mathcal{O}^{(\mathcal{R}_{3})}_{s,p}$, where $p$ is the representation index, dominates the spin-$s$ sector of the lightcone OPE, $\f^{\dagger}_{2,\bar{j}}(\overline{x}_{2})\f_{2,k}(x_{2})$. When we isolate its contribution in this OPE and then calculate the three-point function $\<\bar{\f}_{1,\bar{i}}\mathcal{O}^{(\mathcal{R}_{3})}_{s,p}\f_{1,r}\>$, we produce a product of Clebsch-Gordon coefficients:
\begin{align}
\f^{\dagger}_{2,\bar{j}}(\overline{x}_{2})\f_{2,k}(x_{2})&\supset \sum_{\substack{p}}c_{\bar{2}2\bar{3}}C^{(\bar{2}2\bar{3})}_{\bar{j}k\bar{p}}\mathcal{O}^{(\mathcal{R}_{3})}_{s,p},
\\
\<\bar{\f}_{1,\bar{i}}\mathcal{O}^{(\mathcal{R}_{3})}_{s,p}\f_{1,r}\>&\propto c_{\bar{1}13}C^{(\bar{1}13)}_{\bar{i}rp},
\end{align}
where we have dropped the overall spacetime dependence.

Next, we choose the polarization tensors to project onto a given representation $\mathcal{R}_{4}\in\mathcal{R}_{1}\otimes\mathcal{R}_{2}$, that is we choose $\epsilon^{(\mathcal{R}_{4})}_{\bar{k}\bar{r}q}=C^{(\bar{2}\bar{1}4)}_{\bar{k}\bar{r}q}$. This produces another product of Clebsch-Gordon coefficients, and we are left with the following group theory factors multiplying the OPE coefficients:
\begin{equation}
G\approx \underset{i,j,k,r,p}{\sum}C^{(12\bar{4})}_{ij\bar{q}}C^{(\bar{2}\bar{1}4)}_{\bar{k}\bar{r}q}C^{(\bar{2}2\bar{3})}_{\bar{j}k\bar{p}}C^{(\bar{1}13)}_{\bar{i}rp}c_{\bar{1}13}c_{\bar{2}2\bar{3}}(...), \label{eq:groupfactors}
\end{equation}
where for brevity we suppressed various kinematical factors which are independent of the group representations. Here the $``\approx"$ is because we are studying the lightcone limit of the full four-point function, but the leading lightcone contribution is all we need for HS ANEC. 

It is crucial that HS ANEC should hold for any choice of $(\mathcal{R}_{4},q)$, since these correspond to different Rindler symmetric ways of creating our state and light-ray operator, or different choices of $A$ and $B$ in (\ref{eq:CauchyAB}). Our strategy will then be to sum over $(\mathcal{R}_{4},q)$ and use orthogonality properties of the Clebsch-Gordon coefficients to show charged operators never give a sign-definite contribution in the lightcone OPE. 

The orthogonality properties we need are:
\begin{align}
&\sum_{\mathcal{R}_{4},m_4}C^{(12\bar{4})}_{m_{1}m_{2}\bar{m}_{4}}C^{(\bar{1}\bar{2}4)}_{\bar{m}_1\bar{m}_{2}m_4} = \delta_{m_1\bar{m}_1}\delta_{m_2 \bar{m}_2}, 
\\
&\sum_{\substack{m_{2},m_{3}}}C^{(123)}_{m_{1}m_{2}m_{3}}C^{(\bar{2}\bar{3}\bar{4})}_{\bar{m}_{2}\bar{m}_{3}\bar{m}_{4}}=\delta_{m_{1}\bar{m}_{4}}\delta_{1\bar{4}}.
\end{align}
If we sum (\ref{eq:groupfactors}) over $(\mathcal{R}_{4},q)$ we find:
\begin{align}
\sum_{\mathcal{R}_{4},m_4}G\approx \sum_{\mathcal{R}_{4},m_4} \underset{i,j,k,r,p}{\sum}C^{(12\bar{4})}_{ij\bar{q}}C^{(\bar{2}\bar{1}4)}_{\bar{k}\bar{r}q}C^{(\bar{2}2\bar{3})}_{\bar{j}k\bar{p}}C^{(\bar{1}13)}_{\bar{i}rp}c_{\bar{1}13}c_{2\bar{2}\bar{3}}(...)& \nonumber
\\ =\underset{r,k,p}{\sum}C^{(\bar{2}2\bar{3})}_{\bar{k}k\bar{p}}C^{(\bar{1}13)}_{\bar{r}rp}c_{\bar{1}13}c_{2\bar{2}\bar{3}}(...) = 0& \quad \text{unless} \quad \mathcal{R}_{3}=\mathbb{I},
\end{align}
where in the last step the sums project onto the singlet representation. Put another way, by summing over $(\mathcal{R}_{4},q)$ we are averaging over all polarizations, so only operators in the singlet representation can appear in the $t$-channel.

Therefore if $\mathcal{R}_{3}\neq \mathbb{I}$, the HS ANEC sum rule will fix the OPE coefficients to have either a positive or negative sign depending on how we choose the polarizations. This fixes the OPE coefficients to be zero, unless there is also an operator with the same or smaller twist in the singlet representation. This completes the proof that the twist of a charged operator in a positive OPE is bounded below by the twist of the lightest, uncharged operator with the same spin-$s$ in the same OPE, for all $s\geq2$ and even. We have focused on the four-point function of scalar primaries, but the generalization to spinning operators or systems of four-point functions is straightforward.

It is also clear this bound cannot be improved for general CFTs. From the lightcone bootstrap, we know that if a CFT contains a light, charged scalar, $\f$, with $\Delta_{\f}<d-2$, then the bound is saturated at infinite spin\cite{Li:2015rfa}. By solving crossing for $\<\f^{\dagger}\f\f^{\dagger}\f\>$ we can show there exists double-twist operators for all global symmetry representations which can appear in the $\f^{\dagger}\f$ OPE and they all approach the same twist as $\ell\rightarrow\infty$. We also can use the results of \cite{Li:2015rfa} to check this bound holds at asymptotically large spin.

In addition, if we assume the leading trajectory is composed of double-twist operators $[\f^{\dagger}\f]_{0,s}$, it is not hard to argue that the same leading trajectory appears in every positive OPE, for sufficiently large spin. Using the lightcone bootstrap \cite{light1,light2}, the coupling $\<\mathcal{O}\mathcal{O}^{\dagger}[\f^{\dagger}\f]_{0,s}\>$ at large $s$ is non-zero in interacting CFTs and determined by the operators of minimal twist in the $\O\f$ and $\O\f^{\dagger}$ OPEs. Given the results of \cite{Caron-Huot:2017vep}, which proved the OPE data organizes into analytic families for $s>1$, we also expect the entire trajectory to appear in the $\mathcal{O}^{\dagger}\mathcal{O}$ OPE. However, we cannot rule out the OPE coefficients having accidental zeros at finite spin.

\ssec{HS ANEC Examples}
\label{sec:GeneralBounds}

We will now consider bounds from HS ANEC itself. The simplest bound is $\<\f|\mathcal{E}^{(s)}|\f\>\geq 0$, for scalar $\f$, where we get a single positivity constraint on the OPE coefficient $c_{\f\f s}$:
\bea
\mathcal{E}^{(s,0)}_{\f\f}=c_{\f\f s}\frac{\pi ^{\frac{d}{2}+2} e^{-\frac{1}{2} i \pi  s} 2^{d-2 \Delta_{\f}-s+3}  \Gamma (2 s+\tau_s-1)}{\Gamma \left(\Delta_{\f}-\frac{\tau_s}{2}\right) \Gamma \left(s+\frac{\tau_s}{2}\right)^2 \Gamma \left(-\frac{d}{2}+\Delta_{\f}+s+\frac{\tau_s}{2}\right)}.
\eea
Unitarity implies $\Delta_{\f}\geq\frac{1}{2}(d-2)$ and convexity of the leading trajectory in combination with the lightcone bootstrap imply $\tau_{s}\leq2\Delta_{\f}$, so we have:
\bea
c_{\f\f s}e^{-\frac{1}{2} i \pi  s}\geq0
\eea

One interesting case to study is when $\f$ is the lightest scalar in the theory and $\Delta_{\f}\leq d-2$. Then the leading Regge trajectory is $J^{(s)}= [\f\f]_{0,s}$ and we can write $\tau_{s}=2\Delta_{\f}+\gamma_{s}$. If $\gamma_{s}$ is small, this matrix element is also small due to the factor of $\Gamma^{-1} \left(\Delta_{\f}-\frac{\tau_s}{2}\right)$. This yields strong bounds on the off-diagonal, HS ANEC matrix elements when we consider a more general state. As an example, we can consider,
\bea
|\Phi\>=c_{1}|\f\>+c_{2}|\chi\>,
\eea
where $\f$ and $\chi$ are both scalars. Positivity of HS ANEC for this state gives the matrix condition:
\begin{align}
\<\Phi|\mathcal{E}^{(s)}(n)|\Phi\>=
\left( \begin{array}{cc}
\mathcal{E}^{(s,0)}_{\f\f} &\mathcal{E}^{(s,0)}_{\f\chi}   \\
\mathcal{E}^{(s,0)}_{\chi\f}  &  \mathcal{E}^{(s,0)}_{\chi\chi} 
 \end{array} \right)\succeq 0.
\end{align}

Keeping the twist, $\tau_{s}$, generic we find:
\begin{equation}
\begin{aligned}
\frac{c_{\f\chi s}^{2}}{c_{\f\f s}c_{\chi\chi s}} \ \leq& \ \frac{\Gamma \left(\frac{1}{2} (\Delta_{\f}+\Delta_{\chi}-\tau_s )\right)^2 \Gamma \left(\frac{1}{2} (\Delta_{\f\chi}+2 s+\tau_s )\right)^2 }{\Gamma \left(\Delta_{\f}-\frac{\tau_s }{2}\right) \Gamma \left(\Delta_{\chi}-\frac{\tau_s }{2}\right) \Gamma \left(s+\frac{\tau_s }{2}\right)^4}
\\ &\times \frac{\Gamma \left(\frac{1}{2} (\Delta_{\chi\f}+2 s+\tau_s )\right)^2 \Gamma \left(\frac{1}{2} (-d+\Delta_{\f}+\Delta_{\chi}+2 s+\tau_s )\right)^2}{ \Gamma \left(-\frac{d}{2}+\Delta_{\f}+s+\frac{\tau_s }{2}\right) \Gamma \left(-\frac{d}{2}+\Delta_{\chi}+s+\frac{\tau_s }{2}\right)},
\end{aligned}
\end{equation}
where $\Delta_{ij}=\Delta_i-\Delta_j$. If $\tau_{s}=2\Delta_{\f}+\gamma_{s}$ and $\gamma_{s}$ is small, the ratio of OPE coefficients scales like: 
\bea
\frac{c_{\f\chi s}^{2}}{c_{\f\f s}c_{\chi\chi s}}\lesssim |\gamma_{s}|,
\eea
and this bound becomes stronger as we increase $s$ \cite{light1,light2}. This inequality already gives strong constraints for the Ising CFT where $\gamma_{4}\approx -0.0136$ \cite{ElShowk:2012ht} or in the $\mathcal{N}=1$ Ising model where $\gamma_{4}\approx -.018$ \cite{Atanasov:2018kqw,Rong:2018okz}. Here it is important that the anomalous dimension, $\gamma_s$, is with respect to the generalized free field value, $\tau_{s}^{GFF}=2\Delta_{\f}$, and not the unitarity bound, $\tau_{s}^{free}=d-2$.  

The bound disappears when $\Delta_{\chi}=\Delta_{\f}+2s+\tau+2n$ for $n$ integer, which is consistent with the structure of generalized free field theories. In such theories we can always construct the operator $\chi=J^{(s)}_{\mu_1...\mu_s}\partial^{\mu_1}...\partial^{\mu_s}\partial^{2n}\f$ and the relevant coupling is not suppressed as $\tau_{s}\rightarrow 2\Delta_{\f}$. Finally, at large $\Delta_{\chi}$ the bound becomes:
\bea
\frac{c_{\f\chi s}^{2}}{c_{\f\f s}c_{\chi\chi s}}\leq\frac{\pi ^3  2^{d-2 \Delta_{\f}-2 \Delta_{\chi}-6 s-2 \tau_s +5} \Delta_{\chi}^{-\frac{d}{2}+2 \Delta_{\f}+5 s+2 \tau_s -3} \csc ^2\left(\frac{1}{2} \pi  (\Delta_{\f}-\Delta_{\chi}+2 s+\tau_s )\right)}{\Gamma \left(\Delta_{\f}-\frac{\tau_s }{2}\right) \Gamma \left(s+\frac{\tau_s }{2}\right)^4 \Gamma \left(-\frac{d}{2}+\Delta_{\f}+s+\frac{\tau_s }{2}\right)},
\eea
which decays exponentially for large $\Delta_{\chi}$. This is consistent with results from OPE convergence \cite{Pappadopulo:2012jk} and is similar to what was found using the ANEC \cite{Cordova:2017zej} for other mixed systems. For this system, the ANEC bound is trivial since the stress-tensor Ward identity implies $\<\f_{i}T\f_{j}\>\propto \delta_{ij}$ for scalar $\f_{i}$. We will find similar bounds if we replace $\chi$ by a more general operator.

The next simplest case to consider is an external, conserved current $V^{\mu}$. A similar calculation was also presented for $d=4$ and non-conserved vectors in \cite{Hartman:2016lgu}. Based on symmetries, we find:
\bea
\<e_1\cdot V|\mathcal{E}^{(s)}(n)|e_2\cdot V\>=a^{(s)}_{0} e_1\cdot e_2\left(1+a^{(s)}_{2}\left(\frac{e_1\cdot n \ e_{2}\cdot n}{e_1\cdot e_2}-\frac{1}{d-1}\right)\right),
\eea
which implies:
\bea
a^{(s)}_{0}\geq 0, \quad -\frac{d-1}{d-2}\leq a^{(s)}_{2}\leq d-1.
\eea

The HS ANEC bound is identical in form to the ANEC bound \cite{Hofman:2008ar}. One difference however is $a^{(s=2)}_{0}$ is related to $\<VV\>\propto C_{V}$ by the Ward identity (\ref{eq:WardGen}), but there is no such relation for $s>2$. In general, when $s=2$ conservation can give additional relations between the $a^{(2)}_{i}$ coefficients, but here conservation for the external conserved currents is constraining enough that conservation of the stress-tensor does not yield any additional relations \cite{Costa:2011mg}.

To find bounds on the underlying OPE coefficients, we will parametrize the three point function using the basis introduced in \cite{Costa:2011mg}:
\begin{equation}
\begin{aligned}
\hspace{-.2in}\<\O_{\Delta_1,\ell_1}\O_{\Delta_2,\ell_2}\O_{\Delta_3,\ell_3}\>=\sum_{n_{12},n_{13},n_{23}}c^{(123)}_{n_{23},n_{13},n_{12}}\frac{V_{1}^{\ell_1-n_{12}-n_{13}}V_{2}^{\ell_2-n_{23}-n_{12}}V_{3}^{\ell_3-n_{13}-n_{23}}H_{12}^{n_{12}}H_{13}^{n_{13}}H_{23}^{n_{23}}}{x_{12}^{\Delta_{1}+\Delta_{2}-\Delta_{3}+\ell_{1}+\ell_{2}-\ell_{3}}x_{13}^{\Delta_{1}-\Delta_{2}+\Delta_{3}+\ell_{1}-\ell_{2}+\ell_{3}}x_{23}^{-\Delta_{1}+\Delta_{2}+\Delta_{3}-\ell_{1}+\ell_{2}+\ell_{3}}}
\\
\hfill n_{ij}+n_{ik}\leq \ell_{i}.  \label{eq:ThreePtBasis} 
\end{aligned}
\end{equation}
The definitions of $V_{i}$ and $H_{ij}$ can be found in appendix \ref{app:integrals}.
 
For $\<VJ^{(s)}V\>$, permutation symmetry and conservation imply there are two independent OPE coefficients, which we will take to be $c^{(VsV)}_{000}$ and $c^{(VsV)}_{001}$. The bounds then become:
\bea
e^{-\frac{1}{2} i \pi  s}(2 c^{(VsV)}_{001}-c^{(VsV)}_{000})\geq 0,
\\
e^{-\frac{1}{2} i \pi  s}(c^{(VsV)}_{001} (2 d-\tau_{s}-4)+c^{(VsV)}_{000} (-d+s+\tau_{s}+2))\geq 0.
\eea
which agrees with results found in \cite{Hartman:2016lgu}.

Unlike the case of scalar states, the HS ANEC matrix elements do not vanish in the limit $\tau_{s}\rightarrow 2\tau_{V}$. However, we now have the free parameter $a^{(s)}_{2}$ which can be tuned such that (HS) ANEC is saturated. It has been demonstrated in \cite{zhib,Cordova:2017zej,Meltzer:2017rtf} that saturation of the ANEC bounds yields strong constraints on the CFT data, and it was conjectured in \cite{zhib} that if ANEC is saturated in a state created by the stress-energy tensor, then the theory is free. In section \ref{sec:SatANEC} we will give a proof of this statement.

By requiring positivity for spin-$s$ HS ANEC in the state $|\Theta\>=|e\cdot V\> + c_{\f}|\f\>$, for scalar $\f$, we can also bound the three-point function $\<\f J^{(s)}V\>$ and find similar results as for a system of scalars. Assuming the external current is conserved, the three-point function $\<VJ^{(s)}\f\>$ depends on a single OPE coefficient, $c^{(Vs\f)}_{000}$, and the off-diagonal HS ANEC matrix element is:
\begin{small}
\begin{equation}
\hspace{-.25in}\mathcal{E}^{(s,0)}_{V\f}=\frac{i \pi ^{\frac{d}{2}+2} e^{\frac{i \pi  s}{2}} 2^{-\Delta_{\f}-s+4} \Gamma (2 s+\tau_{s} -1)c^{(Vs\f)}_{000}}{\Gamma \left(\frac{1}{2} (d+\Delta_{\f}-\tau_{s} )\right) \Gamma \left(\frac{1}{2} (\Delta_{\f}+2 s+\tau_{s} -2)\right) \Gamma \left(\frac{1}{2} (d-\Delta_{\f}+2 s+\tau_{s} -2)\right) \Gamma \left(\frac{1}{2} (-d+\Delta_{\f}+2 s+\tau_{s} +2)\right)}.
\end{equation}
\end{small}

The HS ANEC bound then implies:
\begin{equation}
\begin{aligned}
\hspace{-0in}\frac{\big(c^{(Vs\f) \hspace{.0001in}}_{000}\big)^{2}}{c_{\f\f s}\big(2c^{(VsV)}_{001}-c^{(VsV)}_{000}\big)} \ \leq& \  \frac{\Gamma \left(\frac{1}{2} (d+\Delta_{\f}-\tau_{s} )\right)^2 \Gamma \left(\frac{1}{2} (\Delta_{\f}+2 s+\tau_{s} -2)\right)^2 }{ \Gamma \left(d-\frac{\tau_{s} }{2}-1\right) \Gamma \left(\Delta_{\f}-\frac{\tau_{s} }{2}\right) \Gamma \left(s+\frac{\tau_{s} }{2}\right)^3 \Gamma \left(s+\frac{\tau_{s} }{2}+1\right) }
\nonumber \\ &\times \frac{\Gamma \left(\frac{1}{2} (d-\Delta_{\f}+2 s+\tau_{s} -2)\right)^2 \Gamma \left(\frac{1}{2} (-d+\Delta_{\f}+2 s+\tau_{s} +2)\right)^2}{\Gamma \left(\frac{1}{2} (d+2 s+\tau_{s} -4)\right) \Gamma \left(-\frac{d}{2}+\Delta_{\f}+s+\frac{\tau_{s} }{2}\right)}.
\end{aligned}
\end{equation}
As before, if $\Delta_{\f}$ is the lightest operator in the theory, then $\tau_{s}\approx 2\Delta_{\f}+\gamma_{s}$ and this bound becomes stronger as $\gamma_{s}\rightarrow 0$. 

Finally, we will present bounds for systems involving the stress-tensor in $d=3$. Conservation, plus extra degeneracy conditions in $d=3$, implies $\<TJ^{(s)}T\>$ is a function of two OPE coefficients, $c^{(TsT)}_{002}$ and $c^{(TsT)}_{101}$. We will only need the spin-0 matrix element $\mathcal{E}^{(s,0)}_{TT}$:
\begin{small}
\begin{equation}
\hspace{-.25in}0\leq \ \mathcal{E}^{(s,0)}_{TT}=\frac{\pi ^3 e^{\frac{i \pi  s}{2}} 2^{s+\tau_{s}-1} \left(2 c^{(TsT)}_{002} \left(s^2+s (4 \tau_{s}-13)+4 (\tau_{s}-3) (\tau_{s}-2)\right)+c^{(TsT)}_{101} (s (s+3)-(\tau_{s}-9) \tau_{s}-12)\right)}{(s-1) s \left(19 s^2+8 (4 s-5) \tau_{s}-59 s+12 \tau_{s}^2+36\right) \Gamma \left(3-\frac{\tau_{s}}{2}\right) \Gamma \left(s+\frac{\tau_{s}}{2}\right)}.
\end{equation}
\end{small}

After imposing conservation, $\<TJ^{(s)}\f\>$ is a function of a single OPE coefficient, $c^{(Ts\f)}_{000}$, and we find:
\begin{equation}
\hspace{-.25in}\frac{\big(c^{(Ts\f)}_{000}\big)^{2}}{c_{\f\f s}\mathcal{E}^{(s,0)}_{TT}}\leq \frac{e^{-\frac{3}{2}  i \pi  s} 2^{-2 \Delta_{\f}-3 s-2 \tau_{s}+6} \Gamma \left(\frac{1}{2} (\Delta_{\f}-\tau_{s}+5)\right)^2 \Gamma \left(\frac{1}{2} (-\Delta_{\f}+2 s+\tau_{s}+1)\right)^2 \Gamma (\Delta_{\f}+2 s+\tau_{s}-2)^2}{\pi ^{5/2} \Gamma \left(\Delta_{\f}-\frac{\tau_{s}}{2}\right) \Gamma \left(s+\frac{\tau_{s}}{2}\right)^2 \Gamma (2 s+\tau_{s}-1) \Gamma \left(\Delta_{\f}+s+\frac{\tau_{s}}{2}-\frac{3}{2}\right)},
\end{equation}
where the extra phase factor on the right hand side is cancelled by the phase in $\mathcal{E}^{(s,0)}_{TT}$. We see once again that CFTs close to a generalized free field description are strongly constrained by HS ANEC. 

\sec{Spin-2/Spin-4 Mixed Systems}
\label{sec:Spin24_Systems}

We will now study the constraints of ANEC and $s=4$ HS ANEC on the state 
\bea
 |\Psi\>=|\epsilon_{2}\cdot T\>+|\epsilon_{4}\cdot J^{(4)}\>,
 \eea
where $J^{(4)}$ is the lightest spin-4 operator in the $TT$ OPE. Throughout, we will assume HS ANEC is obeyed and that there is a unique, minimal twist, spin-4 operator in the $TT$ OPE. Given the size of the full (HS) ANEC matrices, for general dimensions $d$ we will focus on the implications of ANEC saturation and for $d=3$ we will give new, non-perturbative constraints when the spin-4 operator has a small, but finite, anomalous dimension. To remove clutter, when writing OPE coefficients, $J$, without a superscript, will always refer to the lightest, spin-4 operator. Examples of relevant matrix elements and the solution to the Ward identity for $\<J^{(4)}TJ^{(4)}\>$ can be found in appendices \ref{app:ConsWard} and \ref{app:spin2_4}. 

\ssec{Saturation of ANEC}
\label{sec:SatANEC}
In this section, we will prove that if the lightest spin-$4$ operator has a non-zero anomalous dimension, then the ANEC bounds for $\<TTT\>$ cannot be saturated. Equivalently, if the ANEC bounds on the stress-tensor are saturated, then the theory has a higher-spin symmetry\cite{Maldacena:2011jn,Boulanger:2013zza,Alba:2015upa}.

In order to show this, we will require the following matrices are positive:
\begin{align}
\<\Psi|\mathcal{E}^{(2)}(n)|\Psi\>^{(j)}=
\left( \begin{array}{cc}
\mathcal{E}^{(2,j)}_{TT}  &\mathcal{E}^{(2,j)}_{TJ}  \vspace{.1in} \\
 \mathcal{E}^{(2,j)}_{JT} & \mathcal{E}^{(2,j)}_{JJ}
 \end{array} \right) \succeq 0, \label{eqn:ANEC_24}
 \\
 \nonumber
 \\
 \<\Psi|\mathcal{E}^{(4)}(n)|\Psi\>^{(j)}=
\left( \begin{array}{cc}
\mathcal{E}^{(4,j)}_{TT}  &\mathcal{E}^{(4,j)}_{TJ}   \vspace{.1in}  \\
 \mathcal{E}^{(4,j)}_{JT} & \mathcal{E}^{(4,j)}_{JJ}
 \end{array} \right) \succeq 0, \label{eqn:HSANEC_24}
\end{align}
where for $j>2$ the matrix elements involving $T$ vanish.

In this system of positivity conditions, the OPE coefficients for $\<TJ^{(4)}T\>$ play a double role: they appear in the diagonal elements $\mathcal{E}^{(4,j)}_{TT}$ and the off-diagonal elements $\mathcal{E}^{(2,j)}_{TJ}$. If we only used ANEC we would get upper bounds for $\<TJ^{(4)}T\>$ in terms of $\<TTT\>$ and $\<J^{(4)}TJ^{(4)}\>$, but we would miss additional positivity constraints from spin-$4$ HS ANEC. The OPE coefficients in $\<J^{(4)}TJ^{(4)}\>$ similarly appear in both diagonal and off-diagonal matrix elements.

First, we will use conservation and permutation symmetry to reduce $\<TJ^{(4)}T\>$ to a function of three OPE coefficients. In terms of the structures given in (\ref{eq:ThreePtBasis}), one independent basis of OPE coefficients is $c^{(TJT)}_{002}$, $c^{(TJT)}_{011}$, and $c^{(TJT)}_{101}$. In $d=4$, we find the following inequalities from requiring $\mathcal{E}^{(4,j)}_{TT}\geq 0$:
\begin{align}
0\leq& \ 2 (\tau_{4} (3 \tau_{4}-8)+8) c^{(TJT)}_{002}-\left(\tau_{4}^2+8\right) c^{(TJT)}_{011}-(\tau_{4}-6) \tau_{4} c^{(TJT)}_{101},
\\
\nonumber
\\
0\leq& \  2 (4-\tau_{4}) \left(9 \tau_{4}^2+42 \tau_{4}+8\right) c^{(TJT)}_{002}+3 (\tau_{4}-4) (\tau_{4}+4) (\tau_{4}+6) c^{(TJT)}_{011}
\nonumber \\ & +(\tau_{4}-6) (3 \tau_{4} (\tau_{4}+2)-32) c^{(TJT)}_{101},
\\
\nonumber
\\
0\leq&\ 2 \left(\tau_{4} \left(\tau_{4} \left(9 \tau_{4}^2+6 \tau_{4}-160\right)-240\right)-224\right) c^{(TJT)}_{002}-3 (\tau_{4}+10) \left(\tau_{4}^3-24 \tau_{4}-16\right) c^{(TJT)}_{011} \nonumber
\\ & -(\tau_{4}-6) (\tau_{4}+4) (3 (\tau_{4}-2) \tau_{4}-16) c^{(TJT)}_{101},
\end{align}
which come from the $j=0,1,2$ polarization tensors respectively. We have also dropped overall factors that are positive for $2\leq \tau_{4}\leq 4$. Since we have the same number of OPE coefficients as positivity conditions, it will be convenient to trade $c^{(TJT)}_{ijk}$ for the manifestly positive matrix elements $\mathcal{E}^{(4,j)}_{TT}$. 

To see how the proof works, it is not necessary to calculate $\mathcal{E}^{(2,j)}_{TJ}$ for $j=0,1,2$ in full detail. If $\mathcal{E}^{(2,j)}_{TT}=0$ for some $j$, then (\ref{eqn:ANEC_24}) implies $\mathcal{E}^{(2,j)}_{TJ}=0$. This yields $5-j$ equations while the underlying three-point function, $\<TJ^{(4)}T\>$, is a function of three OPE coefficients. Generically, if we saturate a single ANEC bound for $\<TTT\>$, we get at least as many equations as unknowns. Keeping $\tau_{4}>d-2$, we find there is a unique solution: $\<TJ^{(4)}T\>=0$. Therefore, we see that it is impossible to assume there exists a spin-4 operator in the $TT$ OPE with a non-zero anomalous dimension and that any individual $\<TTT\>$ bound is saturated. We do not need to use HS ANEC for this argument, although we will see that it does allow us to derive stronger conclusions.

As an example, we can consider a $4d$ CFT and set $\mathcal{E}^{(2,2)}_{TT}=0$, or equivalently set $n_{T}=0$ in (\ref{eq:TTTFF}). This imposes the following constraints: 
\begin{align} 
0=&\bigg((\tau_{4} -2) \left(3 \tau_{4}  (\tau_{4} +10) (5 \tau_{4} +42) \mathcal{E}^{(4,0)}_{TT}+4 (\tau_{4}  (\tau_{4}  (7 \tau_{4} +138)+1064)+3312) \mathcal{E}^{(4,1)}_{TT}\right)\nonumber
\\ &\hspace{.1in}+8 (\tau_{4}  (\tau_{4}  (\tau_{4}  (\tau_{4} +22)+308)+2168)+5760) \mathcal{E}^{(4,2)}_{TT}\bigg), \label{eqn:nvSat4dEx1}
\\ \nonumber
\\
0=&(\tau_{4} -2) \left(-9 \tau  (\tau +8)  \mathcal{E}^{(4,0)}_{TT}-48 (\tau +10)  \mathcal{E}^{(4,1)}_{TT}+12 (\tau  (\tau +12)+40)  \mathcal{E}^{(4,2)}_{TT}\right),  \label{eqn:nvSat4dEx2}
\\ \nonumber
\\
0=&(\tau_{4} -2) \bigg(-\frac{3}{4}(\tau -6) \tau  (\tau +10)  \mathcal{E}^{(4,0)}_{TT}+((\tau -4) \tau  (\tau +10)+144)  \mathcal{E}^{(4,1)}_{TT}
\nonumber \\ &\hspace{.7in}+2 \tau  (\tau  (\tau +12)+44)  \mathcal{E}^{(4,2)}_{TT}\bigg).  \label{eqn:nvSat4dEx3}
\end{align} 

We can now also see how free field theories\footnote{The derivation of HS ANEC given in \cite{Hartman:2016lgu} only holds for interacting field theories, but they also showed that it holds in a theory of free scalars.} or theories with a slightly broken higher spin symmetry\footnote{We have focused on parity even three-point functions, although there should be a similar story if we include parity odd structures.} can saturate ANEC and not be ruled out \cite{Hofman:2008ar,Cordova:2017zej,Chowdhury:2017vel}. Here, if we set $\tau_{4}=2$, we only need to set $\mathcal{E}^{(4,2)}_{TT}=0$ and can leave the other matrix elements non-zero. 

In general dimensions we find:
\begin{align}
\mathcal{E}^{(2,j)}_{TT}=0 \quad \Longrightarrow \quad \mathcal{E}^{(2,j)}_{TJ}=0 \quad \Longrightarrow \quad \mathcal{E}^{(4,j)}_{TT}=0 \ \& \ \tau_{4}=d-2, \label{eq:ANECsatGen}.
\end{align}
That is, saturation of ANEC implies saturation of HS ANEC and the presence of a higher spin conserved current. 

If we assume HS ANEC is true, we actually only need to use (\ref{eqn:nvSat4dEx1}) to prove that ANEC saturation implies the theory is free. Each matrix element, $\mathcal{E}^{(4,j)}_{TT}$, in (\ref{eqn:nvSat4dEx1}) is non-negative and multiplies a non-negative coefficient, so either $\tau_{4}=2$ or all the OPE coefficients vanish. Since this off-diagonal element is strictly positive in interacting CFTs, it also gives a strong lower bound on $\mathcal{E}^{(2,2)}_{TT}\mathcal{E}^{(2,2)}_{JJ,00}$. There are analogous results for $j=0, 1$ and also for general $d$ which are given in Appendix \ref{app:GenD}.

At the level of the full correlators, $\<\Psi|\mathcal{E}^{(2)}(n)|\Psi\>$ and $\<\Psi|\mathcal{E}^{(4)}(n)|\Psi\>$, when $\tau_{4}=d-2$ the spin-4 operator is conserved and its longitudinal modes decouple. This decoupling means that instead of obtaining $5-j$ equations from saturation of the spin-$j$ bound, we have a single constraint for each one. This reduction in constraints is what prevents free field theories from being ruled out and explains the factors of $\tau_{4}-2$ in (\ref{eqn:nvSat4dEx1})-(\ref{eqn:nvSat4dEx3}).

We can extend the result (\ref{eq:ANECsatGen}) by using the spin-4 HS ANEC condition (\ref{eqn:HSANEC_24}). In particular, if $\mathcal{E}^{(4,j)}_{TT}=0$ then we also have $\mathcal{E}^{(4,j)}_{TJ}=0$. We can solve the latter constraint in terms of the underlying three-point function, $\<J^{(4)}TJ^{(4)}\>$, which also determines $\mathcal{E}^{(2,i)}_{JJ}$. We find that requiring $\mathcal{E}^{(4,j)}_{TJ}=0$ and the diagonal elements of $\mathcal{E}^{(2,i)}_{JJ}$ are non-negative implies ANEC is saturated in a state created by $J^{(4)}$. We give an example of this phenomenon in Appendix \ref{app:ANEC_Sat_Spin4} for $d=4$.

In total we find the following chain of implications:
\begin{equation}
\begin{aligned}
&\mathcal{E}^{(2,j)}_{TT}=0 \hspace{.12in} \Longrightarrow \hspace{.12in}   \mathcal{E}^{(2,j)}_{TJ}=0 \hspace{.12in} \Longrightarrow \hspace{.12in}   \mathcal{E}^{(4,j)}_{TT}=0 \ \& \ \tau_{4}=d-2  \hspace{.12in} \Longrightarrow \hspace{.12in}  \mathcal{E}^{(4,j)}_{TJ}=0 \hspace{.12in} \Longrightarrow \hspace{.12in}   \mathcal{E}^{(2,q)}_{JJ}=0 \hspace{2in} \nonumber 
\\ &\hspace{5.5in} \text{ for } q=j,\ 3, \ 4. \label{eq:ANEC_sat}
\end{aligned}
\end{equation}
In deriving the last implication, we used $\tau_{4}=d-2$ and the constraints of (HS) ANEC, but did not use the existence of a higher spin symmetry.

The fact $\mathcal{E}^{(2,3)}_{JJ}=\mathcal{E}^{(2,4)}_{JJ}=0$ when we saturate an ANEC bound for the stress-tensor is consistent with the structure of free field theories. For $d\geq 4$ and even, we have three field theories and they only generate the $j=0,1,2$ structures for $\mathcal{E}^{(2,j)}_{JJ}$. 

We can also ask if the reverse situation is possible, that is can we have $\mathcal{E}^{(2,j)}_{JJ}\neq0$ only for $j=3$ or $4$? To see that this is impossible, we can study the following spin-$0$ matrix element\footnote{The polarization tensor used to calculate this matrix element, $\lambda^{(4,0)}_{4}$, is independent of $n$, so the Ward identity (\ref{eq:WardGen}) implies it is proportional to $C^{(4)}_{J}$. We thank David Simmons-Duffin, Petr Kravchuk, and Eric Perlmutter for discussions on this point.}:
\begin{equation}
\mathcal{E}^{(2,0)}_{JJ,44}= C^{(4)}_{J}\frac{2^{d-2 \tau_{4}-7} (d-1) (d+1) \pi ^{\frac{d}{2}+1}(\tau_{4}-d+5) (\tau_{4}-d+4) (\tau_{4}-d+3) (\tau_{4}-d+2)}{(d+2) (d+4) \Gamma (\tau_{4}+8) \Gamma \left(-\frac{d}{2}+\tau_{4}+5\right)}. \label{eq:JTJWardECorrQ}
\end{equation}
If $\tau_4>d-2$ then this matrix element is necessarily non-zero. If $\tau_4=d-2$ this matrix element does vanish, but in that case the theory has a higher-spin symmetry and the correlation functions of the leading trajectory are fixed to coincide with a free field theory of scalars, fermions, or tensors \cite{Boulanger:2013zza,Alba:2015upa}.

Furthermore, if $\mathcal{E}^{(2,j)}_{TT}=\mathcal{E}^{(4,j)}_{TT}=\mathcal{E}^{(2,q)}_{JJ}=0$, then there are also an infinite number of constraints for operators not on the leading Regge trajectory. Using the interference arguments found in \cite{Cordova:2017zej,Meltzer:2017rtf,Afkhami-Jeddi:2018own}, saturation of these bounds implies:
\begin{equation}
\mathcal{E}^{(2,j)}_{T\O}=\mathcal{E}^{(4,j)}_{T\O}=\mathcal{E}^{(2,q)}_{J\O}=0, \quad \text{ for } q=j,\ 3, \ 4,
\end{equation}
where $\O$ is any operator in our CFT.  

When the ANEC bounds are close to being saturated, e.g. $\mathcal{E}^{(2,2)}_{TT}\ll1$, we find the following scalings:
\begin{align}
\left(\mathcal{E}^{(4,0)}_{TT}\gamma_{4}\right)^{2}\lesssim \mathcal{E}^{(2,2)}_{TT}\mathcal{E}^{(2,2)}_{JJ,00},
\\
\left(\mathcal{E}^{(4,1)}_{TT}\gamma_{4}\right)^{2}\lesssim \mathcal{E}^{(2,2)}_{TT}\mathcal{E}^{(2,2)}_{JJ,00},
\\
\left(\mathcal{E}^{(4,2)}_{TT}\right)^{2}\lesssim \mathcal{E}^{(2,2)}_{TT}\mathcal{E}^{(2,2)}_{JJ,00}.
\end{align}
A natural assumption is if $\mathcal{E}^{(2,2)}_{TT}\ll1$, then $\mathcal{E}^{(2,2)}_{JJ,00}\ll1$, but this is not imposed by (HS) ANEC.

One question that remains is the general behavior of $\mathcal{E}^{(4,r)}_{JJ}$, or its underlying three-point function $\<J^{(4)}J^{(4)}J^{(4)}\>$. From (\ref{eqn:HSANEC_24}) we have the positivity condition:
\begin{equation}
\left(\mathcal{E}^{(4,r)}_{TJ,0a}\right)^{2}\leq \mathcal{E}^{(4,r)}_{TT,00}\mathcal{E}^{(4,r)}_{JJ,aa}, \qquad a=0,1,...5-r
\end{equation}
so $\mathcal{E}^{(4,r)}_{TT,00}\mathcal{E}^{(4,r)}_{JJ,aa}$ is bounded from below by $\mathcal{E}^{(4,r)}_{TJ,0a}$. The underlying three-point function for the latter matrix element is $\<J^{(4)}TJ^{(4)}\>$, which depends on $12$ OPE coefficients after imposing conservation. The stress-tensor Ward identity (\ref{eq:WardGen}) yields two additional constraints, so $\<J^{(4)}TJ^{(4)}\>$ depends on $10$ OPE coefficients and the normalization of $J^{(4)}$. 

To completely remove the lower bounds on $\mathcal{E}^{(4,r)}_{TT,00}\mathcal{E}^{(4,r)}_{JJ}$ for all $r$, we need $\mathcal{E}^{(4,r)}_{TJ}=0$ for all $r$ as well. This yields $12$ equations and if $\tau_{4}>d-2$ the only solution is $\<J^{(4)}TJ^{(4)}\>=0$ identically. This is of course inconsistent with the Ward identity, so it is impossible to remove all lower bounds on $\mathcal{E}^{(4,r)}_{TT,00}\mathcal{E}^{(4,r)}_{JJ}$ in interacting CFTs. We will give an example of this in the next section for $d=3$. 

We expect this pattern holds for general $J^{(s)}$, since $\<TJ^{(s)}J^{(s)}\>$ depends on $3s$ independent OPE coefficients \cite{Zhiboedov:2012bm} and setting $\mathcal{E}^{(4,r)}_{TJ}=0$ for all $r$ yields $3s$ equations. The Ward identity for $\<TJ^{(s)}J^{(s)}\>$ gives extra constraints, so we have an over-constrained system and we generically expect there is no solution consistent with the Ward identity. It would be interesting to prove this statement for general $s$.

\ssec{Application to the 3d Ising CFT}
\label{sec:ising}

To be more concrete, we will now specialize to $3d$ CFTs where the lightest spin-4 operator has twist $\tau_{4}=1.02$. This is close to the spectrum of the $3d$ Ising model \cite{ElShowk:2012ht}, but we will not actually use any other information about the theory. The structure of these bounds will carry over to the $O(N)$ models and also to higher dimensions. What is special to $3d$ is we can set any three-point structure proportional to $H_{12}H_{13}H_{23}$ in (\ref{eq:ThreePtBasis}) to zero and we also only have the $j=0, 1$ bounds. We will present approximate, numerical values for simplicity, but of course there exist analytic expressions for all quantities. We continue to denote the lightest spin-$4$ operator by $J$ inside any matrix element or OPE coefficient.

In $d=3$, the matrix elements for $\<T|\mathcal{E}^{(2)}|T\>$, in our normalization, are
\begin{align}
\mathcal{E}^{(2,0)}_{TT}=\frac{n_{B}}{32 \pi },
\qquad
\mathcal{E}^{(2,1)}_{TT}=\frac{n_{F}}{32 \pi }.
\end{align}
We are following the conventions of \cite{Cordova:2017zej} where in free field theories $n_{B,F}$ count the total bosonic and fermionic degrees of freedom, respectively.

We now find positivity for $\<T|\mathcal{E}^{(4)}|T\>$ implies:
\begin{align}
0&\leq \mathcal{E}^{(4,0)}_{TT}=0.315 c^{(TJT)}_{101}-0.311 c^{(TJT)}_{002},
\\
0&\leq \mathcal{E}^{(4,1)}_{TT}=-0.207 c^{(TJT)}_{101}+0.702 c^{(TJT)}_{002}.
\end{align}

To study the full constraints of ANEC, we next consider the spin-0 positivity bounds for $\<J^{(4)}|\mathcal{E}^{(2)}|J^{(4)}\>$:
\begin{align}
&0\leq \mathcal{E}^{(2,0)}_{JJ,00}=-0.0123 c^{(JTJ)}_{000}+0.0275 c^{(JTJ)}_{010}+0.00723 c^{(JTJ)}_{012}-0.0642 c^{(JTJ)}_{020}-0.0671 c^{(JTJ)}_{022}
\nonumber \\ & \hspace{.86in} +0.161 c^{(JTJ)}_{030}-0.575 c^{(JTJ)}_{040}-0.0381 C_{J}^{(4)},
\\ \nonumber
\\
&0\leq \mathcal{E}^{(2,0)}_{JJ,11}=-0.00197 c^{(JTJ)}_{000}+0.00444 c^{(JTJ)}_{010}+0.00118 c^{(JTJ)}_{012}-0.0104 c^{(JTJ)}_{020}-0.0108 c^{(JTJ)}_{022}
\nonumber \\ & \hspace{.86in} +0.0259 c^{(JTJ)}_{030}-0.0929 c^{(JTJ)}_{040}-0.00616 C_{J}^{(4)},
\\ \nonumber
\\
&0\leq \mathcal{E}^{(2,0)}_{JJ,22}=-0.000277 c^{(JTJ)}_{000}+0.000625 c^{(JTJ)}_{010}+0.000162 c^{(JTJ)}_{012}-0.00146 c^{(JTJ)}_{020}
\nonumber \\ & \hspace{.86in}-0.00151 c^{(JTJ)}_{022} +0.00365 c^{(JTJ)}_{030}-0.0131 c^{(JTJ)}_{040}-0.000862 C_{J}^{(4)},
\end{align}
\begin{align}
&0\leq \mathcal{E}^{(2,0)}_{JJ,33}=-0.0000276 c^{(JTJ)}_{000}+0.0000621 c^{(JTJ)}_{010}+0.0000152 c^{(JTJ)}_{012}-0.000145 c^{(JTJ)}_{020}
\nonumber \\ & \hspace{.86in} -0.000146 c^{(JTJ)}_{022}+0.000362 c^{(JTJ)}_{030}-0.00130 c^{(JTJ)}_{040}-0.0000847 C_{J}^{(4)},
\\ \nonumber
\\
&0\leq \mathcal{E}^{(2,0)}_{JJ,44}=1.91\times10^{-8} C^{(4)}_{J}.
\end{align}
where $C_{J}^{(s)}$ is defined by:
\bea
\<J^{(s)}(x_1)J^{(s)}(x_2)\>=C^{(s)}_{J}\frac{H_{12}^s}{x_{12}^{2(\tau_{s}+2s)}}.
\eea
As a reminder, we are using the basis introduced in (\ref{eq:polbasis}) for the polarization tensors. To see why the matrix element $ \mathcal{E}^{(2,0)}_{JJ,44}$ is small, recall from (\ref{eq:JTJWardECorrQ}) that it vanishes when $\tau\rightarrow 1$ in $d=3$.

The spin-1 positivity conditions similarly yield:
\begin{align}
&0\leq \mathcal{E}^{(2,1)}_{JJ,00}=0.0113 c^{(JTJ)}_{000}-0.0253 c^{(JTJ)}_{010}-0.00662 c^{(JTJ)}_{012}+0.0589 c^{(JTJ)}_{020}+0.0616 c^{(JTJ)}_{022}
\nonumber \\ & \hspace{.86in}-0.147 c^{(JTJ)}_{030}+0.528 c^{(JTJ)}_{040}+0.0351 C_{J}^{(4)},
\\ \nonumber
\\
&0\leq \mathcal{E}^{(2,1)}_{JJ,11}=0.00191 c^{(JTJ)}_{000}-0.00429 c^{(JTJ)}_{010}-0.00115 c^{(JTJ)}_{012}+0.0100 c^{(JTJ)}_{020}\nonumber \\ & \hspace{.86in}-0.0250 c^{(JTJ)}_{030}+0.0898 c^{(JTJ)}_{040}+0.00596 C_{J}^{(4)},
\\ \nonumber 
\\
&0\leq \mathcal{E}^{(2,1)}_{JJ,22}=0.000277 c^{(JTJ)}_{000}-0.000625 c^{(JTJ)}_{010}-0.000164 c^{(JTJ)}_{012}+0.00146 c^{(JTJ)}_{020}
\nonumber \\ & \hspace{.86in}+0.00152 c^{(JTJ)}_{022}-0.00364 c^{(JTJ)}_{030}+0.0131 c^{(JTJ)}_{040}+0.000863 C_{J}^{(4)},
\end{align}
\begin{align}
&0\leq \mathcal{E}^{(2,1)}_{JJ,33}=0.0000276 c^{(JTJ)}_{000}-0.0000621 c^{(JTJ)}_{010}-0.0000152 c^{(JTJ)}_{012}+0.000145 c^{(JTJ)}_{020}
\nonumber \\ & \hspace{.86in}+0.000146 c^{(JTJ)}_{022}-0.000362 c^{(JTJ)}_{030}+0.00130 c^{(JTJ)}_{040}+0.0000848 C_{J}^{(4)}.
\end{align}

To make positivity manifest, we will now parameterize $\<TJ^{(4)}T\>$ by $\mathcal{E}^{(4,j)}_{TT}$. In Appendix \ref{app:ising} we similarly parameterize $\<J^{(4)}TJ^{(4)}\>$ by $ \mathcal{E}^{(2,0)}_{JJ,rr}$ for $r=0,1, 2$ and $ \mathcal{E}^{(2,1)}_{JJ,qq}$ for $q=0,...,3$. Of course, we also have non-linear constraints for the matrices $\mathcal{E}^{(2,j)}_{JJ}$ by requiring positivity of all possible principal minors, from size $2\times2$ to $5\times5$, but we will not discuss that here. 

Instead, we will consider positivity of all $2\times2$ principal minors in (\ref{eqn:ANEC_24}) which involve $\<TTT\>$, $\<TJ^{(4)}T\>$ and $\<J^{(4)}TJ^{(4)}\>$. The resulting bounds are:
\bea
\left(0.0597 \mathcal{E}^{(4,0)}_{TT}+0.000439 \mathcal{E}^{(4,1)}_{TT}\right)^{2} \leq \mathcal{E}^{(2,0)}_{TT} \mathcal{E}^{(2,0)}_{JJ,00},
\label{eq:TTJisng0}
\\
\left(0.0000871 \big(\mathcal{E}^{(4,0)}_{TT}-\mathcal{E}^{(4,1)}_{TT}\big)\right)^{2} \leq \mathcal{E}^{(2,0)}_{TT} \mathcal{E}^{(2,0)}_{JJ,11},
\label{eq:TTJisng1}
\\
\left(0.0000128 \big(\mathcal{E}^{(4,0)}_{TT}- \mathcal{E}^{(4,1)}_{TT}\big)\right)^{2} \leq \mathcal{E}^{(2,0)}_{TT} \mathcal{E}^{(2,0)}_{JJ,22},\label{eq:TTJisng2}
\\
\left(3.319\times10^{-6} \big(\mathcal{E}^{(4,0)}_{TT}- \mathcal{E}^{(4,1)}_{TT}\big)\right)^{2} \leq \mathcal{E}^{(2,0)}_{TT} \mathcal{E}^{(2,0)}_{JJ,33}, \label{eq:TTJisng3}
\\
\left(1.002\times10^{-6} \big(\mathcal{E}^{(4,0)}_{TT}-\mathcal{E}^{(4,1)}_{TT}\big)\right)^{2} \leq \mathcal{E}^{(2,0)}_{TT} \mathcal{E}^{(2,0)}_{JJ,44}. \label{eq:TTJisng4}
\eea

The overall coefficients are small in (\ref{eq:TTJisng1}-\ref{eq:TTJisng4}) relative to (\ref{eq:TTJisng0}) because the former should vanish in the limit $\tau_{4}\rightarrow 1$. In fact, even though the right hand side of (\ref{eq:TTJisng4}) is $\sim10^{-8}$, the small prefactor on the left-hand side gives a weak bound on $\mathcal{E}^{(4,0)}_{TT}-\mathcal{E}^{(4,1)}_{TT}$. One nice feature of this particular bound though is it only involves the normalization of $J^{(4)}$, which we can always set to one, and not on the OPE coefficients of $\<J^{(4)}TJ^{(4)}\>$.

Here, the most powerful bound is generically (\ref{eq:TTJisng0}), which says that $\mathcal{E}^{(2,0)}_{TT} \mathcal{E}^{(2,0)}_{JJ,00}$ is bounded below by a strictly positive quantity in interacting CFTs. A similar feature was seen for $d=4$ in (\ref{eqn:nvSat4dEx1}).

Repeating this argument for the spin-1 polarization tensors, we find a similar structure:
\bea
\left(0.0004460 \mathcal{E}^{(4,0)}_{TT}+0.05966 \mathcal{E}^{(4,1)}_{TT}\right)^{2}\leq \mathcal{E}^{(2,1)}_{TT} \mathcal{E}^{(2,1)}_{JJ,00},\label{eq:TTJisngF1}
\\
\left(0.0000854 (\mathcal{E}^{(4,0)}_{TT}-\mathcal{E}^{(4,1)}_{TT})\right)^{2}\leq \mathcal{E}^{(2,1)}_{TT} \mathcal{E}^{(2,1)}_{JJ,11},
\\
\left(0.0000123 (\mathcal{E}^{(4,0)}_{TT}-\mathcal{E}^{(4,1)}_{TT})\right)^{2} \leq \mathcal{E}^{(2,1)}_{TT} \mathcal{E}^{(2,1)}_{JJ,22},
\\
\left(3.32\times10^{-6} (\mathcal{E}^{(4,0)}_{TT}-\mathcal{E}^{(4,1)}_{TT})\right)^{2} \leq \mathcal{E}^{(2,1)}_{TT} \mathcal{E}^{(2,1)}_{JJ,33}.
\eea

Finally, we will consider how spin-4 HS ANEC (\ref{eqn:HSANEC_24}) constrains this same system of three-point functions. The matrix elements for $\mathcal{E}^{(4,j)}_{TJ}$ and $\mathcal{E}^{(4,j)}_{JJ}$ are fairly large, so we will leave their explicit form to Appendix \ref{app:ising}. Instead, we focus on the form of the quadratic bounds which involve the three-point functions $\<TJ^{(4)}T\>$, $\<J^{(4)}TJ^{(4)}\>$ and $\<J^{(4)}J^{(4)}J^{(4)}\>$:
\begin{eqnarray}
\big(\mathcal{E}^{(4,0)}_{TJ,0a}\big)^{2}&\leq \mathcal{E}^{(4,0)}_{TT}\mathcal{E}^{(4,0)}_{JJ,aa}, \qquad &\text{for } a=0,1,...,4 
\\
\big(\mathcal{E}^{(4,1)}_{TJ,0b}\big)^{2}&\leq \mathcal{E}^{(4,1)}_{TT}\mathcal{E}^{(4,1)}_{JJ,bb}, \qquad &\text{for } b=0,1,...,3.
\end{eqnarray}

Unlike $\mathcal{E}^{(2,j)}_{TJ}$, the off-diagonal matrix elements here, $\mathcal{E}^{(4,j)}_{TJ}$, do not have any manifest positivity properties. We do find however that it is impossible to set $\mathcal{E}^{(4,j)}_{TJ}=0$ for all $j$ without violating the Ward identity for $\<J^{(4)}TJ^{(4)}\>$, as was mentioned earlier. As a caveat, there may be some positivity property we are missing by not considering the non-linear constraints more systematically.

Finally, it is possible to rewrite these expressions as lower bounds on $\mathcal{E}^{(4,i)}_{TT}$ and use them inside the bounds (\ref{eq:TTJisng4}) and (\ref{eq:TTJisngF1}) to derive lower bounds on $\<TTT\>$ in terms of $\<J^{(4)}J^{(4)}J^{(4)}\>$ and $\<J^{(4)}TJ^{(4)}\>$. This kind of inequality, which relates $\<J^{(4)}J^{(4)}J^{(4)}\>$ and $\<TTT\>$, cannot be derived from ANEC alone. To derive it through a direct study of crossing symmetry would require studying a mixed system of four-point functions containing $T$ and $J^{(4)}$.

\sec{Comparison to Analytic Bootstrap}
\label{sec:LCBootstrap}
So far, we have focused on how HS ANEC constrains fundamental CFT data, but we will now take the reverse approach and study if predictions from the bootstrap are always compatible with HS ANEC. We will assume the leading trajectory is composed of double-twist operators, $J^{(s)}=[\f\f]_{n=0,s}$, for scalar $\f$. We can then use the analytic bootstrap to extract $\<\psi\psi[\f\f]_{0,s}\>$, for scalar $\p$, from the four-point function $\<\f\f\psi\psi\>$. 

In the examples we consider, the light operators in the $\f\p$ OPE which lead to corrections for $\<\psi\psi[\f\f]_{0,s}\>$ will not be the leading Regge trajectory itself. For example, we can consider light scalars, which do not fall on Regge trajectories, or assume a global symmetry prevents the exchange of $[\f\f]_{0,s}$, e.g. if $\f$ is odd and $\p$ is even under a $\mathbb{Z}_{2}$ symmetry.

One motivation for this analysis is that there are solutions to crossing symmetry at large $N$ which do not obey the spin-$s$ sum rules used in the derivation of (HS) ANEC. For example, in \cite{Hartman:2015lfa} they showed that a scalar EFT\footnote{For this theory there is no gravity in the bulk, so the CFT does not contain a stress-tensor and this is not a violation of the ANEC. However, the spin-two sum rule for the four-point function should still hold.} in AdS with a shift symmetry and a $\lambda (\partial \f)^{4}$ interaction violates the spin-2 sum rule if $\lambda<0$. This is the AdS version of the causality arguments in \cite{Adams:2006sv}. The higher spin sum rules can also be used to bound higher derivative, quartic interactions\footnote{See also \cite{Fitzpatrick:2016thx,Alday:2016htq} for related work.}. One of our goals is to understand if there are similar issues with cubic interactions.

We will show that the exchange of isolated operators or towers of double-twist operators in the $\f\p$ OPE is always consistent with HS ANEC, at finite and asymptotically large spin respectively. For CFTs with a large $N$ description, this corresponds to obeying basic causality constraints at tree and one-loop level, assuming the dual AdS theory only has cubic interactions. To obtain consistent results at finite spin, we also have to include non-perturbative effects in the large spin expansion. We will also present results for large spin OPE coefficients from ladder diagrams in the bulk which may be of interest for the study of large $N$ CFTs in general.

\ssec{$\grSL(2,\mathbb{R})$ Expansion}
To set up the conventions, we use the lightcone coordinates $x=(x^{-},x^{+},x^{i})$ and the metric $ds^{2}=dx^{+}dx^{-}+dx_{i}dx^{i}$. For the four-point function, we choose the standard configuration:
\begin{equation}
x_{1}=0, \quad x_{2}=(z,\bar{z},0), \quad x_{3}=(1,1,0), \quad x_{4}=\infty.
\end{equation}

We will study four-point functions of scalars, $\<\f_{1}\f_{2}\f_{3}\f_{4}\>$, and require the $\f_1\f_2\rightarrow \f_4\f_3$ and $\f_3 \f_2\rightarrow \f_4\f_1$ OPEs agree. This is also known as $s\leftrightarrow t$ crossing symmetry and implies:
\begin{equation}
\begin{aligned}
\left((1-z)(1-\bar{z})\right)^{\frac{\Delta_{2}+\Delta_{3}}{2}}\underset{\O}{\sum}c_{12\O}c_{43\O}&g^{\Delta_{12},\Delta_{34}}_{\O}(z,\bar{z})= 
\\ 
&(z\bar{z})^{\frac{1}{2} (\Delta_{1}+\Delta_{2})}\sum_{\O'}c_{32\O'}c_{41\O'}g^{\Delta_{32},\Delta_{14}}_{\O'}(1-z,1-\bar{z}).
\end{aligned}
\end{equation}
where the blocks are normalized as in \cite{DO3} with $c_{\ell}=1$\footnote{In comparison to \cite{dsdi}, these are the same blocks as $g^{\Delta_{12},\Delta_{34}}_{\Delta,\ell}$, while $G^{\frac{\Delta_{12}}{2},\frac{\Delta_{34}}{2}}_{h,\bar{h}}$ defined there has an extra factor of $v^{-\frac{\Delta_{12}}{2}}$.}.

We will consider the lightcone limit $1-\bar{z}\ll z\ll1$, which allows us to solve for $\<\f_{1}\f_{2}[\f_{3}\f_{4}]_{n,\ell}\>$ and $\<\f_{3}\f_{4}[\f_{1}\f_{2}]_{n,\ell}\>$ for $\ell\gg1$ in terms of the small twist operators in the $\f_1\f_4$ and $\f_2\f_3$ OPE. We also adopt the conventions of \cite{dsdi} and parametrize CFT data in terms of:
\bea
h=\frac{\Delta-\ell}{2}, \quad \bar{h}=\frac{\Delta+\ell}{2}.
\eea
This choice of variables is especially convenient when discussing conformal blocks in $1d$, or the $\grSL(2,\mathbb{R})$ blocks,
\begin{equation}
k^{h_{12},h_{34}}_{2h}(z)=z^{h} _{2}F_{1}(h-h_{12},h+h_{34},2h,z), \quad h_{ij}=h_{i}-h_{j}, \label{eq:SL2R}
\end{equation}
and also in even dimensions where the conformal blocks can be written down in terms of $1d$ blocks\cite{DO1,DO2}. In \cite{dsdi} it was also found how to sum the $\grSL(2,\mathbb{R})$ blocks to reproduce powers of $y=\frac{z}{1-z}$ plus terms which are Casimir-regular\footnote{In the language of \cite{Caron-Huot:2017vep}, these are terms with a vanishing double-discontinuity. See also \cite{Alday:2016njk} for an alternative method to sum degenerate twist trajectories.} in $z$. The expression we will need is:
\begin{align}
&\left(\frac{y}{1 + y}\right)^{-r}\sum_{\substack{h=h_{0}+\ell \\  \ell=0,1,...}}S^{r,s}_{a}(h)k^{r,s}_{2h}(1-z) =y^{a}+\sum\limits_{k=0}^{\infty}\left(B^{r,s}_{a,k}(h_0)y^{k-r}+B^{s,r}_{a,k}(h_0)y^{k-s}\right), \label{eq:mixedblocksum}
\\
&S^{r,s}_{a}(h)=\frac{\Gamma(h-r)\Gamma(h-s)\Gamma(h-a-1)}{\Gamma(-a-r)\Gamma(-a-s)\Gamma(2h-1)\Gamma(h+a+1)}, \label{eq:SCoefs}
\\
&B^{r,s}_{a,k}(h)=\frac{\pi  \Gamma (-a+h-1) \csc (\pi  (r-s)) \Gamma (h+k-r)}{\Gamma (k+1) \Gamma (a+h) (a-k+r) \Gamma (-a-r) \Gamma (-a-s) \Gamma (h-k+r-1) \Gamma (k-r+s+1)}. \label{eq:Bmixed}
\end{align}

We also need to expand the $\grSO(d,2)$ blocks in terms of the $\grSL(2,\mathbb{R})$ blocks:
\begin{align}
g^{2r,2s}_{h,\bar{h}}(z,\bar{z})=\sum\limits_{n=0}^{\infty}\sum\limits_{j=-n}^{n} A^{r,s}_{n,j}(h,\bar{h})y^{h+n}k^{r,s}_{2(\bar{h}+j)}(\bar{z}), \label{eq:dimred}
\end{align}
where the blocks are normalized so $A^{r,s}_{0,0}(h,\bar{h})=1$. Here we will only work to leading order in the $\grSL(2,\mathbb{R})$ expansion.

\ssec{Isolated Operator Exchange}
\label{sec:SingleOP}
Returning to the correlator $\<\f\f\p\p\>$, now that we can perform infinite sums of $\grSL(2,\mathbb{R})$ blocks in the $s$-channel to reproduce a pure power law in the $t$-channel, we can start to solve for $\<\psi\psi [\f\f]_{0,s}\>$. We can also consider $s\leftrightarrow u$ crossing, but it will give the same result up to factors of $(-1)^{s}$. Since we are interested in the case where $s$ is even, this factor is irrelevant and we can ignore the $u$-channel. 

 \begin{figure}
    \centering
       \includegraphics[width = .6\textwidth]{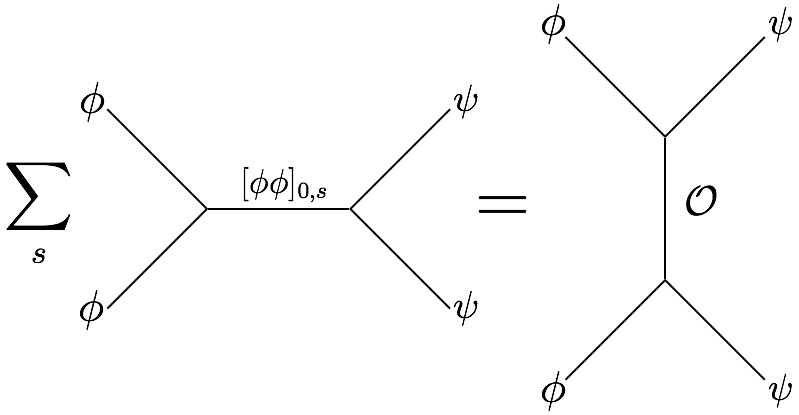}
         \caption{$s=t$ crossing equation for $c_{\f\f[\f\f]_{0,s}}c_{\p\p[\f\f]_{0,s}}$ with isolated operator in $t$-channel.} \label{fig4}
\end{figure}

At large $\bar{h}$, the contribution from a single $t$-channel primary $\O$ in the $\f\p$ OPE to $c_{\f\f[\f\f]_{0,s}}c_{\p\p[\f\f]_{0,s}}$ is \cite{light1,light2,dsdi}:
\bea 
c_{\f\f[\f\f]_{0,s}}(\bar{h})c_{\p\p[\f\f]_{0,s}}(\bar{h})\sim c_{\p\f\O}^{2}\frac{\sqrt{\pi } 2^{3-2\bar{h}} \Gamma (2 \bar{h}_{\O}) \Gamma (2(h_{\p}-h_{\f})) \bar{h}^{2h_{\f}+2h_{\p}-2 h_{\O}-\frac{3}{2}}}{\Gamma \left(h_\p-h_\f+\bar{h}_{\O}\right)^2 \Gamma \left(h_{\f}+h_{\p}-h_{\O}\right)^{2}}, \label{eq:isolatedff}
\eea
so as long as $h_{\p}>h_{\f}$ and unitarity is obeyed, the right hand side is positive and HS ANEC is satisfied. In contrast, $c_{\f\f[\p\p]_{0,s}}c_{\p\p[\p\p]_{0,s}}$ is not sign definite. If we also considered $[\f\f]_{0,s}$ for odd spin, the $u$-channel could give large negative contributions which swamp the positive $t$-channel contribution.

From the perspective of the lightcone bootstrap, if we restrict to isolated operator exchange in the $\p\f\rightarrow \f\p$ channel, it is not clear if this positivity continues to hold at finite $s$. This is of interest for large $N$ CFTs since a $t$-channel, exchange Witten diagram for a field of spin $J$ has a $t$-channel conformal block decomposition that includes the dual single-trace operator, $\O_{\Delta,J}$, plus double-traces $[\f\p]_{n,\ell\leq J}$. The double-trace operators, $[\f\p]_{n,\ell\leq J}$, do not affect $c_{\f\f[\f\f]_{0,s}}c_{\f\f[\p\p]_{0,s}}$ for $s>J$. Therefore, in this class of theories and for $s>J$, we need to understand how to write a single $t$-channel block as a sum of $s$-channel blocks.

The answer was recently found for $d=2$ and $d=4$ in \cite{Liu:2018jhs} using the CFT inversion formula \cite{Caron-Huot:2017vep}. For $d=4$, we can extract OPE data from their $6j$ symbol:
\begin{small}
\begin{align}
&\hspace{-.5in}c_{\f\f[\f\f]_{0,s}}(\bar{h})c_{\p\p[\f\f]_{0,s}}(\bar{h}) = 
\frac{\Gamma (2 (h_{\O}-1)) \csc (2 \pi  (h_{\f}-h_{\p})) \Gamma (-h_{\f}-h_{\p}+\bar{h}_{\O}+2)^2 \sin ^2(\pi  (h_{\f}+h_{\p}-h_{\O})) \Gamma (\bar{h})^2 \Gamma (2 \bar{h}_{\O}) }{\Gamma (2 \bar{h}-1) \Gamma (2 h_{\f}-2 h_{\p}+1) \Gamma (-h_{\f}+h_{\p}+h_{\O}-1)^2}\nonumber
\\&\hspace{-.2in}\times\csc (\pi  (h_{\f}+h_{\p}+\bar{h}-\bar{h}_{\O}))   _4\widetilde{F}_3\left( {\begin{array}{*{20}{c}}
{-h_{\f}-h_{\p}+\bar{h}_{\O}+2,-h_{\f}-h_{\p}+\bar{h}_{\O}+2,h_{\f}-h_{\p}+\bar{h}_{\O},-h_{\f}+h_{\p}+\bar{h}_{\O}}\\
{2 \bar{h}_{\O},-h_{\f}-h_{\p}-\bar{h}+\bar{h}_{\O}+3,-h_{\f}-h_{\p}+\bar{h}+\bar{h}_{\O}+2} 
\end{array};1} \right) \nonumber
\\ \nonumber
\\ &\hspace{-.2in}+\frac{ \csc (2 \pi  (h_{\p}-h_{\f})) \Gamma (2 h_{\O}-2) \Gamma (\bar{h})^4 \Gamma (2 \bar{h}_{\O}) \Gamma (2 h_{\f}+\bar{h}-2) \Gamma (2 h_{\p}+\bar{h}-2) \sin ^2(\pi  (h_{\f}+h_{\p}-h_{\O})) }{\Gamma (2 \bar{h}-1) \Gamma (2 h_{\f}-2 h_{\p}+1) \Gamma (h_{\f}-h_{\p}+\bar{h}_{\O}) \Gamma (-h_{\f}+h_{\p}+\bar{h}_{\O}) \Gamma (-h_{\f}+h_{\p}+h_{\O}-1)^2} \nonumber
\\ &\hspace{-.2in}\times\csc (\pi  (h_{\f}+h_{\p}+\bar{h}-\bar{h}_{\O})) _4\widetilde{F}_3\left( {\begin{array}{*{20}{c}}
{2 h_{\f}+\bar{h}-2,2 h_{\p}+\bar{h}-2,\bar{h},\bar{h}}\\
{h_{\f}+h_{\p}+\bar{h}-\bar{h}_{\O}-1,h_{\f}+h_{\p}+\bar{h}+\bar{h}_{\O}-2,2 \bar{h}} 
\end{array};1} \right) \nonumber
\\ & \hspace{4in}  -\bigg(\bar{h}_{\O}\leftrightarrow h_{\O}-1\bigg).
\end{align}
\end{small}
where $\widetilde{F}$ is the regularized hypergeometric function. 

The first term, and its $(\bar{h}_{\O}\leftrightarrow h_{\O}-1)$ reflection, matches the asymptotic expansion found via the lightcone bootstrap order by order in $\bar{h}^{-1}$. The second term and its reflection, on the other hand, are exponentially suppressed at large $\bar{h}$ in comparison and are only visible through the inversion formula\footnote{We thank David Poland for discussions on this point. See \cite{Liu:2018jhs,Sleight:2018ryu,Cardona:2018qrt,Cardona:2018dov,Albayrak:2019gnz} for examples of this phenomena.}. Neither term individually is positive for all values of the scaling dimensions and spin, but it can be checked extensively that the sum is positive and obeys HS ANEC. This gives a simple example where truncating the large spin expansion, but neglecting possible non-perturbative effects, can give an answer inconsistent with HS ANEC, and therefore causality. 

\ssec{Double-twist Exchange}

Since the exchange of an isolated conformal block is always consistent with HS ANEC, a more non-trivial problem is when the leading contribution in the $t$-channel comes from an infinite tower of operators with degenerate twist. For a generic CFT, this can happen in the $\p\f$ OPE if $\Delta_{\p}+\Delta_{\f}<\tau^{\O}_{min}$, where $\O$ is the isolated operator with minimal twist in this OPE\footnote{Conserved spin-1 currents and the stress-tensor cannot appear in this OPE.}. In this case, we need to sum over the full $[\f\p]_{0,\ell}$ tower in the $t$-channel before matching to the $[\f\f]_{0,s}$ data in the $s$-channel.

 \begin{figure}
    \centering
       \includegraphics[width = .6\textwidth]{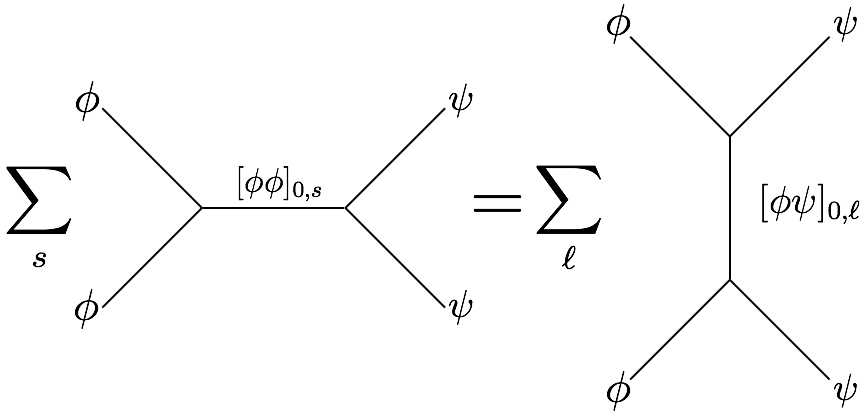}
         \caption{Crossing equation with double-twist dominance in $t$-channel} \label{DtDual}
\end{figure}

For a large $N$ CFT, this can happen if there are no cubic couplings, $\<\f\p\chi\>$, where $\chi$ is any single-trace operator. There are always $s$-channel, exchange Witten diagrams corresponding to $\f\f\rightarrow T \rightarrow \p\p$, or graviton exchange in AdS, but this only affects $\<\p\p[\f\f]_{n,s}\>$ for $s\leq 2$. Boundedness in the Regge limit implies that at tree-level we should only have single-trace operators with spin $J\leq 2$, with similar bounds for quartic interactions \cite{Camanho:2014apa,Maldacena:2015waa}. For large $N$ CFTs, it is therefore possible that $\<\p\p[\f\f]_{n,s>2}\>=0$ at tree level.

We will also assume there exists selection rules such that the $[\f\p]_{0,s}$ tower is dominant over multi-twist trajectories composed of $\f$. For example, if $\f$ is even and $\p$ is odd under a $\mathbb{Z}_{2}$ symmetry, then no $\f$ multi-twist operators can appear in the $\f\p$ OPE. We will also restrict ourselves to large $N$ theories.

In the limit $1-\bar{z}\ll z\ll 1$, the problem of determining $c_{\f\f[\f\f]_{0,s}}c_{\p\p[\f\f]_{0,s}}$ from crossing is now:
\begin{equation}
\left(1-\bar{z}\right)^{h_\f+h_\p}\underset{s}{\sum}c_{\f\f[\f\f]_{0,s}}c_{\p\p[\f\f]_{0,s}}g^{0,0}_{[\f\f]_{0,s}}(z,\bar{z})= z^{2h_{\f}}\sum_{\ell}c_{\f\p[\f\p]_{0,\ell}}^{2}g^{2h_{\p\f},2h_{\f\p}}_{[\f\p]_{0,\ell}}(1-z,1-\bar{z})\bigg|_{z^{2h_\f}}, \label{eq:crossingLCDT}
\end{equation}
where on the right hand side we picked out the piece proportional to $z^{2h_{\f}}$ to match the $[\f\f]_{0,s}$ double-twist states on the left hand side. 

It is not important where we start the sum over $\ell$ since individual conformal blocks always yield an answer consistent with HS ANEC, it is only the infinite sum over $\ell$ which matters. Using (\ref{eq:SL2R}) and (\ref{eq:dimred}) we first expand (\ref{eq:crossingLCDT}) to leading order in $z\ll1-\bar{z}\ll1$ and then expand to second order in the anomalous dimension\footnote{The $0^{th}$ and $1^{st}$ order terms are Casimir-regular and do not contribute to the large spin data in the $s$-channel.}:
\begin{equation}
=z^{2h_{\f}}\sum_{\ell}c_{\f\p[\f\p]_{0,\ell}}^{2}(1-\bar{z})^{h_\f+h_\p}\frac{\Gamma\left(2(h_\p-h_\f)\right) \Gamma\left(2(h_\f+h_\p+\ell)\right)}{\Gamma (2h_\p+\ell)^2}\left(\frac{\gamma_{0,\ell}}{2}\right)^{2}log^{2}(1-\bar{z}). \label{eq:RHS_DT_cross}
\end{equation}

To calculate $c_{\f\p[\f\p]_{0,\ell}}^{2}$ and $\gamma_{0,\ell}$ we now need to solve the crossing problem:
\begin{equation}
\f\p\rightarrow  [\f\p]_{0,\ell} \rightarrow \p\f \hspace{.25in} \Longleftrightarrow  \hspace{.25in}  \f\f\rightarrow \O \rightarrow \p\p
\end{equation}
where we assume isolated operators $\O$ determine the large $\ell$ asymptotics of $[\f\p]_{0,\ell}$. One operator which always appears on the right hand side is the stress-tensor, $T$. 

\begin{figure}
    \centering
       \includegraphics[width = .6\textwidth]{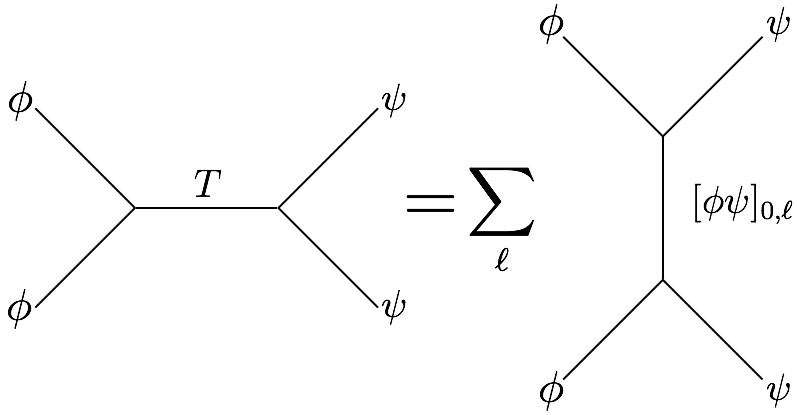}
         \caption{Correction to $[\f\p]$ due to $T$ exchange} \label{Texch}
\end{figure}

This method of iteratively solving crossing in the large spin limit works because the double-twist operators $[\f\f]$ and $[\p\p]$ in the $\f\f\rightarrow  \p\p$ channel give a small contribution to the double-twist operators $[\f\p]$ in the dual channel, $\f\p\rightarrow \p\f$. Generically, the reverse is true and the $[\f\p]$ operators lead to small corrections for $c_{\f\f[\f\f]_{n,s}}c_{\p\p[\f\f]_{n,s}}$, as can be seen explicitly in (\ref{eq:isolatedff}). Here we are considering a fine-tuned example where this small contribution gives the leading order correction to $c_{\f\f[\f\f]_{n,s}}c_{\p\p[\f\f]_{n,s}}$.

The answer for the anomalous dimensions and OPE coefficients of $[\f\p]$ in the lightcone expansion were found in \cite{dsdi}:
\bea
c_{\f\p[\f\p]_{0,\ell}}^{2}(\bar{h})\gamma_{[\f\p]_{0,\ell}}(\bar{h})\sim 2\sum_{\substack{m\geq 0}}c_{\f\f\O}c_{\p\p\O}V^{(0)\f\p\p\f}_{\O,m}(\bar{h}) + \text{u-channel},
\\
c_{\f\p[\f\p]_{0,\ell}}^{2}(\bar{h})\sim \sum_{\substack{m\geq 0}}c_{\f\f\O}c_{\p\p\O}W^{(0)\f\p\p\f}_{\O,m}(\bar{h}) + \text{u-channel},
\eea
where $V$ and $W$ are defined in \cite{dsdi}, but we will not need their full form. Instead, we focus on the $m=0$ terms, which gives the leading order result at large $\bar{h}$. Since we are also at large $N$, we can use the generalized free field values for the OPE coefficients \cite{Fitzpatrick:2011dm}:
\bea
c^{2}_{\f\p[\f\p]_{0,\ell}}=\frac{(2h_{\f})_\ell (2h_{\p})_\ell}{\ell! (2h_{\f}+2h_{\p}+\ell-1)_\ell},
\eea
Focusing on the corrections to the anomalous dimensions, we find for $m=0$:
\bea
\hspace{-.3in}V^{(0)\f\p\p\f}_{\O,0}(\bar{h})=-\frac{\Gamma (2 \bar{h}_{\O}) \Gamma (h_{\f}-h_{\p}+\bar{h}) \Gamma (-h_{\f}+h_{\p}+\bar{h}) \Gamma (h_{\f}+h_{\p}+\bar{h}-h_{\O}-1)}{\Gamma (2 \bar{h}-1) \Gamma (\bar{h}_{\O})^2 \Gamma (2 h_{\f}-h_{\O}) \Gamma (2 h_{\p}-h_{\O}) \Gamma (-h_{\f}-h_{\p}+\bar{h}+h_{\O}+1)}.\quad
\eea

\begin{figure}
    \centering
       \includegraphics[width = .3\textwidth]{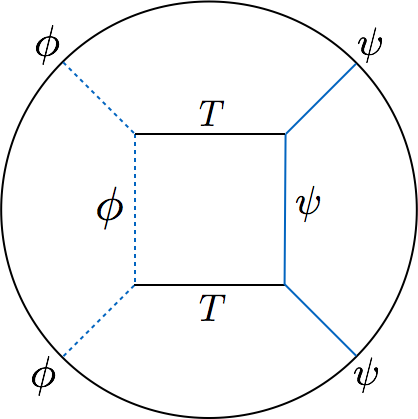}
         \caption{A universal AdS loop diagram for theories with gravity. We label AdS fields by their dual CFT operators.} \label{oneloop}
\end{figure}

This is in fact the full answer when the operator $\O$ has twist exactly $\tau_{\O}=2h_{\O}= d-2$\footnote{For $[\f\p]_{n>0,\ell}$ we have to include $m>0$ terms as well.}. We can now plug in the generalized free field OPE coefficients and the $m=0$ term for the anomalous dimensions into (\ref{eq:RHS_DT_cross}) and perform the sum over $\ell$. The sum converges if $h_{\O}>h_{\f}$ and yields the following correction to the OPE coefficients:
\begin{small}
\begin{equation}
\begin{aligned}
\hspace{-.25in}c_{\f\f[\f\f]_{0,s}}(\bar{h})c_{\p\p[\f\f]_{0,s}}(\bar{h})\bigg|_{\gamma_{[\f\p]_{0,\ell}}^{2}}\hspace{-.2in}\approx2c_{\f\f\O}^{2}c_{\p\p\O}^{2}\frac{\Gamma (2h_{\f})^2 \Gamma (2 \bar{h}_{\O})^2 \Gamma (2h_{\p}-2h_{\f}) \Gamma (2h_{\f}+2h_{\p}) \Gamma (2h_{\f}+2h_{\p}-h_{\O}-1)^2 }{\Gamma (\bar{h}_{\O})^4 \Gamma (h_{\O}+1)^2 \Gamma (2h_{\f}+2h_{\p}-1)^2 \Gamma (2h_{\f}-h_{\O})^2 \Gamma (2h_{\p}-h_{\O})^2}&\nonumber
\\ _6F_5\left( {\begin{array}{*{20}{c}}
{1,1,2h_{\f},h_\f+h_{\p}+\frac{1}{2},2h_{\f}+2h_{\p}-h_{\O}-1,2h_{\f}+2h_{\p}-h_{\O}-1}\\
{h_\f+h_\p-\frac{1}{2},2h_{\p},2h_{\f}+2h_{\p}-1,h_{\O}+1,h_{\O}+1} 
\end{array};1} \right)\partial_{a}^{2}S^{0,0}_{-h_\f-h_\p+a}(\bar{h})\bigg|_{a=0}&. \label{eq:OPEcoeffsLargeNLoop}
\end{aligned}
\end{equation}
\end{small}

If $\O=T$, this gives the large spin, one-loop correction due to gravity for the OPE coefficients $c_{\f\f[\f\f]_{0,s}}c_{\p\p[\f\f]_{0,s}}$. This can also be found via the $s$-channel conformal block decomposition of the Witten diagram in figure \ref{oneloop}. 

As an aside, we can also expand (\ref{eq:crossingLCDT}) to $n^{th}$ order in $\gamma_{[\f\p]_{0,\ell}}^{n}$, isolate the leading $\log$ term in the $t$-channel, sum over $\ell$, and perform the $s$-channel conformal block decomposition to find the OPE coefficients at large spin to all orders in $1/N$. From (\ref{eq:SL2R}) and (\ref{eq:dimred}), the leading $\log$ term comes from expanding $(1-\bar{z})^{h}$ in the anomalous dimension. Assuming the anomalous dimension of $[\f\p]_{0,\ell}$ is due to a single isolated operator, $\O$, we find:
\begin{small}
\begin{align}
&\hspace{-.3in}c_{\f\f[\f\f]_{0,s}}c_{\p\p[\f\f]_{0,s}}\bigg|_{\gamma_{[\f\p]_{0,\ell}}^{n}}\hspace{-.2in}\approx2c_{\f\f\O}^{n}c_{\p\p\O}^{n}\frac{(-1)^n\Gamma (2h_\p-2h_\f) \Gamma (2h_\f)^n \Gamma (2h_\p)^{n-2} \Gamma (2 h_{\O})^n \Gamma (2h_\f+2h_\p-\bar{h}_{\O}-1)^n}{n!\Gamma (\bar{h}_{\O}+1)^n \Gamma (h_{\O})^{2 n} \Gamma (2h_\f+2h_\p-1)^n \Gamma (2h_\f-\bar{h}_{\O})^n \Gamma (2h_\p-\bar{h}_{\O})^n}
\nonumber \\
&\hspace{-0in}  \Gamma (2h_\f+2h_\p)\,_{2n+2}F_{2n+1}\left( {\begin{array}{*{20}{c}}
{2h_{\f},h_\f+h_\p+\frac{1}{2},{}^{n}1,{}^{n}(2h_{\f}+2h_{\p}-\bar{h}_{\O}-1)}  \\
{h_{\f}+h_{\p}-\frac{1}{2},\Delta_{\p},{}^{n-1}(2h_{\f}+2h_{\p}-1),{}^{n}(1+\bar{h}_{\O})} 
\end{array};1} \right)\partial_{a}^{n}S^{0,0}_{-h_\f-h_\p+a}(\bar{h})\bigg|_{a=0}. \label{eq:OPEcoeffsLargeNGenLoop}
\end{align}
\end{small}
where we have introduced the shorthand ${}^{n}b$ for $n$ copies of $b$ in the hypergeometric function. 

\begin{figure}
    \centering
       \includegraphics[width = .3\textwidth]{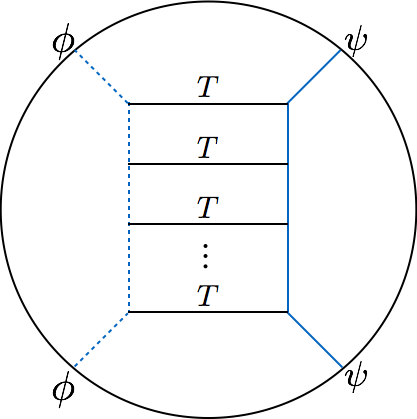}
         \caption{Ladder diagram for graviton exchange in AdS.} \label{ladder}
\end{figure}

In AdS, this gives the leading, large spin contribution for a ladder diagram with $n$ horizontal rungs, each exchanging the same operator, e.g. figure \ref{ladder} when $\O=T$. It should be stressed these results only give finite answers when $h_{\O}>h_{\f}$. In general, we have to be more careful about order of limits when performing the sum over $\ell$ and taking the lightcone limit.

Returning to HS ANEC, we find the OPE coefficients from (\ref{eq:OPEcoeffsLargeNLoop}) are always positive if $h_{\p}>h_{\f}$, which we have assumed from the beginning. This result also follows from the fact the summand in (\ref{eq:RHS_DT_cross}) goes like $\ell^{-4h_{\f}-4h_{\O}-1}$ and converges if $h_{\O}>h_{\f}$. Since individual blocks gave answers consistent with HS ANEC, the positive, convergent sum should as well. The result (\ref{eq:OPEcoeffsLargeNGenLoop}) can be negative for $n=3$, that is at two-loop order, but it is subleading in $1/N$.

The case where $h_{\O}=h_{\f}$ needs to be handled differently and we have to modify our initial assumptions. In the sum over the $[\f\p]_{0,\ell}$ conformal blocks, there will be non-trivial cancellations of divergences and their final contribution to $c_{\f\f[\f\f]_{0,s}}c_{\p\p[\f\f]_{0,s}}$ can in principle be negative. 

The crossing problem for determining the $[\f\p]_{0,\ell}$ data is now:
\begin{equation}
\f\p \rightarrow [\f\p]_{0,\ell}\rightarrow \p\f \quad \Longleftrightarrow \quad \f\f \rightarrow \f \rightarrow \p\p. \label{crossdegpt1}
\end{equation}
Therefore, we need $\<\f\f\f\>\<\p\p\f\>\neq 0$. As a reminder, the original crossing problem is:
\begin{equation}
\f\f \rightarrow [\f\f]_{0,s}\rightarrow \p\p \quad \Longleftrightarrow \quad \f\p \rightarrow \O' \rightarrow \p\f. \label{eq:crossDegpt2}
\end{equation}
If $\<\p\p\f\>\neq0$, then $\O'=\p$ exchange will always give the leading, large $s$ contribution to the $[\f\f]_{0,s}$ data and the effect of the $[\f\p]_{0,\ell}$ family is subleading. Although in this situation the $[\f\p]_{0,\ell}$ operators do yield negative corrections to $c_{\p\p[\f\f]_{0,s}}c_{\f\f[\f\f]_{0,s}}$, at large spin the positive correction from $\p$ is always dominant. Therefore, HS ANEC is still obeyed at asymptotically large spin. For the interested reader, we present the corrections from $[\f\p]_{0,\ell}$ exchange for this case in Appendix \ref{app:RegDT}. 

\sec{Conclusion}
\label{sec:Conclusion}

In this work, we have presented a variety of new results on the structure of unitary, local CFTs through the study of higher spin, positivity conditions. We showed how studying HS ANEC reveals new bounds and restrictions on CFT data which are not manifest in other approaches. In particular, we have shown how HS ANEC puts lower bounds on the dimensions of charged operators and constrains the behavior of CFTs close to either a generalized or genuine free field description. The proof that ANEC saturation implies the theory is free does not require HS ANEC, but the spin-$4$ positivity condition does allow us to derive stronger lower bounds on $\<TTT\>$. Moreover, HS ANEC allows us to relate more directly the observables $\<J^{(4)}J^{(4)}J^{(4)}\>$ and $\<TTT\>$. Finally, we have verified HS ANEC holds in a large class of theories by using the analytic bootstrap.

There are clearly many more avenues that should be explored in this program. For one, it would be useful to rederive some of the results presented here with different methods, especially those that make more direct use of analyticity in spin \cite{Caron-Huot:2017vep,Kravchuk:2018htv}. For example, it would be nice to extend our results to the entire leading Regge trajectory, to continuous spin, and also to find a more direct way to relate $\<TTT\>$ and $\<J^{(s)}J^{(s)}J^{(s)}\>$. Another open question is understanding when HS ANEC saturation also implies the theory is free. To answer these questions, it may help to study the OPE and commutator algebra for light-ray operators \cite{Casini:2017roe,Cordova:2018ygx,MKPK:wip} or the sum rules more directly \cite{Hartman:2016lgu,Caron-Huot:2017vep,ssw,Kravchuk:2018htv}.

Some other possible generalizations would be to include more general representations of the Lorentz group for the external states \cite{Cordova:2017dhq}, include parity odd three-point structures\cite{Chowdhury:2017vel}, or perform a more thorough analysis of OPE coefficients in large $N$ theories\cite{Liu:2018jhs}. It may also be interesting to study the constraints of slightly-broken higher spin symmetries on the correlation functions of HS ANEC operators, especially when considering higher-point functions \cite{Turiaci:2018nua}.

We have also shown that HS ANEC is not always obeyed in the asymptotic, lightcone expansion. At finite spin, we need to include terms exponentially suppressed at large spin which are only visible with the inversion formula. One effect we have not discussed however is mixing between multi-twist families, which can lead to large corrections at finite spin. It would be interesting to explore if the effects of mixing always yields results consistent with HS ANEC or if there is a preferred way to include mixing effects such that HS ANEC is manifestly obeyed. This should be testable using the twist Hamiltonian \cite{dsdi} or by studying large $N$ CFTs at one-loop and higher \cite{Aharony:2016dwx,Yuan:2017vgp,Liu:2018jhs}. Understanding how and when to truncate the large spin expansion is an important open question to see if the beautiful matching between the analytic and numerical bootstrap demonstrated in \cite{Alday:2015ewa,dsdi,Cornagliotto:2017snu} holds more broadly.

Finally, the study of causality and positivity conditions in CFTs is also crucial to understanding the AdS/CFT correspondence. It is well known that CFTs with $N\gg1$ and no light higher spin, single-trace operators, i.e. $\Delta_{gap}\gg1$, are dual to weakly-coupled AdS theories of Einstein gravity plus matter \cite{Heemskerk:2009pn,Heemskerk:2010ty,Camanho:2014apa,Afkhami-Jeddi:2016ntf,Caron-Huot:2017vep,KPZ2017,Li:2017lmh,Afkhami-Jeddi:2017rmx,Costa:2017twz,Afkhami-Jeddi:2018own,Meltzer:2017rtf}. What is less studied is how the higher spin trajectory must behave such that causality is not violated at finite coupling \cite{Amati:1987wq,Amati:1987uf,DAppollonio:2015fly,Camanho:2014apa,Costa:2017twz}. Since we studied the exact, leading trajectory of the CFT, we can make contact with this program when $\Delta_{gap}\ll1$. When the gap scale is small, the CFT is weakly coupled and the AdS theory contains many light string states. For this class of CFTs, our trajectory coincides with the leading single-trace trajectory for a large range of spin. On the other hand, when $\Delta_{gap}\gg1$, our trajectory consists solely of double-trace operators and is distinct from the single-trace trajectory. The new results we have presented here, when applied to large $N$, weakly coupled CFTs, can hopefully shed light on the behavior of holographic CFTs at finite coupling.

\sec*{Acknowledgements}
I would like to thank Nima Afkhami-Jeddi, Soner Albayrak, Dean Carmi, Clay C\'ordova, Tom Hartman, Amir Hossein Tajdini, Zohar Komargodski, Filip Kos, Petr Kravchuk, Arvin Shahbazi Moghaddam, David Poland, David Simmons-Duffin, Gustavo Turiaci, and Alessandro Vichi for discussions. I also grateful to Soner Albayrak and David Simmons-Duffin for comments on the draft and Eric Perlmutter for collaboration during the early stages of this project. This work was supported by NSF grant PHY-1350180, Simons Foundation grant 488651, and a Sherman Fairchild Postdoctoral Fellowship.

\appendix
\section{Integrals of Three-Point Functions} 
\label{app:integrals}
Our method for calculating (HS) ANEC matrix elements will follow the approach of \cite{Komargodski:2016gci}. To simplify the calculations, we replace the polarization tensors for symmetric, traceless operators with null polarization vectors, $\lambda^{\mu_{1}...\mu_{s}}\rightarrow z^{\mu_{1}}...z^{\mu_{s}}$. Then the conformally invariant tensor structures, $V_{i,jk}$ and $H_{ij}$ are \cite{Costa:2011mg}:
\begin{align}
V_{i,jk}=\frac{x_{ij}^{2}x_{ik}\cdot z_{i}-x_{ik}^{2}x_{ij}\cdot z_{i}}{x_{jk}^{2}},
\\
H_{ij}=x_{ij}^{2}z_{i}\cdot z_{j}-2x_{ij}\cdot z_{i}x_{ij}\cdot z_{j}.
\end{align}

For convenience, we define $V_{1}=V_{1,23}$, plus cyclic permutations. The general three-point function tensor structure is:
\begin{align}
&Q_{\vec{n},\vec{m}}=\frac{V_{1}^{n_{1}}V_{2}^{n_{2}}V_{3}^{n_{3}}H_{12}^{m_{12}}H_{13}^{m_{13}}H_{23}^{m_{23}}}{x_{12}^{2\bar{h}_{123}}x_{23}^{2\bar{h}_{231}}x_{31}^{2\bar{h}_{312}}},
\\
&\bar{h}_{ijk}=\bar{h}_{i}+\bar{h}_{j}-\bar{h}_{k}, \quad \bar{h}_{i}=\frac{1}{2}(\Delta_{i}+\ell_{i}),
\\ 
& \vec{n}=(n_{1},n_{2},n_{3}), \quad \vec{m}=(m_{23},m_{13},m_{12}).
\end{align}
In $d=3$, there is a degeneracy among the $V_{i,jk}$ and $H_{ij}$ structures and we can drop any three-point function proportional to $H_{12}H_{13}H_{23}$.

In the collider limit, $x_{2}\cdot \bar{n}\rightarrow\infty$, this reduces to
\bea
Q_{\vec{n},\vec{m}}\approx \frac{\widehat{V}_{1}^{n_{1}}\widehat{V}_{2}^{n_{2}}\widehat{V}_{3}^{n_{3}}\widehat{H}_{12}^{m_{12}}\widehat{H}_{13}^{m_{13}}\widehat{H}_{23}^{m_{23}}}{(-1)^{h_{2}}(x_{2}\cdot\bar{n})^{h_{2}} (x_{12}\cdot n)^{h_{123}+n_{3}}(x_{23}\cdot n)^{h_{231}+n_{1}}(x_{13}^{2})^{h_{312}+n_{2}}},
\eea

where the hatted tensor structures are \cite{Komargodski:2016gci}:
\begin{align}
\widehat{V}_{1}&=-\frac{x_{13}\cdot z_{1}x_{12}\cdot n-z_{1}\cdot n \frac{x_{13}^{2}}{2}}{x_{23}\cdot n},
\qquad
\widehat{V}_{2}=\frac{x_{13}\cdot n}{x_{13}^{2}},
\\
\widehat{V}_{3}&=-\frac{x_{13}\cdot z_{3}x_{23}\cdot n- z_{3}\cdot n \frac{x_{13}^{2}}{2}}{x_{12}\cdot n},
\qquad 
\widehat{H}_{12}=-z_{1}\cdot n,
\\
\widehat{H}_{13}&=z_{1}\cdot z_{3}x_{13}^{2}-2x_{13}\cdot z_{1}x_{13}\cdot z_{3},
\qquad
\widehat{H}_{23}=-z_{3}\cdot n.
\end{align}

To perform the light-ray integrals, we use:
\begin{align}
\int\limits_{-\infty}^{\infty}d(x_{2}\cdot n)\frac{1}{(x_{12}\cdot n)^{a}(x_{23}\cdot n)^{b}}=\frac{2\pi i \Gamma(a+b-1)}{(x_{13}\cdot n)^{a+b-1}\Gamma(a)\Gamma(b)}.
\end{align}

The general Fourier integrals we need to compute are:
\bea
I_{FT}(q,n,a_1,a_2,b_1,b_2)=\int d^{d}x e^{-iq\cdot x}\frac{(x\cdot z_{1})^{a_1}(x\cdot z_{3})^{a_3}}{(x\cdot n)^{b_{1}}(x^{2})^{b_{2}}}. 
\eea
The case where $a_{1}=a_{3}=0$ is simple to compute:
\bea
I_{FT}(q,n,0,0,b_1,b_2)= \frac{e^{\frac{i \pi  b_{1}}{2}} (-n\cdot q)^{-b_{1}} \left(\pi ^{\frac{d}{2}+1} 2^{-b_{1}-2 b_{2}+d+1}\right) (-q^{2})^{b_{1}+b_{2}-\frac{d}{2}}}{\Gamma (b_{2}) \Gamma \left(b_{1}+b_{2}-\frac{d}{2}+1\right)}.
\eea
so the general answer is given recursively by:
\begin{align}
I_{FT}(q,n,a_1,a_2,b_1,b_2)=e^{\frac{-i\pi}{2}(a_{1}+a_{3})}\partial_{t_{1}}^{a_{1}}\partial_{t_{3}}^{a_{3}}I_{FT}(q+t_{1}z_{1}+t_{3}z_{3},n,0,0,b_1,b_2)\bigg|_{t_{i}\rightarrow 0}.
\end{align}
For the calculations presented here, this recursive definition is sufficient\footnote{See also \cite{Karateev:2018oml} for efficient methods to calculate Fourier transforms of translationally invariant kernels in CFTs.}. As a reminder, we will always choose $n\cdot q=q^{2}=-1$.

\section{Conservation and Ward Identities}
\label{app:ConsWard}
In this appendix we will include some details on conservation conditions and Ward identities which we used in the body of the paper. To impose conservation we will use the embedding space technology of \cite{Costa:2011mg}. The simplest three-point function to study is $\<VJ^{(s)}\f\>$ for a conserved current $V$ and a scalar $\f$. We have two OPE coefficients, $\{c^{(Vs\f)}_{000}, c^{(Vs\f)}_{001}\}$ which are related by:
\bea
c^{(Vs\f)}_{001}=-\frac{c^{(Vs\f)}_{000} (-\Delta_{\f}+s+\tau_{s})}{d+\Delta_{\f}-\tau_{s}-2}.
\eea 
The next simplest case is $\<VJ^{(s)}V\>$ for a conserved vector. After imposing permutation symmetry we have four OPE coefficients, $\{c^{(VsV)}_{000},c^{(VsV)}_{001},c^{(VsV)}_{101},c^{(VsV)}_{010}\}$ and conservation yields \cite{Costa:2011mg}:
\begin{align}
&c^{(VsV)}_{101}= \frac{s c^{(VsV)}_{000} (-d+s+\tau_{s}+1)-c^{(VsV)}_{001} \left(\tau_{s} (-d+4 s+2)+2 s (-2 d+s+3)+\tau_{s}^2\right)}{(2 d-\tau_{s}-4) (2 s+\tau_{s})},
\\
&c^{(VsV)}_{010}= \frac{c^{(VsV)}_{000} (-d+s+\tau_{s}+1)+c^{(VsV)}_{001} (2 d-\tau_{s}-2)}{2 s+\tau_{s}}.
\end{align}

Conservation for $\<TJ^{(s)}\f\>$ implies the constraints:
\begin{align}
&c^{(Ts\f)}_{001}= \frac{2 c^{(Ts\f)}_{000} ((d-1) \Delta_{\f}-d (s+\tau_{s})+\tau_{s})}{(d-2) (d+\Delta_{\f}-\tau_{s})},
\\
&c^{(Ts\f)}_{002}= \frac{c^{(Ts\f)}_{000} \left((d-1) \Delta_{\f}^2-2 (d-1) \Delta_{\f} (s+\tau_{s})+2 (d-1) s \tau_{s}+(d-2) (s-1) s+(d-1) \tau_{s}^2\right)}{(d-2) (d+\Delta_{\f}-\tau_{s}-2) (d+\Delta_{\f}-\tau_{s})}.
\end{align}

The result for $\<TJ^{(s)}T\>$ can be found in appendix A of \cite{Costa:2011mg}, so we will not reproduce it here. The next case to consider is $\<J^{(4)}TJ^{(4)}\>$. The full solution to conservation was worked out in \cite{Zhiboedov:2012bm}, as well as the more general case $\<T\O_{\Delta,\ell}\O_{\Delta,\ell}\>$. For $\<J^{(4)}TJ^{(4)}\>$ we find:
\begin{align}
&c^{(JTJ)}_{001}= \frac{1}{16} (2 d (d+6) c^{(JTJ)}_{002}-(d-2) (d+8) c^{(JTJ)}_{101}+8 c^{(JTJ)}_{000}),
\\
&c^{(JTJ)}_{011}= \frac{1}{12} (8 d c^{(JTJ)}_{002}+2 d (d+4) c^{(JTJ)}_{012}-3 (d-2) c^{(JTJ)}_{101}-(d-2) (d+6) c^{(JTJ)}_{111}+6 c^{(JTJ)}_{010}),
\\
&c^{(JTJ)}_{021}= \frac{1}{8} (6 d c^{(JTJ)}_{012}+2 d (d+2) c^{(JTJ)}_{022}-2 (d-2) c^{(JTJ)}_{111}-(d-2) (d+4) c^{(JTJ)}_{121}+4 c^{(JTJ)}_{020}),
\\
&c^{(JTJ)}_{031}= \frac{1}{4} \left(-\left(d^2-4\right) c^{(JTJ)}_{131}+4 d c^{(JTJ)}_{022}-(d-2) c^{(JTJ)}_{121}+2 c^{(JTJ)}_{030}\right).
\end{align}

As expected, $\<J^{(4)}TJ^{(4)}\>$ is a function of 12 OPE coefficients. We can next solve the Ward identities by computing:
\bea
\int_{S^{d-2}}d\vec{n} \<\lambda_{4}\cdot J^{(4)}|\mathcal{E}^{(2)}(n)|\lambda_{4}\cdot J^{(4)}\>=2^{d}q^{0}\<\lambda_{4}\cdot J^{(4)}|\lambda_{4}\cdot J^{(4)}\>.
\eea

To keep things compact, we introduce a vector of OPE coefficients:
\bea
C^{JTJ}=\left(C^{(4)}_{J},c^{(JTJ)}_{000}, c^{(JTJ)}_{002}, c^{(JTJ)}_{010}, c^{(JTJ)}_{012}, c^{(JTJ)}_{020}, c^{(JTJ)}_{022}, c^{(JTJ)}_{030}, c^{(JTJ)}_{040}, c^{(JTJ)}_{121}, c^{(JTJ)}_{131}\right).
\eea

Then the Ward identity implies:
\begin{align}
&c^{(JTJ)}_{101}= C_{(JTJ)}\cdot A_{101}, \qquad c^{(JTJ)}_{111}= C^{(JTJ)}\cdot A_{111}
\end{align}
\begin{align}
\hspace{-.25in}A_{101}=
\left(
\begin{array}{c}
\frac{4 \pi ^{\frac{1}{2}-\frac{d}{2}} ((d+4) \Delta_{J}-8 (d+1)) \Gamma \left(\frac{d}{2}-1\right) \Gamma \left(\frac{d}{2}+4\right)}{3 (d-2) \Gamma \left(\frac{d-1}{2}\right)}
\\ \frac{4 (d+1)}{d-2}
\\ \frac{2 (d-6)}{3}
\\-\frac{(d+1) (d+6)}{d-2}
\\-\frac{(d-2) (d+6)}{3} 
\\ \frac{(d+1) (d+4) (d+6)}{3 (d-2)}
\\ \frac{1}{3} d (d+4) (d+6)
\\ -\frac{(d+1) (d+2) (d+4) (d+6)}{6 (d-2)}
\\ \frac{(d-1) (d+2) (d+4)^2 (d+6)}{6 (d-2)}
\\ \frac{(d+4) (d+6)}{6} 
\\ -\frac{d (d+2) (d+4) (d+6)}{6} 
\end{array}
\right), \qquad
\vec{A}_{111}=\left(
\begin{array}{c}
\frac{2 \pi ^{\frac{1}{2}-\frac{d}{2}} ((d+6) \Delta_{J}-8 (d+2)) \Gamma \left(\frac{d}{2}-1\right) \Gamma \left(\frac{d}{2}+3\right)}{(d-2) \Gamma \left(\frac{d-1}{2}\right)}
\\ \frac{12 (d+2)}{(d-2) (d+6)}
\\ \frac{3 (d-2)}{d+6}
\\ -\frac{3 (d+2)}{d-2}
\\ -\frac{3d}{2}
\\ \frac{(d+2) (d+4)}{d-2}
\\ \frac{(d+4) (3 d+2)}{2} 
\\ -\frac{(d+2)^2 (d+4)}{2 (d-2)}
\\ \frac{(d-1) (d+2) (d+4) (d+6)}{2 (d-2)}
\\ d+4
\\ -\frac{d (d+2) (d+4)}{4}
\end{array}
\right).
\end{align}

We have choosen to keep the normalization  of the spin-4 operators, $C^{(4)}_{J}$, arbitrary, although it can be fixed to any convenient value. This approach for studying Ward identities is also useful for non-conserved, spin-two operators and operators with even higher spin.

\sec{Examples of Spin-2 $\&$ 4 Matrix Elements}
\label{app:spin2_4}
In this appendix, we list some matrix elements used in deriving the results of section \ref{sec:SatANEC}. Results for $d=3$ can be found by setting all OPE coefficients $c_{ijk}$ with $i,j,k \geq 1$ to zero.
\subsection{General Dimensions}
\label{app:GenD}
First, we will calculate $\mathcal{E}^{(s,j)}_{TT}$ for arbitrary $s$, $j$, and $d$. We define the vector:
\bea
C^{TsT}=\left(c^{(TsT)}_{002},c^{(TsT)}_{011},c^{(TsT)}_{101}\right),
\eea
and then the matrix elements are:
\begin{align}
&\mathcal{E}^{(s,0)}_{TT}=B^{(4,0)}_{TT}\cdot C_{TsT}\frac{\pi ^{\frac{d+3}{2}} i^s 2^{-d+s+\tau_{s}} (d+2 s+\tau_{s}-2)\Gamma \left(s+\frac{\tau_{s}}{2}+\frac{1}{2}\right)}{\Gamma \left(d-\frac{\tau_{s}}{2}\right) \Gamma \left(s+\frac{\tau_{s}}{2}\right) \Gamma \left(\frac{1}{2} (d+2 s+\tau_{s})\right)}\mathcal{Q}(d,s,\tau_{s})^{-1},
\\
&\mathcal{E}^{(s,1)}_{TT}=B^{(4,1)}_{TT}\cdot C_{TsT}\frac{\pi ^{\frac{d+3}{2}} i^s 2^{-d+s+\tau_{s}} (d+2 s+\tau_{s}-2)\Gamma \left(s+\frac{\tau_{s}}{2}+\frac{1}{2}\right)}{\Gamma \left(d-\frac{\tau_{s}}{2}\right) \Gamma \left(s+\frac{\tau_{s}}{2}\right) \Gamma \left(\frac{1}{2} (d+2 s+\tau_{s})\right)}\mathcal{Q}(d,s,\tau_{s})^{-1},
\\
&\mathcal{E}^{(s,2)}_{TT}=B^{(4,2)}_{TT}\cdot C_{TsT}\frac{\pi ^{\frac{d+3}{2}} i^s 2^{-d+s+\tau_{s}}\Gamma \left(s+\frac{\tau_{s}}{2}+\frac{1}{2}\right)}{\Gamma \left(d-\frac{\tau_{s}}{2}\right) \Gamma \left(s+\frac{\tau_{s}}{2}\right) \Gamma \left(\frac{1}{2} (d+2 s+\tau_{s})\right)}\mathcal{Q}(d,s,\tau_{s})^{-1},
\end{align}
where $\mathcal{Q}(d,s,\tau_{s})$ is a polynomial:
\begin{small}
\begin{align}
\mathcal{Q}(d,s,\tau_{s})=&s (2 s+\tau_{s}-1) \bigg((d-6) (d-1)\tau_{s}^3-2\tau_{s}^2 (d (d (d-2 s-9)+11 s+14)-8 s-2)
\nonumber \\ &+\tau_{s}\left((d (5 d-22)+16) s^2-(d-2) d (6 d-29) s+d ((d-10) (d-3) d-12)+8 (s-1)\right)
\nonumber \\ &+2 (d-2) \left((d-2) s^3-2 (d-3) d s^2+(d ((d-7) d+7)+2) s+(d-2) d (d+1)\right)\bigg).
\end{align}
\end{small}

The vectors $B^{(s,j)}_{TT}$ are in general too large to include here, but $B^{(s,0)}_{TT}$, which is used for deriving bounds on $\<TJ^{(s)}\f\>$, does take a simple form:

\begin{align}
&B^{(s,0)}_{TT}=\left(
\begin{array}{c}
 4 (d-1) \tau_{s}^2-8 s \tau_{s}+6 d (-2 d+s+4) \tau_{s}+4 (d-2) \left(2 d^2-2 (s+1) d+(s-1) s\right) \\
 (d-6) \tau_{s}^2-((d-12) d+4 (s+4)) \tau_{s}-2 (d-2) (2 d+(s-5) s) \\
 -(d-2) (2 d-\tau_{s}-2) (d-s-\tau_{s}) \\
\end{array}
\right).
\end{align}
If we set $s=4$, the vectors $B^{(4,1)}_{TT}$ and $B^{(4,2)}_{TT}$ simplify slightly:
\begin{scriptsize}
\begin{align}
&B^{(4,1)}_{TT}=\left(
\begin{array}{c}
 4 \big(2 d^5-(5 \tau_{4}+28) d^4+(\tau_{4} (4 \tau_{4}+47)+114) d^3-(\tau_{4}+2) (\tau_{4} (\tau_{4}+19)+68) d^2
 \\ \hspace{1in}+(\tau_{4} (\tau_{4} (2 \tau_{4}+19)+32)+8) d-\tau_{4}^2 (\tau_{4}+2)\big) \\ \\
 -(d-6) (d-1) \tau_{4}^3+2 (d-1) ((d-14) d+46) \tau_{4}^2-d (d ((d-27) d+230)-648) \tau_{4}
 \\ \hspace{1in}-4 (d-6) d ((d-15) d+46)-160 (3 \tau_{4}+4) 
 \\ \\
 -(d-2) (2 d-\tau_{4}-2) \left((d-1) \tau_{4}^2-2 (d-6) d \tau_{4}+(d-6) (d-5) d-10 (\tau_{4}+4)\right) \\
\end{array}
\right),
\\ \nonumber \\
&B^{(4,2)}_{TT}=\left(
\begin{array}{c}
 4 \big((d-1)^2 \tau_{4}^4-(d-1) (d (5 d-22)+6) \tau_{4}^3+(3 d-2) ((d-3) d (3 d-17)+4) \tau_{4}^2
 \\ \hspace{.1in}+d (72-(d-6) d (d (7 d-53)+86)) \tau_{4}+2 d (d (d (d ((d-18) d+113)-276)+136)+208)-16 (5 \tau_{4}+6)\big)
  \\ \\
 (d-6) (d-1) \tau_{4}^4-(d-1) (3 (d-14) d+140) \tau_{4}^3+(d (d (d (3 d-73)+614)-1748)+1216) \tau_{4}^2
 \\ \hspace{.1in} -d (d (d ((d-39) d+578)-3392)+7296) \tau_{4}-4 d ((d-20) d ((d-15) d+82)+2728)+64 (70 \tau_{4}+89) 
 \\
 \\
 -(d-2) (2 d-\tau_{4}-2) \big(d^4-3 (\tau_{4}+5) d^3+(\tau_{4} (3 \tau_{4}+31)+74) d^2
 \\ \hspace{.5in}-(\tau_{4} (\tau_{4} (\tau_{4}+17)+88)+160) d+\tau_{4} (\tau_{4} (\tau_{4}+14)+88)+224\big) \\
\end{array}
\right).
\end{align}
\end{scriptsize}

Next, we will consider the matrices $\mathcal{E}^{(2,j)}_{TJ}$ for $j=0, 1, 2$. We will focus on the matrix elements $\mathcal{E}^{(2,j)}_{TJ,00}$ since these give the strongest lower bounds on $\mathcal{E}^{(2,j)}_{TT}\mathcal{E}^{(2,j)}_{JJ}$ and are sufficient to show that ANEC saturation implies there is a higher spin symmetry. To make positivity manifest we will use $\mathcal{E}^{(4,j)}_{TT}$ as a basis rather than the OPE coefficients:
\begin{align}
\mathcal{E}^{(2,i)}_{TJ}=\sum\limits_{j=0}^{2}\mathcal{C}^{i}_{j}\mathcal{E}^{(4,j)}_{TT}.
\end{align}
We find:
\begin{scriptsize}
\begin{align}
&\mathcal{C}^{0}_{i}=\left(
\begin{array}{c}
 \frac{2^{d-\tau_{4}-6} d\Gamma (d)}{(d-2) (d+2) (d+4) \Gamma (\tau_{4}+7)} \bigg(4 d^7-4 \tau_{4} d^6+(5 \tau_{4} (\tau_{4}+10)+148) d^5-2 (\tau_{4} (\tau_{4} (\tau_{4}+15)+76)+148) d^4
 \\ \hspace{.9in} +(\tau_{4} (\tau_{4} (\tau_{4} (\tau_{4}+24)+186)+524)+264) d^3-2 (\tau_{4} (\tau_{4} (2 \tau_{4}+43)+266)+488) d^2
 \\ \hspace{.9in}  -2 (\tau_{4} (\tau_{4} (\tau_{4} (\tau_{4}+18)+94)+84)-320) d-2 \tau_{4} (\tau_{4}+2) (\tau_{4}+4) (\tau_{4}+6)\bigg) 
 \\
 \\
 \frac{2^{d-\tau_{4}-4} (d-2) d (\tau_{4}-d+2) \left(4 d^4+16 d^3+(\tau_{4} (\tau_{4}+12)+60) d^2+(\tau_{4}+10) (\tau_{4} (\tau_{4}+9)+22) d+(\tau_{4}+4) (\tau_{4}+6)^2\right) \Gamma (d)}{(d+2) (d+4) \Gamma (\tau_{4}+7)} 
 \\
 \\
 \frac{2^{d-\tau_{4}-5} (d-3) d^2 (\tau_{4}-d+4) (\tau_{4}-d+2) \left(4 d^2+4 (\tau_{4}+8) d+\tau_{4} (\tau_{4}+14)+56\right) \Gamma (d)}{(d+2) (d+4) \Gamma (\tau_{4}+7)} \\
\end{array}
\right),
\\
&\hspace{-0in}\mathcal{C}^{1}_{i}=\left(
\begin{array}{c}
 \frac{2^{d-\tau_{4}-8} (d-1) d (\tau_{4}-d+2) \left(2 (\tau_{4}+10) d^4-(\tau_{4} (\tau_{4}+12)+12) d^3+(\tau_{4}+4) (\tau_{4}+6) (\tau_{4}+12) d^2-2 (\tau_{4} (\tau_{4}+10)+40) d-2 (\tau_{4}+4) (\tau_{4}+6)^2\right) \Gamma (d)}{(d-2)^2 (d+4) \Gamma (\tau_{4}+7)} 
 \\
 \\
 \frac{2^{d-\tau_{4}-7} d \left(4 (\tau_{4}+10) d^4-2 (\tau_{4}-2) (\tau_{4}+14) d^3+(\tau_{4} (5 \tau_{4}-18)+32) d^2+2 (\tau_{4} (\tau_{4} (\tau_{4} (\tau_{4}+19)+143)+462)+560) d+\tau_{4} (\tau_{4}+2) (\tau_{4}+4) (\tau_{4}+6)\right) \Gamma (d)}{(d+4) \Gamma (\tau_{4}+7)} 
 \\
 \\
 \frac{2^{d-\tau_{4}-7} (d-3) d^2 (\tau_{4}-d+2) \left(2 (\tau_{4}+10) d^2+(\tau_{4} (3 \tau_{4}+34)+128) d+\tau_{4} (\tau_{4} (\tau_{4}+16)+76)+128\right) \Gamma (d)}{(d-2) (d+4) \Gamma (\tau_{4}+7)} \\
\end{array}
\right),
\\
&\hspace{-0in}\mathcal{C}^{2}_{i}=\left(
\begin{array}{c}
 \frac{2^{d-\tau_{4}-8} (d-1) d (\tau_{4}-d+4) (\tau_{4}-d+2) \left((\tau_{4}+8) (\tau_{4}+10) d^2+2 (\tau_{4} (\tau_{4}+18)+84) d-4 (\tau_{4} (\tau_{4}+16)+68)\right) \Gamma (d)}{3 (d-2)^2 (d+4) \Gamma (\tau_{4}+7)} 
 \\ \\
 \frac{2^{d-\tau_{4}-6} d (\tau_{4}-d+2) \left((\tau_{4}+8) (\tau_{4}+10) d^2+(\tau_{4} (\tau_{4} (\tau_{4}+19)+138)+408) d+\tau_{4} (\tau_{4} (3 \tau_{4}+46)+224)+400\right) \Gamma (d)}{3 (d+4) \Gamma (\tau_{4}+7)} 
 \\ \\
 \frac{2^{d-\tau_{4}-7} d\Gamma (d)}{3 (d-2) (d+4) \Gamma (\tau_{4}+7)}\bigg(2 (\tau_{4} (\tau_{4} (\tau_{4}+17)+102)+244) d^3-8 (\tau_{4} (\tau_{4} (\tau_{4} (\tau_{4}+16)+100)+280)+336)
\\ -4 (\tau_{4} (5 \tau_{4} (\tau_{4}+12)+232)+440) d+(\tau_{4} (\tau_{4} (\tau_{4} (\tau_{4}+16)+112)+352)+256) d^2+(\tau_{4}+8) (\tau_{4}+10) d^4\bigg)  \\
\end{array}
\right).
\end{align}
\end{scriptsize}

If $\tau_{4}>d-2$ and $d\geq3$ then $\mathcal{C}^{i}_{j}>0$ and ANEC can not be saturated. If we set $\tau_{4}=d-2$ it is possible to saturate the spin-$j$ ANEC bound by only setting $\mathcal{E}^{(4,j)}_{TT}=0$. This positivity is not obvious in the expressions given above, but if we set $d=4$ they simplify to:

\begin{align}
&\mathcal{C}^{(2,0)}_{TJ}=\left(
\begin{array}{c}
 \frac{9\ 2^{-\tau_{4}-3} (\tau_{4} (\tau_{4} (\tau_{4} (3 \tau_{4}+44)+396)+1120)+8064)}{\Gamma (\tau_{4}+7)} \\
 \frac{2^{-\tau_{4}} (\tau_{4}-2) (\tau_{4} (\tau_{4} (5 \tau_{4}+108)+724)+4032)}{\Gamma (\tau_{4}+7)} \\
 \frac{2^{-\tau_{4}} (\tau_{4}-2) \tau_{4} (\tau_{4} (\tau_{4}+30)+248)}{\Gamma (\tau_{4}+7)} \\
\end{array}
\right), \label{eq:cCoeffsTJ}
\\
&\mathcal{C}^{(2,1)}_{TJ}=\left(
\begin{array}{c}
 \frac{9\ 2^{-\tau_{4}-5} (\tau_{4}-2) (\tau_{4} (\tau_{4} (7 \tau_{4}+124)+900)+4176)}{\Gamma (\tau_{4}+7)} \\
 \frac{2^{-\tau_{4}-3} (3 \tau_{4} (\tau_{4} (\tau_{4} (9 \tau_{4}+164)+1140)+2944)+56448)}{\Gamma (\tau_{4}+7)} \\
 \frac{3\ 2^{-\tau_{4}-2} (\tau_{4}-2) (\tau_{4} (\tau_{4} (\tau_{4}+28)+244)+960)}{\Gamma (\tau_{4}+7)} \\
\end{array}
\right),
\\
&\mathcal{C}^{(2,2)}_{TJ}=\left(
\begin{array}{c}
 \frac{3\ 2^{-\tau_{4}-4} (\tau_{4}-2) \tau_{4} (\tau_{4}+10) (5 \tau_{4}+42)}{\Gamma (\tau_{4}+7)} \\
 \frac{2^{-\tau_{4}-2} (\tau_{4}-2) (\tau_{4} (\tau_{4} (7 \tau_{4}+138)+1064)+3312)}{\Gamma (\tau_{4}+7)} \\
 \frac{2^{-\tau_{4}-1} (\tau_{4} (\tau_{4} (\tau_{4} (\tau_{4}+22)+308)+2168)+5760)}{\Gamma (\tau_{4}+7)} \\
\end{array}
\right),
\end{align}
and the desired properties hold.

\subsection{$d=3$ and Ising-like spectrum}
\label{app:ising}
In this section, we will give spin-$4$ HS ANEC matrix elements for $d=3$ CFTs when $\tau_{4}=1.02$, which are relevant to section \ref{sec:ising}. 

The off-diagonal spin-$4$ matrix elements $\mathcal{E}^{(4,j)}_{TJ}$ are:
\begin{footnotesize}
\begin{align}
\hspace{-.35in}
\mathcal{E}^{(4,0)}_{TJ}=
\left(
\begin{array}{c}
0.00399 C^{(4)}_{J}-9.61 \mathcal{E}^{(2,0)}_{JJ,00}+316 \mathcal{E}^{(2,0)}_{JJ,11}-3860 \mathcal{E}^{(2,0)}_{JJ,22}-26.6 \mathcal{E}^{(2,1)}_{JJ,00}+269\mathcal{E}^{(2,1)}_{JJ,11}-3440 \mathcal{E}^{(2,1)}_{JJ,22}-7470\mathcal{E}^{(2,1)}_{JJ,33} 
\vspace{.2in}
\\
0.00347 C^{(4)}_{J}-23.6 \mathcal{E}^{(2,0)}_{JJ,00}+254 \mathcal{E}^{(2,0)}_{JJ,11}-1970\mathcal{E}^{(2,0)}_{JJ,22}-23.6 \mathcal{E}^{(2,1)}_{JJ,00}+162\mathcal{E}^{(2,1)}_{JJ,11}-1760\mathcal{E}^{(2,1)}_{JJ,22}-2800 \mathcal{E}^{(2,1)}_{JJ,33} 
\vspace{.2in}
\\
0.000315 C^{(4)}_{J}-2.10 \mathcal{E}^{(2,0)}_{JJ,00}-166\mathcal{E}^{(2,0)}_{JJ,11}+593\mathcal{E}^{(2,0)}_{JJ,22}-2.10 \mathcal{E}^{(2,1)}_{JJ,00}-190\mathcal{E}^{(2,1)}_{JJ,11}+224\mathcal{E}^{(2,1)}_{JJ,22}+287\mathcal{E}^{(2,1)}_{JJ,33} 
\vspace{.2in}
\\
-0.0000478 C^{(4)}_{J}+0.102 \mathcal{E}^{(2,0)}_{JJ,00}-301\mathcal{E}^{(2,0)}_{JJ,11}+1830\mathcal{E}^{(2,0)}_{JJ,22}+0.102 \mathcal{E}^{(2,1)}_{JJ,00}-303\mathcal{E}^{(2,1)}_{JJ,11}+1230\mathcal{E}^{(2,1)}_{JJ,22}+1800\mathcal{E}^{(2,1)}_{JJ,33} 
\vspace{.2in} 
\\
-0.0000863 C^{(4)}_{J}+0.198 \mathcal{E}^{(2,0)}_{JJ,00}-340\mathcal{E}^{(2,0)}_{JJ,11}+2350\mathcal{E}^{(2,0)}_{JJ,22}+0.198 \mathcal{E}^{(2,1)}_{JJ,00}-333\mathcal{E}^{(2,1)}_{JJ,11}+1670\mathcal{E}^{(2,1)}_{JJ,22}+2470\mathcal{E}^{(2,1)}_{JJ,33}\\
\end{array}
\right), \nonumber
\\ &  \vspace{-.1in} \hspace{1in}
\end{align}
\end{footnotesize}
\begin{footnotesize}
\begin{align}
\hspace{-.35in}
\mathcal{E}^{(4,1)}_{TJ}=
\left(
\begin{array}{c}
-0.00148 C^{(4)}_{J}+11.8 \mathcal{E}^{(2,0)}_{JJ,00}+25.5 \mathcal{E}^{(2,0)}_{JJ,11}-3410\mathcal{E}^{(2,0)}_{JJ,22}+28.7 \mathcal{E}^{(2,1)}_{JJ,00}+53.8 \mathcal{E}^{(2,1)}_{JJ,11}-3170\mathcal{E}^{(2,1)}_{JJ,22}-7300\mathcal{E}^{(2,1)}_{JJ,33} 
\vspace{.2in} 
\\
-0.00325 C^{(4)}_{J}+23.6 \mathcal{E}^{(2,0)}_{JJ,00}-288\mathcal{E}^{(2,0)}_{JJ,11}-984\mathcal{E}^{(2,0)}_{JJ,22}+23.7 \mathcal{E}^{(2,1)}_{JJ,00}-195\mathcal{E}^{(2,1)}_{JJ,11}-1270\mathcal{E}^{(2,1)}_{JJ,22}-2430\mathcal{E}^{(2,1)}_{JJ,33} 
\vspace{.2in} 
\\
-0.000268 C^{(4)}_{J}+2.25 \mathcal{E}^{(2,0)}_{JJ,00}-21.3 \mathcal{E}^{(2,0)}_{JJ,11}-753\mathcal{E}^{(2,0)}_{JJ,22}+2.25 \mathcal{E}^{(2,1)}_{JJ,00}+11.8 \mathcal{E}^{(2,1)}_{JJ,11}-816\mathcal{E}^{(2,1)}_{JJ,22}-1200\mathcal{E}^{(2,1)}_{JJ,33}
\vspace{.2in} 
\\
0.0000194 C^{(4)}_{J}+0.0727 \mathcal{E}^{(2,0)}_{JJ,00}+53.9 \mathcal{E}^{(2,0)}_{JJ,11}-677\mathcal{E}^{(2,0)}_{JJ,22}+0.0726 \mathcal{E}^{(2,1)}_{JJ,00}+67.5 \mathcal{E}^{(2,1)}_{JJ,11}-650\mathcal{E}^{(2,1)}_{JJ,22}-837\mathcal{E}^{(2,1)}_{JJ,33}\\
\end{array}
\right). \nonumber
\\ &  \vspace{-.1in} \hspace{1in}
\end{align}
\end{footnotesize}
One can check that if we try to set $\mathcal{E}^{(4,0)}_{TJ}=\mathcal{E}^{(4,1)}_{TJ}=0$, then we are forced to set $\<TJ^{(4)}J^{(4)}\>=0$. This implies, for this class of theories, we can not remove all the lower bounds on $\mathcal{E}^{(4,j)}_{JJ}$ while obeying the Ward identity. 

Since $\<J^{(4)}|\mathcal{E}^{(4)}|J^{(4)}\>$ is not constrained by any Ward identities or conservation conditions, its full form is generally complex. To complete our discussion of $3d$ CFTs though, we will present bounds for $d=3$, $\tau_{4}=1.02$ and the polarization tensors $\lambda^{(4,0)}_{0}$ and $\lambda^{(4,1)}_{0}$, i.e. those which do not involve any $q$ vectors:
\begin{small}
\begin{align}
\hspace{-.25in}0\leq \mathcal{E}^{(4,0)}_{JJ,00}=&0.0000113 c^{(JJJ)}_{000}-0.0000192 c^{(JJJ)}_{001}+0.0000146 c^{(JJJ)}_{002}-0.0000163 c^{(JJJ)}_{003}+0.0000246 c^{(JJJ)}_{004}\nonumber
\\&+0.0000103 c^{(JJJ)}_{011}-0.0000150 c^{(JJJ)}_{012}+0.0000160 c^{(JJJ)}_{013}+5.31\times10^{-6} c^{(JJJ)}_{022},
\\ \nonumber
\\
\hspace{-.25in}0\leq \mathcal{E}^{(4,1)}_{JJ,00}=& -8.42\times10^{-6} c^{(JJJ)}_{000}+0.0000109 c^{(JJJ)}_{001}-2.06\times10^{-6} c^{(JJJ)}_{002}-5.04\times10^{-6} c^{(JJJ)}_{003}+0.0000143 c^{(JJJ)}_{004}\nonumber
\\ &-4.38\times10^{-6} c^{(JJJ)}_{011}+1.48\times10^{-7} c^{(JJJ)}_{012}+7.60\times 10^{-6} c^{(JJJ)}_{013}+1.42\times 10^{-6} c^{(JJJ)}_{022}.
\end{align}
\end{small}
\subsection{Spin-$4$ ANEC Saturation}
\label{app:ANEC_Sat_Spin4}
Here, we will give an example of how saturating the ANEC bound for $T$ implies ANEC is also saturated for $J^{(4)}$. To be specific, we consider $d=4$ and consider saturation of the spin-0 bound. First, we see from (\ref{eq:cCoeffsTJ}) that setting $\mathcal{E}^{(2,0)}_{TT}=0$ implies $\tau_{4}=2$ and $\mathcal{E}^{(4,0)}_{TT}=0$. Saturation of HS ANEC then implies $\mathcal{E}^{(4,0)}_{TJ}=0$, which in terms of the underlying OPE coefficients gives:
\begin{align}
c^{(JTJ)}_{022}&= \frac{23}{192} c^{(JTJ)}_{000}-\frac{1}{8} c^{(JTJ)}_{002}-\frac{7}{48} c^{(JTJ)}_{010}+\frac{1}{3} c^{(JTJ)}_{012}+\frac{1}{8} c^{(JTJ)}_{020}-\frac{6 C^{(4)}_{J}}{\pi^{2} },
\\
c^{(JTJ)}_{030}&= \frac{25}{96} c^{(JTJ)}_{000}-\frac{11}{24} c^{(JTJ)}_{010}+\frac{3}{4} c^{(JTJ)}_{020}-\frac{4C^{(4)}_{J}}{\pi^{2} },
\\
c^{(JTJ)}_{040}&= \frac{13}{192} c^{(JTJ)}_{000}-\frac{5}{48} c^{(JTJ)}_{010}+\frac{1}{8} c^{(JTJ)}_{020}-\frac{2C^{(4)}_{J}}{ \pi^{2} },
\\
c^{(JTJ)}_{121}&= \frac{67}{480} c^{(JTJ)}_{000}-\frac{7}{20} c^{(JTJ)}_{002}-\frac{1}{6} c^{(JTJ)}_{010}+\frac{7}{6} c^{(JTJ)}_{012}+\frac{1}{4} c^{(JTJ)}_{020}-\frac{12 C^{(4)}_{J}}{\pi },
\\
c^{(JTJ)}_{131}&= \frac{169 c^{(JTJ)}_{000}}{1440}-\frac{1}{15} c^{(JTJ)}_{002}-\frac{23}{144} c^{(JTJ)}_{010}+\frac{5}{36} c^{(JTJ)}_{012}+\frac{1}{6}c^{(JTJ)}_{020}-\frac{4C^{(4)}_{J}}{\pi }.
\end{align}
We can then take this answer and calculate the diagonal matrix elements $\mathcal{E}^{(2,j)}_{JJ,aa}$ for $j=0,1,2$ and $a=0,1,...,5-j$. This gives 12 positivity conditions on 6 variables, the 5 OPE coefficients and the normalization, $C^{(4)}_{J}$. The positivity requirements completely fix 4 of the OPE coefficients: 
\begin{align}
&c^{(JTJ)}_{002}= \frac{3}{16} c^{(JTJ)}_{000}-\frac{9 C^{(4)}_{J}}{7 \pi }, \qquad c^{(JTJ)}_{010}= \frac{1}{8} c^{(JTJ)}_{000}-\frac{30 C^{(4)}_{J}}{7 \pi },
\\ &c^{(JTJ)}_{012}= \frac{3 C^{(4)}_{J}}{2 \pi }-\frac{3}{32} c^{(JTJ)}_{000}, \qquad c^{(JTJ)}_{020}= -\frac{3}{8} c^{(JTJ)}_{000}.
\end{align}

The final answer for the matrix elements has the expected structure:
\begin{align}
&\mathcal{E}^{(2,0)}_{JJ}=0,
\\ &\mathcal{E}^{(2,1)}_{JJ,ij}\propto\delta_{i0}\delta_{j0}\pi ^3 (7 \pi  c^{(JTJ)}_{000}+144 C^{(4)}_{J}),
\\ &\mathcal{E}^{(2,2)}_{JJ,ij}\propto \delta_{i0}\delta_{j0}\pi ^3 (16 C^{(4)}_{J}-7 \pi  c^{(JTJ)}_{000}).
\end{align}

This is also a nice consistency check on the formulas presented here. For $\mathcal{E}^{(2,1)}_{JJ}$ and $\mathcal{E}^{(2,2)}_{JJ}$, all but the first, diagonal entry must vanish by conservation. Similar results hold for all $d$ and if we saturate the other $\<TTT\>$ collider bounds.

\sec{Sums of Double-Twist Operators}
\label{app:RegDT}
In this appendix we give more details on how the $[\f\p]_{0,\ell}$ double-twist operators affect $c_{\f\f[\f\f]_{0,s}}c_{\p\p[\f\f]_{0,s}}$ if $\f$ exchange gives the dominant contribution to the $[\f\p]_{0,\ell}$ anomalous dimensions. 

The first issue is the sum in (\ref{eq:RHS_DT_cross}) diverges logarithmically. To obtain a finite result, we replace the exchanged $\f$ with a generic $\O$ and then expand $\frac{1}{4}c^{2}_{\f\p[\f\p]_{0,\ell}}\gamma_{[\f\p]_{0,\ell}}^{2}$ in terms of the functions $S^{h_{\p\f},h_{\f\p}}_{a}(\bar{h})$ by matching the asymptotic expansions:
\begin{align}
&\frac{1}{4}c^{2}_{\f\p[\f\p]_{0,\ell}}(\bar{h})\gamma_{[\f\p]_{0,\ell}}(\bar{h})^{2}\sim \frac{\left(c_{\f\f\O}c_{\p\p\O}V^{(0)\f\p\p\f}_{\O,0}(\bar{h})\right)^{2}}{S^{h_{\p\f},h_{\f\p}}_{-h_{\f}-h_{\p}}(\bar{h})}\sim \sum_{\substack{a=0,1,...}}b_{a}S^{h_{\p\f},h_{\f\p}}_{2h_{\O}-h_{\f}-h_{\p}+a}(\bar{h}), \label{eq:cgammaExpand}
\\ \nonumber 
\\
&b_0= \frac{\Gamma (2 h_{\f}) \Gamma (2 h_{\p}) \Gamma (2 \bar{h}_{\O})^2 \Gamma (2 (h_{\f}- h_{\O})) \Gamma (2 (h_{\p}-h_{\O}))}{\Gamma (\bar{h}_{\O})^4 \Gamma (2 h_{\f}-h_{\O})^2 \Gamma (2 h_{\p}-h_{\O})^2},  \label{eq:b0expand}
\\
&b_1= \frac{2 h_{\O}^2 \Gamma (2 h_{\f}) \Gamma (2 h_{\p}) \Gamma (2 \bar{h}_{\O})^2 (h_{\f}+h_{\p}-h_{\O}-1) \Gamma (2 h_{\f}-2 h_{\O}-1) \Gamma (2 h_{\p}-2 h_{\O}-1)}{\Gamma (\bar{h}_{\O})^4 \Gamma (2 h_{\f}-h_{\O})^2 \Gamma (2 h_{\p}-h_{\O})^2},
\\ 
&b_2=  \Gamma (2 h_{\f}-2 h_{\O}-2) \Gamma (2 h_{\p}-2 h_{\O}-2) (-h_{\f}-h_{\p}+h_{\O}+1) (-2 h_{\f}-2 h_{\p}+2 h_{\O}+3)
\nonumber \\ &\hspace{3.35in}\times \frac{h_{\O}^2 (h_{\O}+1)^2 \Gamma (2 h_{\f}) \Gamma (2 h_{\p}) \Gamma (2 \bar{h}_{\O})^2}{\Gamma (\bar{h}_{\O})^4 \Gamma (2 h_{\f}-h_{\O})^2 \Gamma (2 h_{\p}-h_{\O})^2}.
\end{align}
In the limit $h_{\O}\rightarrow h_{\f}$, the $b_{i}$'s diverge while the $S^{h_{\p\f},h_{\f\p}}_{2h_{\O}-h_{\f}-h_{\p}+a}$ vanish. If we set $h_{\O}=h_{\f}+\frac{\epsilon}{2}$, first perform the sum over $\ell$, and then let $\epsilon\rightarrow 0$ we find \cite{dsdi}:
\begin{equation}
\underset{\epsilon\rightarrow 0}{\lim}\sum_{\substack{h=h_{0}+\ell \\ \ell=0,1,...}}\Gamma(-\epsilon)S^{r,-r}_{-r+\epsilon}(h) k^{r,-r}_{2h}(1-z)\bigg|_{z^{0}}=\frac{1}{h_{0}-r}+H_{2 r-1} -H_{h_{0}-r}-H_{h_{0}+r-2},\label{eq:infSum}
\end{equation}
where $H_{a}$ is the harmonic number. Keeping only the $b_{0}$ term in (\ref{eq:cgammaExpand}), expanding (\ref{eq:crossingLCDT}) to leading order in $1-\bar{z}$, and using (\ref{eq:infSum}) to perform the sum, we find:
\begin{align} 
&c_{\f\f[\f\f]_{0,s}}(\bar{h})c_{\p\p[\f\f]_{0,s}}(\bar{h})\supset\frac{4^{3 h_{\f}-1} \Gamma \left(h_{\f}+\frac{1}{2}\right)^3 \Gamma (2 h_{\p}) \Gamma (2 h_{\p}-2 h_{\f}) }{\pi ^{3/2} \Gamma (h_{\f})^3 \Gamma (2 h_{\p}-h_{\f})^2}\partial_{a}^{2}S^{0,0}_{-h_\f-h_\p+a}(\bar{h})\bigg|_{a=0}
\nonumber \\ & \hspace{1.75in} \times\left(H_{-2 h_{\f}+2 h_{\p}-1}-H_{2 h_{\f}+\ell_{0}}-H_{2 h_{\p}+\ell_{0}-2}+\frac{1}{2 h_{\f}+\ell_{0}}\right),\label{eq:NegOPEdt}
\end{align}
where the sum starts at $\ell_{0}$\footnote{We are ignoring the effect of having to reparameterize the sum in generic CFTs \cite{dsdi}, which gives a subleading effect if we take $\ell_{0}$ large and/or assume $\gamma_{[\f\p]_{0,\ell}}\ll1$.}. Due to the harmonic numbers, this correction is generically negative, especially as we increase $\ell_0$. For example, at large $\ell_0$ we have:
\begin{equation}
c_{\f\f[\f\f]_{0,s}}(\bar{h})c_{\f\f[\f\f]_{0,s}}(\bar{h})\supset -log(\ell_{0})c_{\f\f\f}^{2}c_{\p\p\f}^{2}\frac{2^{6 h_{\f}-1} \Gamma \left(h_{\f}+\frac{1}{2}\right)^3 \Gamma (2 h_{\p}) \Gamma (2 h_{\p}-2 h_{\f})}{\pi ^{3/2} \Gamma (h_{\f})^3 \Gamma (2 h_{\p}-h_{\f})^2}\partial_{a}^{2}S^{0,0}_{a}(\bar{h})\bigg|_{a=0},
\end{equation}
which is manifestly negative. Of course, as we increase $\ell_{0}$ we need to specify information about the double-twist operators with spin $\ell<\ell_{0}$.

\bibliographystyle{utphys}
\bibliography{biblio}

\providecommand{\href}[2]{#2}\begingroup\raggedright\begin{thebibliography}{10}

\bibitem{Cornalba:2007fs}
L.~Cornalba, ``{Eikonal methods in AdS/CFT: Regge theory and multi-reggeon
  exchange},''
\href{http://arxiv.org/abs/0710.5480}{{\ttfamily arXiv:0710.5480 [hep-th]}}.

\bibitem{Costa:2012cb}
M.~S. Costa, V.~Goncalves, and J.~Penedones, ``{Conformal Regge theory},''
  \href{http://dx.doi.org/10.1007/JHEP12(2012)091}{{\em JHEP} {\bfseries 1212}
  (2012) 091},
\href{http://arxiv.org/abs/1209.4355}{{\ttfamily arXiv:1209.4355 [hep-th]}}.

\bibitem{Caron-Huot:2017vep}
S.~Caron-Huot, ``{Analyticity in Spin in Conformal Theories},''
\href{http://arxiv.org/abs/1703.00278}{{\ttfamily arXiv:1703.00278 [hep-th]}}.

\bibitem{Hofman:2008ar}
D.~M. Hofman and J.~Maldacena, ``{Conformal collider physics: Energy and charge
  correlations},'' \href{http://dx.doi.org/10.1088/1126-6708/2008/05/012}{{\em
  JHEP} {\bfseries 05} (2008) 012},
\href{http://arxiv.org/abs/0803.1467}{{\ttfamily arXiv:0803.1467 [hep-th]}}.

\bibitem{Hartman:2016lgu}
T.~Hartman, S.~Kundu, and A.~Tajdini, ``{Averaged Null Energy Condition from
  Causality},''
\href{http://arxiv.org/abs/1610.05308}{{\ttfamily arXiv:1610.05308 [hep-th]}}.

\bibitem{Faulkner:2016mzt}
T.~Faulkner, R.~G. Leigh, O.~Parrikar, and H.~Wang, ``{Modular Hamiltonians for
  Deformed Half-Spaces and the Averaged Null Energy Condition},''
  \href{http://dx.doi.org/10.1007/JHEP09(2016)038}{{\em JHEP} {\bfseries 09}
  (2016) 038},
\href{http://arxiv.org/abs/1605.08072}{{\ttfamily arXiv:1605.08072 [hep-th]}}.

\bibitem{Komargodski:2016gci}
Z.~Komargodski, M.~Kulaxizi, A.~Parnachev, and A.~Zhiboedov, ``{Conformal Field
  Theories and Deep Inelastic Scattering},''
  \href{http://dx.doi.org/10.1103/PhysRevD.95.065011}{{\em Phys. Rev.}
  {\bfseries D95} no.~6, (2017) 065011},
\href{http://arxiv.org/abs/1601.05453}{{\ttfamily arXiv:1601.05453 [hep-th]}}.

\bibitem{Hofman:2016awc}
D.~M. Hofman, D.~Li, D.~Meltzer, D.~Poland, and F.~Rejon-Barrera, ``{A Proof of
  the Conformal Collider Bounds},''
  \href{http://dx.doi.org/10.1007/JHEP06(2016)111}{{\em JHEP} {\bfseries 06}
  (2016) 111},
\href{http://arxiv.org/abs/1603.03771}{{\ttfamily arXiv:1603.03771 [hep-th]}}.

\bibitem{light1}
A.~L. Fitzpatrick, J.~Kaplan, D.~Poland, and D.~Simmons-Duffin, ``{The Analytic
  Bootstrap and AdS Superhorizon Locality},''
  \href{http://dx.doi.org/10.1007/JHEP12(2013)004}{{\em JHEP} {\bfseries 12}
  (2013) 004},
\href{http://arxiv.org/abs/1212.3616}{{\ttfamily arXiv:1212.3616 [hep-th]}}.

\bibitem{light2}
Z.~Komargodski and A.~Zhiboedov, ``{Convexity and Liberation at Large Spin},''
  \href{http://dx.doi.org/10.1007/JHEP11(2013)140}{{\em JHEP} {\bfseries 11}
  (2013) 140},
\href{http://arxiv.org/abs/1212.4103}{{\ttfamily arXiv:1212.4103 [hep-th]}}.

\bibitem{Poland:2018epd}
D.~Poland, S.~Rychkov, and A.~Vichi, ``{The Conformal Bootstrap: Theory,
  Numerical Techniques, and Applications},''
\href{http://arxiv.org/abs/1805.04405}{{\ttfamily arXiv:1805.04405 [hep-th]}}.

\bibitem{Simmons-Duffin:2016gjk}
D.~Simmons-Duffin, \href{http://dx.doi.org/10.1142/9789813149441_0001}{``{The
  Conformal Bootstrap},''} in {\em {Proceedings, Theoretical Advanced Study
  Institute in Elementary Particle Physics: New Frontiers in Fields and Strings
  (TASI 2015): Boulder, CO, USA, June 1-26, 2015}}, pp.~1--74.
\newblock 2017.
\newblock
\href{http://arxiv.org/abs/1602.07982}{{\ttfamily arXiv:1602.07982 [hep-th]}}.
\newblock

\bibitem{Li:2015itl}
D.~Li, D.~Meltzer, and D.~Poland, ``{Conformal Collider Physics from the
  Lightcone Bootstrap},'' \href{http://dx.doi.org/10.1007/JHEP02(2016)143}{{\em
  JHEP} {\bfseries 02} (2016) 143},
\href{http://arxiv.org/abs/1511.08025}{{\ttfamily arXiv:1511.08025 [hep-th]}}.

\bibitem{krav}
A.~Dymarsky, F.~Kos, P.~Kravchuk, D.~Poland, and D.~Simmons-Duffin, ``{The 3d
  Stress-Tensor Bootstrap},''
\href{http://arxiv.org/abs/1708.05718}{{\ttfamily arXiv:1708.05718 [hep-th]}}.

\bibitem{Dymarsky:2017xzb}
A.~Dymarsky, J.~Penedones, E.~Trevisani, and A.~Vichi, ``{Charting the space of
  3D CFTs with a continuous global symmetry},''
\href{http://arxiv.org/abs/1705.04278}{{\ttfamily arXiv:1705.04278 [hep-th]}}.

\bibitem{Maldacena:2011jn}
J.~Maldacena and A.~Zhiboedov, ``{Constraining Conformal Field Theories with A
  Higher Spin Symmetry},''
  \href{http://dx.doi.org/10.1088/1751-8113/46/21/214011}{{\em J. Phys.}
  {\bfseries A46} (2013) 214011},
\href{http://arxiv.org/abs/1112.1016}{{\ttfamily arXiv:1112.1016 [hep-th]}}.

\bibitem{Boulanger:2013zza}
N.~Boulanger, D.~Ponomarev, E.~D. Skvortsov, and M.~Taronna, ``{On the
  uniqueness of higher-spin symmetries in AdS and CFT},''
  \href{http://dx.doi.org/10.1142/S0217751X13501625}{{\em Int. J. Mod. Phys.}
  {\bfseries A28} (2013) 1350162},
\href{http://arxiv.org/abs/1305.5180}{{\ttfamily arXiv:1305.5180 [hep-th]}}.

\bibitem{Alba:2015upa}
V.~Alba and K.~Diab, ``{Constraining conformal field theories with a higher
  spin symmetry in $d> 3$ dimensions},''
\href{http://arxiv.org/abs/1510.02535}{{\ttfamily arXiv:1510.02535 [hep-th]}}.

\bibitem{Aharony:2011jz}
O.~Aharony, G.~Gur-Ari, and R.~Yacoby, ``{d=3 Bosonic Vector Models Coupled to
  Chern-Simons Gauge Theories},''
  \href{http://dx.doi.org/10.1007/JHEP03(2012)037}{{\em JHEP} {\bfseries 03}
  (2012) 037},
\href{http://arxiv.org/abs/1110.4382}{{\ttfamily arXiv:1110.4382 [hep-th]}}.

\bibitem{Giombi:2011kc}
S.~Giombi, S.~Minwalla, S.~Prakash, S.~P. Trivedi, S.~R. Wadia, and X.~Yin,
  ``{Chern-Simons Theory with Vector Fermion Matter},''
  \href{http://dx.doi.org/10.1140/epjc/s10052-012-2112-0}{{\em Eur. Phys. J.}
  {\bfseries C72} (2012) 2112},
\href{http://arxiv.org/abs/1110.4386}{{\ttfamily arXiv:1110.4386 [hep-th]}}.

\bibitem{Maldacena:2012sf}
J.~Maldacena and A.~Zhiboedov, ``{Constraining conformal field theories with a
  slightly broken higher spin symmetry},''
  \href{http://dx.doi.org/10.1088/0264-9381/30/10/104003}{{\em
  Class.Quant.Grav.} {\bfseries 30} (2013) 104003},
\href{http://arxiv.org/abs/1204.3882}{{\ttfamily arXiv:1204.3882 [hep-th]}}.

\bibitem{Aharony:2018npf}
O.~Aharony, L.~F. Alday, A.~Bissi, and R.~Yacoby, ``{The Analytic Bootstrap for
  Large $N$ Chern-Simons Vector Models},''
  \href{http://dx.doi.org/10.1007/JHEP08(2018)166}{{\em JHEP} {\bfseries 08}
  (2018) 166},
\href{http://arxiv.org/abs/1805.04377}{{\ttfamily arXiv:1805.04377 [hep-th]}}.

\bibitem{Turiaci:2018nua}
G.~J. Turiaci and A.~Zhiboedov, ``{Veneziano Amplitude of Vasiliev Theory},''
  \href{http://dx.doi.org/10.1007/JHEP10(2018)034}{{\em JHEP} {\bfseries 10}
  (2018) 034},
\href{http://arxiv.org/abs/1802.04390}{{\ttfamily arXiv:1802.04390 [hep-th]}}.

\bibitem{ssw}
D.~Simmons-Duffin, D.~Stanford, and E.~Witten, ``{A spacetime derivation of the
  Lorentzian OPE inversion formula},''
\href{http://arxiv.org/abs/1711.03816}{{\ttfamily arXiv:1711.03816 [hep-th]}}.

\bibitem{Maldacena:2015waa}
J.~Maldacena, S.~H. Shenker, and D.~Stanford, ``{A bound on chaos},''
  \href{http://dx.doi.org/10.1007/JHEP08(2016)106}{{\em JHEP} {\bfseries 08}
  (2016) 106},
\href{http://arxiv.org/abs/1503.01409}{{\ttfamily arXiv:1503.01409 [hep-th]}}.

\bibitem{Hartman:2015lfa}
T.~Hartman, S.~Jain, and S.~Kundu, ``{Causality Constraints in Conformal Field
  Theory},'' \href{http://dx.doi.org/10.1007/JHEP05(2016)099}{{\em JHEP}
  {\bfseries 05} (2016) 099},
\href{http://arxiv.org/abs/1509.00014}{{\ttfamily arXiv:1509.00014 [hep-th]}}.

\bibitem{Kravchuk:2018htv}
P.~Kravchuk and D.~Simmons-Duffin, ``{Light-ray operators in conformal field
  theory},''
\href{http://arxiv.org/abs/1805.00098}{{\ttfamily arXiv:1805.00098 [hep-th]}}.

\bibitem{Maldacena:1997re}
J.~M. Maldacena, ``{The Large N limit of superconformal field theories and
  supergravity},'' \href{http://dx.doi.org/10.1023/A:1026654312961,
  10.4310/ATMP.1998.v2.n2.a1}{{\em Int. J. Theor. Phys.} {\bfseries 38} (1999)
  1113--1133}, \href{http://arxiv.org/abs/hep-th/9711200}{{\ttfamily
  arXiv:hep-th/9711200 [hep-th]}}.
[Adv. Theor. Math. Phys.2,231(1998)].

\bibitem{Witten:1998qj}
E.~Witten, ``{Anti-de Sitter space and holography},''
  \href{http://dx.doi.org/10.4310/ATMP.1998.v2.n2.a2}{{\em Adv. Theor. Math.
  Phys.} {\bfseries 2} (1998) 253--291},
\href{http://arxiv.org/abs/hep-th/9802150}{{\ttfamily arXiv:hep-th/9802150
  [hep-th]}}.

\bibitem{Gubser:1998bc}
S.~S. Gubser, I.~R. Klebanov, and A.~M. Polyakov, ``{Gauge theory correlators
  from noncritical string theory},''
  \href{http://dx.doi.org/10.1016/S0370-2693(98)00377-3}{{\em Phys. Lett.}
  {\bfseries B428} (1998) 105--114},
\href{http://arxiv.org/abs/hep-th/9802109}{{\ttfamily arXiv:hep-th/9802109
  [hep-th]}}.

\bibitem{Vasiliev:1990en}
M.~A. Vasiliev, ``{Consistent equation for interacting gauge fields of all
  spins in (3+1)-dimensions},''
\href{http://dx.doi.org/10.1016/0370-2693(90)91400-6}{{\em Phys. Lett.}
  {\bfseries B243} (1990) 378--382}.

\bibitem{Vasiliev:1999ba}
M.~A. Vasiliev, ``{Higher spin gauge theories: Star product and AdS space},''
\href{http://arxiv.org/abs/hep-th/9910096}{{\ttfamily arXiv:hep-th/9910096
  [hep-th]}}.

\bibitem{Sezgin:2002rt}
E.~Sezgin and P.~Sundell, ``{Massless higher spins and holography},''
  \href{http://dx.doi.org/10.1016/S0550-3213(02)00739-3,
  10.1016/S0550-3213(03)00267-0}{{\em Nucl. Phys.} {\bfseries B644} (2002)
  303--370}, \href{http://arxiv.org/abs/hep-th/0205131}{{\ttfamily
  arXiv:hep-th/0205131 [hep-th]}}.
[Erratum: Nucl. Phys.B660,403(2003)].

\bibitem{Vasiliev:2003ev}
M.~A. Vasiliev, ``{Nonlinear equations for symmetric massless higher spin
  fields in (A)dS(d)},''
  \href{http://dx.doi.org/10.1016/S0370-2693(03)00872-4}{{\em Phys. Lett.}
  {\bfseries B567} (2003) 139--151},
\href{http://arxiv.org/abs/hep-th/0304049}{{\ttfamily arXiv:hep-th/0304049
  [hep-th]}}.

\bibitem{Klebanov:2002ja}
I.~R. Klebanov and A.~M. Polyakov, ``{AdS dual of the critical O(N) vector
  model},'' \href{http://dx.doi.org/10.1016/S0370-2693(02)02980-5}{{\em Phys.
  Lett.} {\bfseries B550} (2002) 213--219},
\href{http://arxiv.org/abs/hep-th/0210114}{{\ttfamily arXiv:hep-th/0210114
  [hep-th]}}.

\bibitem{Casini:2010bf}
H.~Casini, ``{Wedge reflection positivity},''
  \href{http://dx.doi.org/10.1088/1751-8113/44/43/435202}{{\em J. Phys.}
  {\bfseries A44} (2011) 435202},
\href{http://arxiv.org/abs/1009.3832}{{\ttfamily arXiv:1009.3832 [hep-th]}}.

\bibitem{Heemskerk:2009pn}
I.~Heemskerk, J.~Penedones, J.~Polchinski, and J.~Sully, ``{Holography from
  Conformal Field Theory},''
  \href{http://dx.doi.org/10.1088/1126-6708/2009/10/079}{{\em JHEP} {\bfseries
  0910} (2009) 079},
\href{http://arxiv.org/abs/0907.0151}{{\ttfamily arXiv:0907.0151 [hep-th]}}.

\bibitem{Costa:2017twz}
M.~S. Costa, T.~Hansen, and J.~Penedones, ``{Bounds for OPE coefficients on the
  Regge trajectory},''
\href{http://arxiv.org/abs/1707.07689}{{\ttfamily arXiv:1707.07689 [hep-th]}}.

\bibitem{Simmons-Duffin:2016wlq}
D.~Simmons-Duffin, ``{The Lightcone Bootstrap and the Spectrum of the 3d Ising
  CFT},'' \href{http://dx.doi.org/10.1007/JHEP03(2017)086}{{\em JHEP}
  {\bfseries 03} (2017) 086},
\href{http://arxiv.org/abs/1612.08471}{{\ttfamily arXiv:1612.08471 [hep-th]}}.

\bibitem{zhib}
A.~Zhiboedov, ``{On Conformal Field Theories With Extremal a/c Values},''
  \href{http://dx.doi.org/10.1007/JHEP04(2014)038}{{\em JHEP} {\bfseries 04}
  (2014) 038},
\href{http://arxiv.org/abs/1304.6075}{{\ttfamily arXiv:1304.6075 [hep-th]}}.

\bibitem{Osborn:1993cr}
H.~Osborn and A.~Petkou, ``{Implications of conformal invariance in field
  theories for general dimensions},''
  \href{http://dx.doi.org/10.1006/aphy.1994.1045}{{\em Annals Phys.} {\bfseries
  231} (1994) 311--362},
\href{http://arxiv.org/abs/hep-th/9307010}{{\ttfamily arXiv:hep-th/9307010
  [hep-th]}}.

\bibitem{Li:2015rfa}
D.~Li, D.~Meltzer, and D.~Poland, ``{Non-Abelian Binding Energies from the
  Lightcone Bootstrap},'' \href{http://dx.doi.org/10.1007/JHEP02(2016)149}{{\em
  JHEP} {\bfseries 02} (2016) 149},
\href{http://arxiv.org/abs/1510.07044}{{\ttfamily arXiv:1510.07044 [hep-th]}}.

\bibitem{ElShowk:2012ht}
S.~El-Showk, M.~F. Paulos, D.~Poland, S.~Rychkov, D.~Simmons-Duffin, and
  A.~Vichi, ``{Solving the 3D Ising Model with the Conformal Bootstrap},''
  \href{http://dx.doi.org/10.1103/PhysRevD.86.025022}{{\em Phys.Rev.}
  {\bfseries D86} (2012) 025022},
\href{http://arxiv.org/abs/1203.6064}{{\ttfamily arXiv:1203.6064 [hep-th]}}.

\bibitem{Atanasov:2018kqw}
A.~Atanasov, A.~Hillman, and D.~Poland, ``{Bootstrapping the Minimal 3D
  SCFT},''
\href{http://arxiv.org/abs/1807.05702}{{\ttfamily arXiv:1807.05702 [hep-th]}}.

\bibitem{Rong:2018okz}
J.~Rong and N.~Su, ``{Bootstrapping minimal $\mathcal{N}=1$ superconformal
  field theory in three dimensions},''
\href{http://arxiv.org/abs/1807.04434}{{\ttfamily arXiv:1807.04434 [hep-th]}}.

\bibitem{Pappadopulo:2012jk}
D.~Pappadopulo, S.~Rychkov, J.~Espin, and R.~Rattazzi, ``{OPE Convergence in
  Conformal Field Theory},''
  \href{http://dx.doi.org/10.1103/PhysRevD.86.105043}{{\em Phys.Rev.}
  {\bfseries D86} (2012) 105043},
\href{http://arxiv.org/abs/1208.6449}{{\ttfamily arXiv:1208.6449 [hep-th]}}.

\bibitem{Cordova:2017zej}
C.~C{\'o}rdova, J.~Maldacena, and G.~J. Turiaci, ``Bounds on ope coefficients
  from interference effects in the conformal collider,''
  \href{http://dx.doi.org/10.1007/JHEP11(2017)032}{{\em Journal of High Energy
  Physics} {\bfseries 2017} (2017) 32},
  \href{http://arxiv.org/abs/1710.03199}{{\ttfamily arXiv:1710.03199
  [hep-th]}}.
\url{https://doi.org/10.1007/JHEP11(2017)032}.

\bibitem{Costa:2011mg}
M.~S. Costa, J.~Penedones, D.~Poland, and S.~Rychkov, ``{Spinning Conformal
  Correlators},'' \href{http://dx.doi.org/10.1007/JHEP11(2011)071}{{\em JHEP}
  {\bfseries 1111} (2011) 071},
\href{http://arxiv.org/abs/1107.3554}{{\ttfamily arXiv:1107.3554 [hep-th]}}.

\bibitem{Meltzer:2017rtf}
D.~Meltzer and E.~Perlmutter, ``{Beyond $a = c$: gravitational couplings to
  matter and the stress tensor OPE},''
  \href{http://dx.doi.org/10.1007/JHEP07(2018)157}{{\em JHEP} {\bfseries 07}
  (2018) 157},
\href{http://arxiv.org/abs/1712.04861}{{\ttfamily arXiv:1712.04861 [hep-th]}}.

\bibitem{Chowdhury:2017vel}
S.~D. Chowdhury, J.~R. David, and S.~Prakash, ``{Constraints on parity
  violating conformal field theories in $d=3$},''
\href{http://arxiv.org/abs/1707.03007}{{\ttfamily arXiv:1707.03007 [hep-th]}}.

\bibitem{Afkhami-Jeddi:2018own}
N.~Afkhami-Jeddi, S.~Kundu, and A.~Tajdini, ``{A Conformal Collider for
  Holographic CFTs},''
\href{http://arxiv.org/abs/1805.07393}{{\ttfamily arXiv:1805.07393 [hep-th]}}.

\bibitem{Zhiboedov:2012bm}
A.~Zhiboedov, ``{A note on three-point functions of conserved currents},''
\href{http://arxiv.org/abs/1206.6370}{{\ttfamily arXiv:1206.6370 [hep-th]}}.

\bibitem{Adams:2006sv}
A.~Adams, N.~Arkani-Hamed, S.~Dubovsky, A.~Nicolis, and R.~Rattazzi,
  ``{Causality, analyticity and an IR obstruction to UV completion},''
  \href{http://dx.doi.org/10.1088/1126-6708/2006/10/014}{{\em JHEP} {\bfseries
  0610} (2006) 014},
\href{http://arxiv.org/abs/hep-th/0602178}{{\ttfamily arXiv:hep-th/0602178
  [hep-th]}}.

\bibitem{Fitzpatrick:2016thx}
A.~L. Fitzpatrick and J.~Kaplan, ``{A Quantum Correction To Chaos},''
\href{http://arxiv.org/abs/1601.06164}{{\ttfamily arXiv:1601.06164 [hep-th]}}.

\bibitem{Alday:2016htq}
L.~F. Alday and A.~Bissi, ``{Unitarity and positivity constraints for CFT at
  large central charge},''
\href{http://arxiv.org/abs/1606.09593}{{\ttfamily arXiv:1606.09593 [hep-th]}}.

\bibitem{DO3}
F.~Dolan and H.~Osborn, ``{Conformal Partial Waves: Further Mathematical
  Results},''
\href{http://arxiv.org/abs/1108.6194v2}{{\ttfamily arXiv:1108.6194v2
  [hep-th]}}.

\bibitem{dsdi}
D.~Simmons-Duffin, ``{The Lightcone Bootstrap and the Spectrum of the 3d Ising
  CFT},'' \href{http://dx.doi.org/10.1007/JHEP03(2017)086}{{\em JHEP}
  {\bfseries 03} (2017) 086},
\href{http://arxiv.org/abs/1612.08471}{{\ttfamily arXiv:1612.08471 [hep-th]}}.

\bibitem{DO1}
F.~Dolan and H.~Osborn, ``{Conformal four point functions and the operator
  product expansion},''
  \href{http://dx.doi.org/10.1016/S0550-3213(01)00013-X}{{\em Nucl.Phys.}
  {\bfseries B599} (2001) 459--496},
\href{http://arxiv.org/abs/hep-th/0011040}{{\ttfamily arXiv:hep-th/0011040
  [hep-th]}}.

\bibitem{DO2}
F.~Dolan and H.~Osborn, ``{Conformal partial waves and the operator product
  expansion},'' \href{http://dx.doi.org/10.1016/j.nuclphysb.2003.11.016}{{\em
  Nucl.Phys.} {\bfseries B678} (2004) 491--507},
\href{http://arxiv.org/abs/hep-th/0309180}{{\ttfamily arXiv:hep-th/0309180
  [hep-th]}}.

\bibitem{Alday:2016njk}
L.~F. Alday, ``{Large Spin Perturbation Theory for Conformal Field Theories},''
  \href{http://dx.doi.org/10.1103/PhysRevLett.119.111601}{{\em Phys. Rev.
  Lett.} {\bfseries 119} no.~11, (2017) 111601},
\href{http://arxiv.org/abs/1611.01500}{{\ttfamily arXiv:1611.01500 [hep-th]}}.

\bibitem{Liu:2018jhs}
J.~Liu, E.~Perlmutter, V.~Rosenhaus, and D.~Simmons-Duffin, ``{$d$-dimensional
  SYK, AdS Loops, and $6j$ Symbols},''
\href{http://arxiv.org/abs/1808.00612}{{\ttfamily arXiv:1808.00612 [hep-th]}}.

\bibitem{Sleight:2018ryu}
C.~Sleight and M.~Taronna, ``{A Note on Anomalous Dimensions from Crossing
  Kernels},''
\href{http://arxiv.org/abs/1807.05941}{{\ttfamily arXiv:1807.05941 [hep-th]}}.

\bibitem{Cardona:2018qrt}
C.~Cardona, S.~Guha, S.~K. Kanumilli, and K.~Sen, ``{Resummation at finite
  conformal spin},'' \href{http://dx.doi.org/10.1007/JHEP01(2019)077}{{\em
  JHEP} {\bfseries 01} (2019) 077},
  \href{http://arxiv.org/abs/1811.00213}{{\ttfamily arXiv:1811.00213
  [hep-th]}}.
[JHEP19,077(2020)].

\bibitem{Cardona:2018dov}
C.~Cardona and K.~Sen, ``{Anomalous dimensions at finite conformal spin from
  OPE inversion},''
\href{http://arxiv.org/abs/1806.10919}{{\ttfamily arXiv:1806.10919 [hep-th]}}.

\bibitem{Albayrak:2019gnz}
S.~Albayrak, D.~Meltzer, and D.~Poland, ``{More Analytic Bootstrap:
  Nonperturbative Effects and Fermions},''
\href{http://arxiv.org/abs/1904.00032}{{\ttfamily arXiv:1904.00032 [hep-th]}}.

\bibitem{Camanho:2014apa}
X.~O. Camanho, J.~D. Edelstein, J.~Maldacena, and A.~Zhiboedov, ``{Causality
  Constraints on Corrections to the Graviton Three-Point Coupling},''
  \href{http://dx.doi.org/10.1007/JHEP02(2016)020}{{\em JHEP} {\bfseries 02}
  (2016) 020},
\href{http://arxiv.org/abs/1407.5597}{{\ttfamily arXiv:1407.5597 [hep-th]}}.

\bibitem{Fitzpatrick:2011dm}
A.~L. Fitzpatrick and J.~Kaplan, ``{Unitarity and the Holographic S-Matrix},''
  \href{http://dx.doi.org/10.1007/JHEP10(2012)032}{{\em JHEP} {\bfseries 1210}
  (2012) 032},
\href{http://arxiv.org/abs/1112.4845}{{\ttfamily arXiv:1112.4845 [hep-th]}}.

\bibitem{Casini:2017roe}
H.~Casini, E.~Teste, and G.~Torroba, ``{Modular Hamiltonians on the null plane
  and the Markov property of the vacuum state},''
  \href{http://dx.doi.org/10.1088/1751-8121/aa7eaa}{{\em J. Phys.} {\bfseries
  A50} no.~36, (2017) 364001},
\href{http://arxiv.org/abs/1703.10656}{{\ttfamily arXiv:1703.10656 [hep-th]}}.

\bibitem{Cordova:2018ygx}
C.~Cordova and S.-H. Shao, ``{Light-ray Operators and the BMS Algebra},''
\href{http://arxiv.org/abs/1810.05706}{{\ttfamily arXiv:1810.05706 [hep-th]}}.

\bibitem{MKPK:wip}
M.~Kologlu, P.~Kravchuk, D.~Simmons-Duffin, and A.~Zhiboedov
  \href{http://arxiv.org/abs/To appear}{{\ttfamily To appear}}.

\bibitem{Cordova:2017dhq}
C.~Cordova and K.~Diab, ``{Universal Bounds on Operator Dimensions from the
  Average Null Energy Condition},''
  \href{http://dx.doi.org/10.1007/JHEP02(2018)131}{{\em JHEP} {\bfseries 02}
  (2018) 131},
\href{http://arxiv.org/abs/1712.01089}{{\ttfamily arXiv:1712.01089 [hep-th]}}.

\bibitem{Aharony:2016dwx}
O.~Aharony, L.~F. Alday, A.~Bissi, and E.~Perlmutter, ``{Loops in AdS from
  Conformal Field Theory},''
\href{http://arxiv.org/abs/1612.03891}{{\ttfamily arXiv:1612.03891 [hep-th]}}.

\bibitem{Yuan:2017vgp}
E.~Y. Yuan, ``{Loops in the Bulk},''
\href{http://arxiv.org/abs/1710.01361}{{\ttfamily arXiv:1710.01361 [hep-th]}}.

\bibitem{Alday:2015ewa}
L.~F. Alday and A.~Zhiboedov, ``{An Algebraic Approach to the Analytic
  Bootstrap},'' \href{http://dx.doi.org/10.1007/JHEP04(2017)157}{{\em JHEP}
  {\bfseries 04} (2017) 157},
\href{http://arxiv.org/abs/1510.08091}{{\ttfamily arXiv:1510.08091 [hep-th]}}.

\bibitem{Cornagliotto:2017snu}
M.~Cornagliotto, M.~Lemos, and P.~Liendo, ``{Bootstrapping the $(A_1,A_2)$
  Argyres-Douglas theory},''
  \href{http://dx.doi.org/10.1007/JHEP03(2018)033}{{\em JHEP} {\bfseries 03}
  (2018) 033},
\href{http://arxiv.org/abs/1711.00016}{{\ttfamily arXiv:1711.00016 [hep-th]}}.

\bibitem{Heemskerk:2010ty}
I.~Heemskerk and J.~Sully, ``{More Holography from Conformal Field Theory},''
  \href{http://dx.doi.org/10.1007/JHEP09(2010)099}{{\em JHEP} {\bfseries 1009}
  (2010) 099},
\href{http://arxiv.org/abs/1006.0976}{{\ttfamily arXiv:1006.0976 [hep-th]}}.

\bibitem{Afkhami-Jeddi:2016ntf}
N.~Afkhami-Jeddi, T.~Hartman, S.~Kundu, and A.~Tajdini, ``{Einstein gravity
  3-point functions from conformal field theory},''
\href{http://arxiv.org/abs/1610.09378}{{\ttfamily arXiv:1610.09378 [hep-th]}}.

\bibitem{KPZ2017}
M.~Kulaxizi, A.~Parnachev, and A.~Zhiboedov, ``{Bulk Phase Shift, CFT Regge
  Limit and Einstein Gravity},''
\href{http://arxiv.org/abs/1705.02934}{{\ttfamily arXiv:1705.02934 [hep-th]}}.

\bibitem{Li:2017lmh}
D.~Li, D.~Meltzer, and D.~Poland, ``{Conformal Bootstrap in the Regge Limit},''
\href{http://arxiv.org/abs/1705.03453}{{\ttfamily arXiv:1705.03453 [hep-th]}}.

\bibitem{Afkhami-Jeddi:2017rmx}
N.~Afkhami-Jeddi, T.~Hartman, S.~Kundu, and A.~Tajdini, ``{Shockwaves from the
  Operator Product Expansion},''
\href{http://arxiv.org/abs/1709.03597}{{\ttfamily arXiv:1709.03597 [hep-th]}}.

\bibitem{Amati:1987wq}
D.~Amati, M.~Ciafaloni, and G.~Veneziano, ``{Superstring Collisions at
  Planckian Energies},''
\href{http://dx.doi.org/10.1016/0370-2693(87)90346-7}{{\em Phys. Lett.}
  {\bfseries B197} (1987) 81}.

\bibitem{Amati:1987uf}
D.~Amati, M.~Ciafaloni, and G.~Veneziano, ``{Classical and Quantum Gravity
  Effects from Planckian Energy Superstring Collisions},''
\href{http://dx.doi.org/10.1142/S0217751X88000710}{{\em Int. J. Mod. Phys.}
  {\bfseries A3} (1988) 1615--1661}.

\bibitem{DAppollonio:2015fly}
G.~D'Appollonio, P.~Di~Vecchia, R.~Russo, and G.~Veneziano, ``{Regge behavior
  saves String Theory from causality violations},''
  \href{http://dx.doi.org/10.1007/JHEP05(2015)144}{{\em JHEP} {\bfseries 05}
  (2015) 144},
\href{http://arxiv.org/abs/1502.01254}{{\ttfamily arXiv:1502.01254 [hep-th]}}.

\bibitem{Karateev:2018oml}
D.~Karateev, P.~Kravchuk, and D.~Simmons-Duffin, ``{Harmonic Analysis and Mean
  Field Theory},''
\href{http://arxiv.org/abs/1809.05111}{{\ttfamily arXiv:1809.05111 [hep-th]}}.

\end{thebibliography}\endgroup

\end{document}